\documentclass[11pt,a4]{article}
\pdfoutput=1
\usepackage{graphicx,amssymb,amsmath,amsfonts}
\usepackage{hyperref}
\usepackage[margin=1.2in]{geometry}
\usepackage[noadjust]{cite}
\usepackage[utf8]{inputenc}

\def\be{\begin{equation}}
\def\ee{\end{equation}}

\numberwithin{equation}{section}

\def\bea{\begin{eqnarray}}
\def\eea{\end{eqnarray}}

\newcommand{\non}{\nonumber \\}
\newcommand{\CR}{\non\cr}

\newcommand{\alp}{\ensuremath{\alpha^\prime}}

\newcommand{\ssb}{s\bar{s}}
\newcommand{\ccb}{c\bar{c}}

\newcommand{\bbb}{b\bar{b}}
\newcommand{\MEV}{\text{ MeV}}
\newcommand{\GEVm}{\text{ GeV}^{-2}}
\newcommand{\GEV}{GeV\(^{-2}\)}

\newcommand{\jp}[2]{\ensuremath{\frac{#1}{2}^{#2}}}
\newcommand{\jph}[2]{\ensuremath{#1/2^{#2}}}

\newcommand{\plm}{\ensuremath{\pm}}

\newcommand{\pa}{\partial}

\begin{document}
\begin{titlepage}
\title{\textbf{The decay width of stringy hadrons}}

\author{\textbf{Jacob Sonnenschein} \\ \href{mailto:cobi@post.tau.ac.il}{cobi@post.tau.ac.il} \and \textbf{Dorin Weissman} \\ \href{mailto:dorinw@mail.tau.ac.il}{dorinw@mail.tau.ac.il}}

\date{\emph{The Raymond and Beverly Sackler School of Physics and Astronomy},\\
	\emph{Tel Aviv University, Ramat Aviv 69978, Israel} \\ \today}
	

\maketitle

\begin{abstract}In this paper we further develop a string model of hadrons by computing their strong decay widths and comparing them to experiment. The main decay mechanism is that of a string splitting into two strings. The corresponding total decay width behaves as $\Gamma=\frac\pi2 ATL$ where $T$ and $L$ are the tension and length of the string and $A$ is a dimensionless universal constant. We show that this result holds for a bosonic string not only in the critical dimension. The partial width of a given decay mode is given by $\Gamma_i/\Gamma=\Phi_i\exp(-2\pi Cm_{sep}^2/T)$ where $\Phi_i$ is a phase space factor, $m_{sep}$ is the mass of the "quark" and "antiquark" created at the splitting point, and $C$ is a dimensionless coefficient close to unity. Based on the spectra of hadrons we observe that their (modified) Regge trajectories are characterized by a negative intercept. This implies a repulsive Casimir force that gives the string a "zero point length".

We fit the theoretical decay width to experimental data for mesons on the trajectories of $\rho$, $\omega$, $\pi$, $\eta$, $K^*$, $\phi$, $D$, and $D^*_s$, and of the baryons $N$, $\Delta$, $\Lambda$, and $\Sigma$. We examine both the linearity in $L$ and the exponential suppression factor. The linearity was found to agree with the data well for mesons but less for baryons. The extracted coefficient for mesons $A=0.095\pm0.015$ is indeed quite universal. The exponential suppression was applied to both strong and radiative decays. We discuss the relation with string fragmentation and jet formation. We extract the quark-diquark structure of baryons from their decays. A stringy mechanism for Zweig suppressed decays of quarkonia is proposed and is shown to reproduce the decay width of $\Upsilon$ states. The dependence of the width on spin and flavor symmetry is discussed. We further apply this model to the decays of glueballs and exotic hadrons.
\end{abstract}

\end{titlepage}

\tableofcontents
\flushbottom
\section{Introduction} \label{sec:intro}
The spectra of hadrons, both mesons and baryons, fit nicely a stringy description. This is an observation that has already been made in the early days of string  theory \cite{Collins:book} for hadrons made out of light quarks. Recently it  was shown  to hold also for hadrons  made out of medium and heavy quarks \cite{Sonnenschein:2014jwa,Sonnenschein:2014bia}. Furthermore, the stringy nature of hadrons was proposed as an important tool in the search for glueballs \cite{Sonnenschein:2015zaa} and exotic hadrons like tetraquarks \cite{Sonnenschein:2016ibx}.   

In this paper we move on one step ahead and consider the strong decays of hadrons as stringy processes. The basic underlying assumption   is that  hadrons can be  described as various configurations of strings and hence their decays relate to the process of splitting of strings.  The strings can  be viewed as residing   in holographic backgrounds or be mapped to a four dimensional flat space time. The latter setup which is referred to as the HISH (Holography inspired stringy hadron) model is reviewed in  \cite{Sonnenschein:2016pim}. Essentially it describes hadrons as string with massive endpoints residing in four dimensional flat spacetime. In this paper we use both descriptions depending on which one is more convenient for a given issue. 

Using several arguments of the  string model we have concluded that the strong decay width of a hadron takes the form 
\be\label{Lindw}
\Gamma= \frac{\pi}{2}A T L(M, m_1,m_2,T)
\ee
  where  $T$ is the string tension and $L$ is the length of the string. The length can be expressed in terms of the mass of the hadron $M$, and $m_1$ and $m_2$, the two string endpoint masses. The constant $A$ is dimensionless, and it is equal to the asymptotic ratio of $\Gamma/M$ for large $L$.
The linearity with the length was first derived explicitly in \cite{Dai:1989cp} for a bosonic string in the critical dimension $D = 26$ . On the other hand according to \cite{Mitchell:1989uc,Wilkinson:1989tb,Mitchell:1988qe}, considering only the string transverse modes, the decay width behaves as $L^{-5/6}$ in $D = 4$. Using the treatment of the intercept in a non-critical dimension \cite{Hellerman:2013kba} we reconcile between the two results and show that in fact also in any non-critical dimension the linearity property holds. 	
	
To check the linearity with the length we write the latter in terms of the observable quantities $J$ and $M$, the angular momentum and mass of the hadron. The relation between $L$ and $J$ must also involve the quantum  intercept or what we refer  to as  the ``zero point length'' of the string. The latter follows from the fact that for all the trajectories   both of mesons and baryons the intercept is negative, where  the trajectories are defined through the relation of the mass squared $M^2$ and the orbital angular momentum, rather than the total angular momentum.  The fact that the intercept is negative  implies that the there is a repulsive Casimir force which serves to balance the string tension. Therefore there is a non-zero string length, and hence a finite positive mass and width, even when the string is not rotating.
	
	The partial  width to decay  into a particular channel $i$ is shown to be 
	\be
	\frac{\Gamma_i}{\Gamma} = \Phi_i \exp[{-2\pi C\frac{m_{sep}^2}{T}}]
		\ee
		where $\Phi_i$ is a phase space factor,  $m_{sep}$ is the mass of the ``quark''  and ``antiquark'' generated at the splitting point and $C$ is a dimensionless  coefficient that can be approximated as $C(T_{\text{eff}},L) \sim 1 + c_c\frac{M^2}{ T_{\text{eff}}}   +\sum_{j=1}^2 c_{m_j} \frac{m_j}{M}$. For a holographic background with small curvature   around the ``holographic wall'', and with small masses to the string endpoint particles $m_j$ $j=1,2$, the dimensionless coefficients $c_c$ and $c_{m_j}$ are small. This result is the holographic dual of  the Schwinger mechanism of pair creation in a color flux tube \cite{Casher:1978wy,Gurvich:1979nq}, and was first developed in \cite{Peeters:2005fq}. In this paper we first review that derivation, then generalize it  to  a holographic string in any (curved) confining background. Furthermore, we incorporate the impact of the  massive endpoints of the string.  The  process of multi-breaking and its relation to string fragmentation and jet formation is briefly addressed.
 

The mechanism of transforming  a single hadron initial state into two hadrons in the final state, which  is as mentioned above  basically the breaking of a string into two strings, is quite universal. It shows up in the decay of both mesons and baryons. We assume, based on our analysis of the baryon spectra \cite{Sonnenschein:2014bia}, that the structure of a baryon is that of a single string with a ``quark'' on one end and a baryonic vertex connected by two short strings forming a ``diquark'' at the other end. For certain baryons we do not know a priori which  pair of  quarks form the diquark. We show that  this information can be extracted from the decay processes. 

Glueballs, which in the HISH model \cite{Sonnenschein:2015zaa} are rotating folded closed strings, have several mechanisms of decay: (i) By crossing and splitting into two glueballs.
(ii) By crossing and a breakup of one of the two closed strings into one closed and one open, hence a glueball and a meson. (iii) A break up and attachment to a flavor brane to form a meson. (iv) A double breaking which leads to a two meson final state. We describe and discuss these modes of decay.

A special class of decays with typically narrow width is that of ``Zweig suppressed decays'' of quarkonia. The latter, due to kinematical constraints, may be unable to decay via breaking of the string. We suggest a mechanism  for the  decay of the stringy meson  by an ``annihilation'' of the two ends of the quarkonia string to form a closed string (glueball), which then breaks up into two open strings (mesons). We derive an suppression factor for this process which reads 
 \be \Gamma \propto \exp(-T_{Z}L^2/2) \label{eq:Zweig_Li}\ee
where $T_Z$ is the relevant tension (as discussed in section \ref{sec:decay_Zweig}).

We also argue that there is a class of exotic hadrons built of a string that connects a diquark and an anti-diquark so that by breaking the string the decay is predominantly into a baryon and anti-baryon. We propose that this mechanism can be useful in identifying exotic hadrons. 

In this paper we discuss the strong decays of hadrons. A natural question, of great phenomenological significance, is how to incorporate weak decays. This topic has not been investigated in this research work. However, it seems that it is related to the decay of the string endpoint ``quarks'' and not of the string  of the HISH model itself. We mention one mechanism that resembles this type of decay, and that is the break up of the vertical segments of the holographic strings. From the point of view of the four dimensional physics this looks like a decay of the quarks but whether or not it can really be associated with weak decay requires further investigation.

The fact that both the parent hadron and its decay products are on (massive modified) Regge trajectories, implies various constraints on the decay and in particular renders certain decay channels to be forbidden. For instance the S-wave decay between consecutive states in a Regge trajectory does not take place for any type of meson.

In holography flavor symmetry is associated with the corresponding local symmetry on probe flavor branes. In fact the latter is also a stringy effect since it is being created by the open strings that connect the flavor branes. Thus if the $u$ and $d$ flavor branes are coinciding in the holographic radial direction then there is a $U(2)$ symmetry of isospin in addition to a baryon number symmetry. If the branes are slightly separated isospin turns into an approximated symmetry. Thus in a very natural manner isospin constrains on decay processes show up also in the holographic stringy setup. We demonstrate this for both mesons and baryons. Having said that,  it is also known that there are isospin breaking effects associated with the strong interactions. The obvious case is that of the difference between a charged and a neutral hadron which follows partially from the fact that $m_u\neq m_d$. However, there are also strong decays that do not conserve isospin. In a stringy framework, this can be attributed to a string splitting with virtual pair creation. In certain cases we can also use an approximated $U(3)$ flavor symmetry. In addition there are constraints due to symmetry requirements on the total wave function of the product hadrons. For a meson the requirement of a Bose symmetry of the wave function holds for the stringy picture just as for the particle one. However, the totally antisymmetric wave function of the baryon, as realized for instance in the quark model, is not manifest for the stringy baryons since we take the latter to have a quark and a diquark on its ends.  

In this paper we place special emphasis on performing a comparison between the predictions of the theoretical model and experimental data. We have tried to fit, as many  as possible, measured decay widths using the stringy model. In the included fits, we check the linear dependence of the width on the string length for mesons  on the trajectories of $\rho/a,\  \omega/f,\ \pi/b,\ \eta/h,\  K^*,\  \phi,\  D,\  D^*$ and baryons   
on   $N,\  \Delta,\  \Lambda$ trajectories. We used $\chi^2$ tests for the goodness of the fits. The latter is not as nice as for the fits of the hadronic spectra but still we found reasonable agreement between the model and the measured values. For mesons the states that do not admit a good fit were  found to be, $\rho$, $h_1(1170)$, and $\pi_2(1670)$. They are all significantly wider than other states on their trajectories. The fits to the decay widths for the baryons are significantly worse, especially for the $N$ baryons. The decay widths of the excited $N$ baryons are much larger than one would expect. Their decay width seems closer to being  linear in $J$ or $M^2$  rather than linear in the length.
We found that for hadrons with ``short'' strings an improvement in fitting the total width is achieved by modifying \ref{Lindw} with a phase space factor of the dominant channel. This result is  probably a manifestation of the impact  of the string endpoint masses on the  string decay width of short strings.   

We also checked the exponential suppression for strong decays of certain mesons and nucleons and for the radiative decays of $J/\Psi$ and $\Upsilon$. Unfortunately, there is almost no exponential suppression factor for a creating of a light  quark pair so it is impossible to check it. For  the creation of  a strange pair it is $\approx$ 0.3--0.4, and we can check the ratio of creating a light versus strange pair for several decays.    

There is another type of experimental phenomenon for which the mechanism of string breaking  is relevant and this is string  fragmentation and the creation of jets, via multi-breaking of the string. It is well known that event generators such as Pythia incorporate the exponential suppression factors \cite{Sjostrand:2006za}.  Though  the exponential suppression   is  common to the  different multi-breaking mechanisms, there are also major differences.

We apply the suppression factor of Zweig  suppressed  decay   to  the  decays of the \(\Upsilon\) mesons, where
the three \(\Upsilon(nS)\) states that are below the threshold to decay by tearing the string and forming a pair into a pair of \(B\) mesons. The special property of the total decay width of the first three \(\Upsilon\) mesons is the  decrease with the mass. The next state is above threshold and is significantly wider. The experimental data is well fitted by this mechanism.  
As for the isospin breaking decays we consider the decay process of $D_s^*\rightarrow D_s + \pi_0$ and show that this can follow from a virtual strange anti-strange pair creation that accompanies the decay of $D_s^*$. We also briefly address the newly discovered \cite{Aaij:2017nav} five narrow resonances of $\Omega_c$ and their quark diquark structure. 

The study of the strong decay processes of hadrons has a long history and includes various different approaches. Here we mention a sample of relevant papers that include further references. Decays in the quark model were studied in \cite{Micu:1968mk,Rosner:1974sn,Koniuk:1979vy}, via pair creation in \cite{LeYaouanc:1972vsx}, and a breaking of a flux tube in \cite{Kokoski:1985is}. Furthermore, we have decays in the context of chiral Lagrangian \cite{Amundson:1992yp,Falk:1992cx}, in the the Skyrme model \cite{Dorey:1994fk}, and lattice QCD \cite{Aoki:2011yj}.

As was mentioned above the stringy description of hadrons dates back to the early days of  string theory. The reincarnation of the idea in the era  of the string/gauge duality was investigated in many papers, including \cite{Baker:2002km,Faddeev:2003aw,Schreiber:2004ie,Pons:2004dk,Meyer:2004jc,deTeramond:2005su,BoschiFilho:2005yh,Kirsch:2005uy,Cotrone:2005fr,Erdmenger:2006bg,Peeters:2007ab,Shifman:2007xn,Aharony:2009gg,Gursoy:2009jd,Armoni:2009zq,AliAkbari:2009pf,Imoto:2010ef,Aharony:2010db,Sadeghi:2011cf,Zahn:2013yma,Aharony:2013ipa,Braga:2014pxa,Ballon-Bayona:2014oma,Qian:2014rda,Dubovsky:2015zey,Rossi:2016szw,Caron-Huot:2016icg,Veneziano:2017cks} and references therein.

The paper is organized as follows. Section \ref{sec:HISH} is devoted to a brief review of the HISH model \cite{Sonnenschein:2016pim}. We spell out the basic ingredients of these models and present the stringy configurations associated with mesons, baryons and glueballs. Then in section \ref{sec:rotating_string}  we describe the physics of rotating strings. We start with the classical rotation of a string with massive particles on its endpoints. We then discuss how to introduce the quantum corrections to the classical motion. We emphasize the role of the repulsive Casimir that is responsible for the fact that the string does not shrink to zero size even without any rotation. Section \ref{sec:decaywidth} deals with the decay process of a long holographic string. We start with a review of the calculation of the 
decay of an open string in flat spacetime in critical dimension. Next we generalize this to the cases of a rotating string, an excited string and a string with  massive endpoints. We further discuss the decay width in non-critical dimensions and then we elaborate on what we refer as the decay of a stringy hadron.

In section \ref{sec:decay_exponent} we review the derivation of the exponential suppression of the width associated with the  mechanism of creating a ``quark antiquark'' pair  at the two endpoints that emerge  when a string is torn apart. We start in \ref{sec:decay_Schwinger} in a review of the field theory picture of pair creation in Schwinger mechanism. The suppression factor for stringy holographic hadrons is described in \ref{sec:decay_holoexp}. This includes the calculation of the probability to hit a flavor brane in flat spacetime in a discretised string bit model, and a continuous string. We then describe the curved spacetime and the string with massive endpoints corrections. We finally summarize the holographic suppression factor.

In subsection \ref{sec:decay_multi} multi string breaking and string fragmentation are discussed. The decays of baryons, glueballs and exotic hadrons are brought in sections \ref{sec:decay_baryons}, \ref{sec:decay_glueball}, \ref{sec:decayexotic} respectively. Zweig suppressed decay channels are described in section \ref{sec:decay_Zweig}. We end up this section with holographic stringy decays via breaking of the vertical segments of the stringy hadron in section \ref{sec:decay_vertical}, and section \ref{sec:decay_Regge}, where we determine certain constraints on possible decay modes bases on the fact that the decaying states are on Regge trajectories. This is done for the linear trajectories to those   with  first order massive corrections and for heavy quarks. Section \ref{sec:decay_spin} is devoted to the dependence of the decay processes on the spin and favor symmetry. In particular we discuss the spin structure of the stringy decays in \ref{sec:decay_spinc}. We describe the natural appearance of baryon number and isospin and $SU(3)$ flavor symmetry constraints on decays of holographic stringy mesons  and baryons. Finally we also address the general symmetry properties of the total wave functions.

Section \ref{sec:pheno} is devoted to the phenomenology of hadron decay width and in particular to confronting the theoretical results with the experimental data. First in subsection \ref{sec:fitting_model} we define the fitting model. This includes the relation of the string length to the phenomenological intercept, and the introduction of phase space and time dilation factors to the decay width. In \ref{sec:pheno_mesons} we present the fits to total decay width of mesons to see their linearity with the length. We also examine the length dependence of the Zweig suppressed decays. In \ref{sec:pheno_baryons} we present the fits of the decay of baryons and the lessons about the quark-diquark structure one extracts from the baryons' decays. Section \ref{sec:pheno_lambdas} is devoted to a comparison of the   exponential suppression of pair creation with data points of decays of hadrons both strong and radiative decays. The implication of the decay mechanism on the search of exotic hadrons is presented in section \ref{sec:pheno_exotic} for glueballs as well as for tetraquarks. We summarize the result and list several open questions in section \ref{sec:summary}. In appendix \ref{app:spectrum} we have listed the different hadrons used in the fits of section \ref{sec:pheno} and their calculated masses and widths in the HISH model.

\section {A brief review of  the HISH model \label{sec:HISH}}
The holographic duality is an equivalence between certain bulk string theories and boundary field theories. The original duality was between the ${\cal N}=4$ SYM theory and string theory in $AdS_5\times S^5$. Obviously the  ${\cal N}=4$ theory is not the right framework to describe hadrons that resemble those found in nature. Instead we need a stringy dual of a four dimensional gauge dynamical system which is  non-supersymmetric, non-conformal, and confining. The  two main requirements on the desired string background is that it admits confining Wilson lines, and that it is dual to a boundary that  includes a matter sector, which is invariant under a chiral flavor symmetry that is spontaneously broken. There are by now  several ways to get a string background which is dual to a confining boundary field theory. For a review paper and a list of relevant references see for instance \cite{Aharony:2002up}.

Practically most of the applications of holography of both conformal and non-conformal  backgrounds are based on relating bulk fields (not strings) and operators on the dual  boundary field theory. This is based on the usual limit of $\alp\rightarrow 0$  with which we go, for instance, from a closed string theory to a theory of gravity.

However, to describe realistic hadrons we need strings rather than bulk fields since in nature the string tension, which is related  to $\alp$ via $T=(2\pi \alp)^{-1}$, is not very large. In gauge dynamical terms the IR region is characterized by $\lambda= g^2 N_c$ of order one rather than very large.

It is well known that in holography there is a wide sector of gauge invariant  physical observables  which cannot be faithfully described by bulk fields but rather require dual stringy phenomena. This is the case for  Wilson, 't Hooft, and  Polyakov lines.

In the  holography inspired stringy hadron (HISH) model \cite{Sonnenschein:2016pim}  we argue that in fact also  the  description of the spectra, decays and other properties of hadrons - mesons, baryons, glueballs and exotics - should be recast as a description in terms of holographic stringy configurations only, and not fields. The major argument against describing the hadron spectra in terms of fluctuations of fields, like bulk fields or modes on probe flavor branes, is that they generically do not reproduce the Regge trajectories in the spectra, namely, the (almost) linear relation between $M^2$ and the angular momentum $J$. Moreover, for top-down models with the assignment of mesons as fluctuations of flavor branes  one can get mesons only with $J=0$ or $J=1$. Higher $J$ states will have to be related to strings, but then there is a big gap of order $\lambda$ (or some fractional power of $\lambda$ depending on the model) between the low and  high $J$ mesons. Similarly  the attempts to get the observed linearity between $M^2$ and the excitation number $n$ are problematic whereas for strings it is a natural property.

\begin{figure}[t!] \centering
\includegraphics[width= 0.9\textwidth]{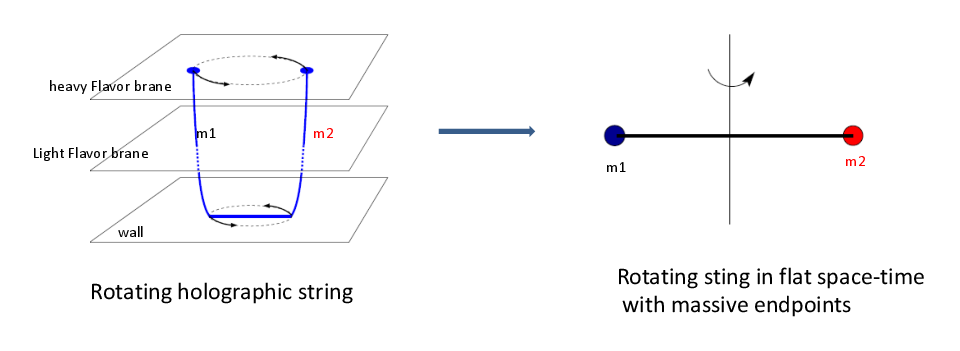}
					\caption{\label{fig:mapholflat} \textbf{Left:} Rotating holographic open string. \textbf{Right:} The corresponding open string with massive endpoints in flat spacetime. In this case $m_1=m_2$.\label{hishmap}}
		\end{figure}

The construction of the HISH model is based on the following steps. (i) Analyzing string configurations in holographic string models that correspond to hadrons. (ii) Devising a  transition from the holographic regime of large $N_c$ and large $\lambda$ to the real world that bypasses formally taking the limits of $\frac{1}{N_c}$ and $\frac{1}{\lambda}$ expansions. (iii) Proposing a model of stringy hadrons in four flat dimensions that is inspired by the corresponding holographic strings. (iv) Confronting the outcome of the models with the experimental data (as was done in \cite{Sonnenschein:2014jwa,Sonnenschein:2014bia,Sonnenschein:2015zaa}).

Confining holographic models are characterized  by a ``wall'' that truncates in one way or another the range of the holographic radial direction. A common feature to all the holographic stringy hadrons is that there is a segment of the string that stretches along a constant radial coordinate in the vicinity of the ``wall'', as in figure \ref{hishmap}. For a stringy glueball it is the whole folded closed string that rotates there and for an open string it is part of the string, the horizontal segment, that connects with vertical segments either to the boundary for a Wilson line or to flavor branes for a meson or baryon. This fact that the classical solutions of the flatly  rotating strings reside at fixed radial direction is a main rationale behind the map between rotating strings in curved spacetime and rotating strings in flat spacetime described in figure \ref{fig:mapholflat}. 

A key ingredient of the map is the ``string endpoint mass'', $m_{sep}$, that provides in the flat spacetime description the dual of the vertical string segments. It is important to note that this mass is neither the QCD mass parameter (current quark mass) nor the constituent quark mass, and that the massive endpoint as a map of an exactly vertical segment is an approximation that is more accurate the longer the horizontal string.

The stringy picture of mesons has been thoroughly investigated in the past (see \cite{Collins:book} and references therein). One can find also that in recent years there have been attempts to describe hadrons in terms of strings. Papers on the subject that have certain overlap with our approach are \cite{Baker:2002km,Schreiber:2004ie,Hellerman:2013kba,Zahn:2013yma}. A somewhat different approach to the stringy nature of QCD is the approach of low-energy effective theory on long strings reviewed in \cite{Aharony:2013ipa}. 

\begin{figure}[t!] \centering
					\includegraphics[width=.90\textwidth]{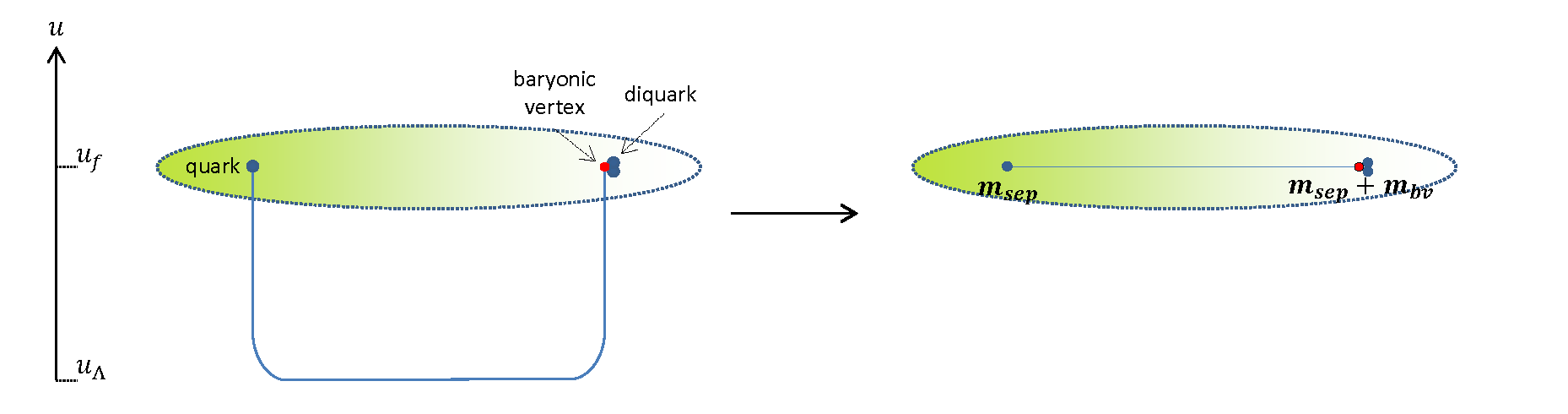}
					\caption{\label{holtoflat2}  The holographic setup (for \(N_c=3\)) of a quark and a diquark is mapped to a similar configuration in flat spacetime. The vertical segments of the holographic string are mapped into masses of the endpoints. The mass of the baryonic vertex also contributes to the endpoint mass.}
\end{figure}

The HISH model describes hadrons in terms of the following basic ingredients:
\begin{itemize}
\item
Open strings which are characterized by a tension $T$, or equivalently a slope $\alp=(2\pi T)^{-1}$). The open string generically has a given energy/mass and angular momentum associated with its  rotation. The latter gets contribution from the classical configuration and in addition there is also a quantum contribution from the intercept $a$. An essential property of the HISH intercept is that it must always be negative, $a<0$. When considering trajectories of the orbital angular momentum and $M^2$ it is an experimental fact that all the trajectories of hadrons are characterized by negative intercept. Otherwise, the ground state would be a tachyon. This, as will be discussed in section \ref{sec:casimir_force}, is what is behind the repulsive Casimir force that guarantees that even a non-rotating stringy hadron has a finite length. Any hadronic open string can be in its ground state or in a stringy  excited state. The latter are determined by the quantization of the string with the appropriate boundary conditions. 

\item
Massive particles - or ``quarks'' - attached to the ends of the open strings which can have four\footnote{There is no a priori reasoning behind assuming \(m_u = m_d\) in the HISH model, but the difference between the two mass is too small to be relevant (or measurable) in the current work.} different values $m_{u/d}$, $m_s$, $m_c$, $m_b$. The latter are determined by fitting the experimental spectra of hadrons. These particles naturally contribute to the energy and angular momentum of the hadron of which they are part. Moreover, in the HISH the endpoint particles of the string can carry electric charge, flavor charges and spin.
These properties affect the value of the intercept as is reflected by the differences of the values of the intercept obtained for trajectories of hadrons with different quark content and spin.

\item
``Baryonic vertices'' (BV) which are connected to a net number of \(N_c=3\) strings. In holography the BV is built from a $D_p$ brane wrapping a \(p\)-cycle connected to flavor branes by $N_c$ strings. A priori there could be different metastable configurations
of the $N_c$ strings. In the HISH we take that the preferable string configuration is that of a single long string and two very short ones. Since the endpoints of the two short strings are one next to the other we can consider them as forming a diquark. There are two arguments in favor of this string setup. It was shown that the Y-shape configuration, which is the most symmetric form when $N_c=3$, is in fact unstable and an introduction of a small perturbation to a rotating Y-shaped string solution will cause it to evolve into the string between quark and diquark. Secondly, had the Y-shape string been stable, the baryon trajectory slope $\alp_B$ should have been $\alp_B= (2/3) \alp_m$, where $\alp_m$ is that of a meson. However as was shown in \cite{Sonnenschein:2014bia} the slope of the baryonic trajectory is within $5\%$ the same as that of a meson trajectory. 

The setup of a holographic baryon and its map to HISH model are depicted in figure \ref{holtoflat2}. We emphasize that unlike other models which have quarks and diquarks as elementary particles (such as \cite{Friedmann:2009mx}), in the HISH model the diquark is always attached to a baryonic vertex. A BV can be also connected to a combination of quarks and anti-baryonic vertices as will be discussed in section \ref{sec:decayexotic} \cite{Sonnenschein:2016ibx}.

\item Closed strings which have an effective tension that is twice the tension of the open string ($\alp_{closed}= \frac12 \alp_{open}$). They can have non-trivial angular momentum by taking a configuration of a folded rotating string. The excitation number of a closed string is necessarily even, and so is the angular momentum (on the leading Regge trajectory). The close string intercept, the quantum contribution to the angular momentum, is also twice that of an open string ($a_{closed}=  2 a_{open}$).
	\end{itemize}
	
Hadrons, namely mesons, baryons and glueballs are being constructed in the simplest manner from the HISH building blocks.
 \begin{itemize}
\item
A single open string attached to two massive endpoint particles corresponds to a meson.
\item
A single string that  connects on one end to a quark and on the other hand a baryonic vertex with a diquark attached to it is the HISH description of a baryon.
\item
A single closed string is a glueball.
\item
A  single string connecting  a baryonic vertex with a diquark on one end and an anti-baryonic vertex and anti-diquark, is a stringy description of a tetraquark which may be realized in nature.
\end{itemize}

\section{Rotating strings with massive endpoints} \label{sec:rotating_string}
As discussed in the previous section the basic ansatz of the HISH models is that hadrons are described by rotating and excited strings with massive particles on their ends in flat four dimensional spacetime. We now briefly review the properties of such strings. 
\subsection{The classical  rotating string } \label{sec:rotating_string_cl}
The HISH string with massive endpoints is described by the Nambu-Goto action for the string in four dimensional flat spacetime,
\be S_{NG} = -T\int d\tau d\sigma \sqrt{-\det h_{\alpha\beta}} \ee
with
\be h_{\alpha\beta} \equiv \eta_{\mu\nu}\pa_\alpha X^\mu \pa_\beta X^\nu\,, \ee
and the added massive point particle action at the endpoints \(\sigma=\pm\ell\),
\be S_{pp}\label{particle} = -m\int d\tau \sqrt{-\dot{X}^2}\,, \ee
where \(\dot X^\mu = \partial_\tau X^\mu\).
Here we assume that the masses of the two endpoints are the same.
The rotating configuration can be written as
\be X^0 = \tau\,,\qquad X^1 = \sigma \cos(\omega\tau)\,,\qquad X^2 = \sigma\sin(\omega\tau)\,, \label{eq:rotating_solution}\ee
and it must satisfy the boundary condition at \(\sigma = \pm\ell\)
\be \frac{T}{\gamma} = \frac{\gamma m \beta^2}{\ell} \label{eq:boundary_cl}\ee
where \(\beta \equiv \omega \ell\) is the endpoint velocity and \(\gamma = (1-\beta^2)^{-1/2}\) the usual relativistic \(\gamma\) factor.
The string length in target space is
\be L = \int_{-\ell}^{\ell} d\sigma \lvert \frac{dX^i}{d\sigma} \rvert = 2\ell \ee
Using the previous equation it can expressed as a function of \(m\), \(T\), and the velocity \(\beta\) as
\be L = \frac{2m \gamma^2 \beta^2}{T}\,. \label{eq:L_cl}\ee
The energy and angular momentum are functions of \(m\), \(T\), \(\beta\), and \(L\).
	\be E = 2\gamma m + TL\frac{\arcsin{\beta}}{\beta}\,, \ee
	\be J = \gamma m \beta L + \frac14T L^2\left(\frac{\arcsin\beta-\beta\sqrt{1-\beta^2}}{\beta^2}\right)\,.\ee
The last three equations, for the energy, angular momentum, and length as functions of \((T,m,\beta)\) define the classical relation between \(J\) and \(E\), or the Regge trajectory of the string with massive endpoints. The linear trajectory \(J = E^2/2\pi T\) can be obtained by taking the massless limit \(\gamma m \to 0\), \(\beta \to 1\). With endpoint masses, we can apply this relation to much of the observed hadronic spectrum \cite{Sonnenschein:2014jwa,Sonnenschein:2014bia}.

We can write approximate relations between the length and \(E\) or \(J\) in the two opposing limits, \(\beta \ll 1\) and \(\beta \approx 1\). In the ultra-relativistic case we have,
\be L = \frac{2 E}{\pi T} - \frac{m}{T} - \frac{4}{3\sqrt\pi}\frac{m^{3/2}E^{-1/2}}{T} + \mathcal O(E^{-3/2})\,,\label{non-rel} \ee
\be L = \frac{2 \sqrt{\frac{2}{\pi }}}{\sqrt{T}}\sqrt J - \frac{m}{T} - \frac{2 \left(\frac{2}{\pi }\right)^{3/4} m^{3/2}}{3 T^{5/4}} J^{-1/4} + \mathcal O(J^{-1/2})\,. \ee
In the non-relativistic case, for small \(J\) and \(E-2m\ll 2m\),
\be L = \frac{2}{3T}(E-2m) - \frac{1}{162 m T}(E-2m)^2 + \ldots \ee
\be L = (\frac{2}{mT})^{1/3}J^{2/3}-\frac19 2^{2/3}(\frac{T}{m^5})^{1/3} J^{4/3} + \ldots \,. \ee
The quantity we are ultimately interested in here is the decay width, which will be proportional to the string length, so the equations above also show the dependence of the width on mass and angular momentum, up to a constant - and quantum corrections, which we move on to describe in the following subsection. But first, we write the general classical expressions in the case the string is not symmetric.

\subsubsection{Generalization to the asymmetric case} \label{sec:rotating_string_asym}
When we have two different endpoint masses, the length can be written as
\be L = \ell_1+\ell_2\,,\ee
where the length \(\ell_i\) is the distance of the mass \(\ell_i\) from the center of mass. The two boundary conditions are then
\be \ell_i = \frac{m_i\gamma_i^2\beta_i^2}{T}\,. \ee
The angular momentum and energy are
\be E = \sum_{i=1,2}\left(\gamma_i m_i + T\ell_i\frac{\arcsin{\beta_i}}{\beta_i}\right)\,, \ee
\be J = \sum_{i=1,2}\left[\gamma_i m_i \beta_i \ell_i + \frac12T\ell_i^2\left(\arcsin\beta_i-\beta_i\sqrt{1-\beta_i^2}\right)\right]\,.\ee
The endpoint velocities \(\beta_i\) are related to each other from the condition that the angular velocity is the same for both arms of the string, implying
\be \omega = \frac{\beta_1}{\ell_1} = \frac{\beta_2}{\ell_2}\,.\ee
Using this equation, the energy, length, and angular momentum can all be related through a single parameter.

\subsection{Quantum corrections to the rotating string} \label{sec:quantum_string}
For the relativistic string without massive endpoints we have the well known linearity of \(E^2\) in \(J\),
\be J = \alp E^2\,, \label{eq:JE_classical} \ee
with the slope \(\alp \equiv (2\pi T)^{-1}\). This can be obtained from the rotating solution described above (in eq. \ref{eq:rotating_solution}). When the endpoints are massless, the boundary condition is that the endpoints move at the speed of light, \(\beta = 1\). The energy and angular momentum are then functions of the tension \(T\) and the string length \(L\),
\be\label{classical} E = \frac{\pi}{2} T L\,,\qquad J = \frac{\pi}{8}TL^2\,.\ee
Next we briefly summarize the quantum correction to the classical trajectory. We start with the well known string with no massive endpoints and then the case of a string with massive endpoints.
\subsubsection{The quantum trajectory of a string with no massive endpoints} 
The quantum trajectory follows from the quantization of the Nambu-Goto action of the  open string with Dirichlet boundary conditions that  was performed in \cite{Arvis:1983fp} and yielded the following expression for the energy of the string
\be\label{Arvis}
E_{stat} = \sqrt{ (TL)^2 + 2\pi T \left ( n- \frac{D-2}{24}\right )}
\ee
If instead of the static string we take the rotating one, we can use the classical solution \ref{classical} to verify that the correct transformation is by replacing $TL\rightarrow \sqrt{2\pi TJ}$, so that we find for the rotating case
\be
E_{rot} = \sqrt{ 2\pi T  \left(J+  n- \frac{D-2}{24}\right )}\,,
\ee
which can be expressed as the well known  quantum Regge trajectory 
\be\label{Retra}
   n+J =\alpha' M^2 + a
   \ee
The intercept $a$ which was taken here to be $a=\frac{D-2}{24}$ is defined by
\be\label{interceptnomass} 
a\equiv -\frac{D-2}{2\pi}\sum_{n=1}^{\infty} \omega_n = -\frac{ (D-2)}{2}\sum_{n=1}^{\infty} n= \frac{(D-2)}{24} 
\ee
where $\omega_n$ are the eigenfrequencies of the oscillations of the string in the $(D-2)$ transverse directions given by $\omega_n=\pi n$, and we have performed the zeta function regularization of the sum to get $\sum_n n=-\frac{1}{12}$. A different definition of the intercept can be made via the Casimir energy which reads 
\be
E_{\text{Casimir}}\equiv  -a\frac{\pi}{L}
\ee
In spite of the fact that \ref{interceptnomass} is written for any $D$ spacetime dimensions, it is true only in the critical dimension $D=26$ for which $a=1$. In any $D$ non-critical dimensions, one has to take into account also  the Liouville mode.
 This can be done by adding to the string action a  non-critical term  as was proposed by  Polchinski and Strominger  in \cite{Polchinski:1991ax}.
In the orthogonal gauge 
it reads
\be\label{PolStr} {\cal S}_{ps} =
\frac{26-D}{24\pi}\int d\tau \int_{-\delta}^\delta d\sigma
\frac{(\pa_+^2X\cdot \pa_-X)(\pa_-^2X\cdot \pa_+X)}{(\pa_+X\cdot
\pa_-X)^2}= \ee
where the derivatives are done with respect to the coordinates \(\sigma^\pm = \sigma \pm \tau\), and we have denoted the range of $\sigma$ to be $-\delta\leq \sigma\leq \delta$. For the case of no massive endpoints $\delta=\frac{\pi}{2}$.
This term is divergent for the classical rotating string configuration and thus has to be regularized and renormalized. This was done in   \cite{Hellerman:2013kba}. The outcome of that analysis was that the dependence on \(D\) cancels out between the two contributions to the intercept, as
\be
a= \frac{D-2}{24} + \frac{26-D}{24} = 1\,,
\ee
where the first term is the contribution of the ordinary transverse modes and the second is from the Liouville mode. 

\subsubsection{ The quantum trajectory  of a string with massive endpoints} \label{sec:quantum_massive}
As we have seen above already the classical trajectory  for the massive case is quite different from that of the massless case. To decipher the quantum picture  of the massive case we first need to determine the intercept for this case. In the massive case there are  fluctuations in directions transverse to the plane of rotation  and in addition there is also a planar  mode. The eigenfrequencies of the transverse modes are given by the transcendental equation 
\be\label{transcendental1}
\tan( \omega_n) =\frac{ 2 q \omega_n}{q^2 \omega_n^2 -1}\,,
\ee
where $q\equiv \frac{m}{TL}$. In section \ref{sec:decay_exp_mass} we will further discussed this condition, and in particular the effect of the mass on the frequency of the first excited states is drawn in figure \ref{massiveendpointsupp}. In the two limits of two massless endpoints \(q = 0\), or two infinitely heavy endpoints, the Dirichlet case of $q\rightarrow\infty$, then $\omega_n=n\pi$. In \cite{Lambiase:1995st} it was found that, for the string with two identical masses on its ends, the contribution of a single mode in a direction transverse to the plane of rotation to the intercept is given by
\be
\hat a(q) = \frac{1}{2\pi^2}\int_0^\infty dz \log\left[1-e^{-2 z}\left (\frac{q-z}{q+z}\right )^2\right ]\,.
\ee
This result is obtained after converting the infinite sum over the eigenfrequencies into a contour integral, and renormalizing the result by subtracting the contribution from an infinitely long string. It is easy to check that for the case of $q=0$ or $q=\infty$ indeed $\hat a = \frac{1}{24}$.

\begin{figure}[t!h] \centering
					\includegraphics[width=.48\textwidth]{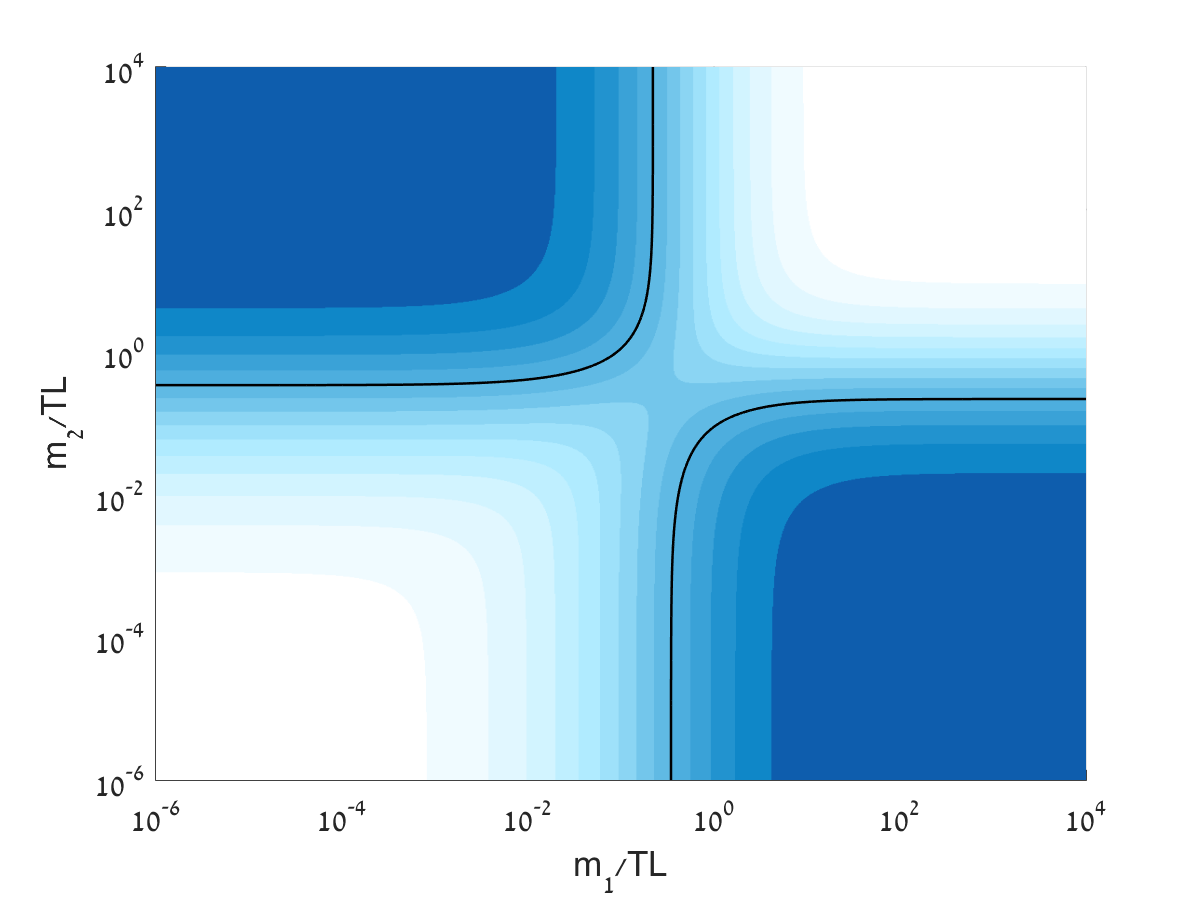}
					\includegraphics[width=.48\textwidth]{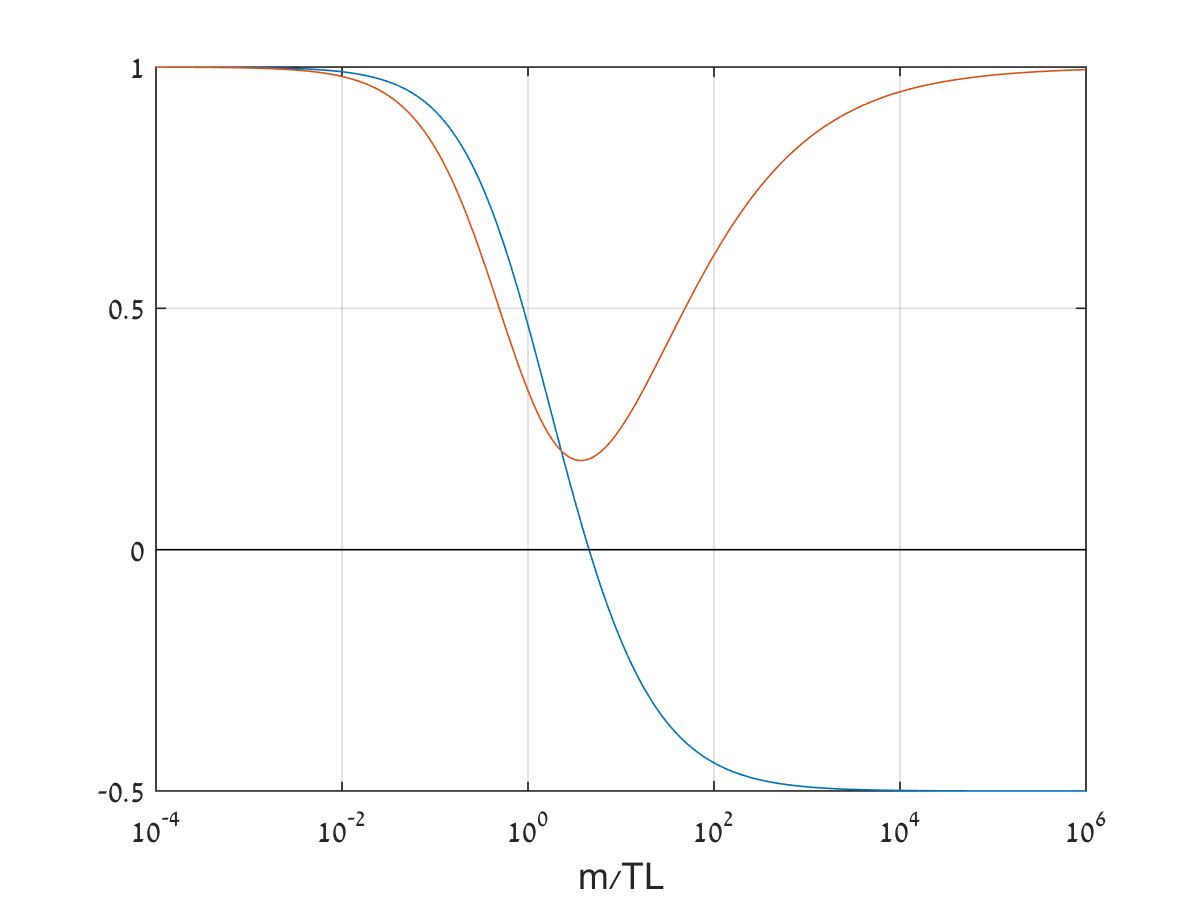}
					\caption{\label{etaq} The transverse coordinates' contribution to the intercept for string with two massive endpoints as a function of the masses. \textbf{Left:} As a function of both masses - in the lightest area \(0.9\leq \hat a \leq 1\), while in the darkest \(-0.5\leq \hat a \leq -0.4\). The black contour line is \(\hat a = 0\). \textbf{Right:} As a function of a single mass for the symmetric case \(m_1 = m_2 = m\) (red line), and the asymmetric case \(m_1 = 0\), \(m_2 = m\) (blue line).}
\end{figure}

The generalization of this result to the case of two different masses is simply
\be
\hat a(q_1,q_2) = \frac{1}{2\pi^2}\int_0^\infty dz \log\left[1-e^{-2 z}\left (\frac{q_1-z}{q_1+z}\right)\left(\frac{q_2-z}{q_2+z}\right)\right]\,.
\ee
In figure \ref{etaq} we draw the function \(\hat a(q_1,q_2)\). One can see that for the asymmetric case the result can change sign \cite{Kleinert:1996mv}, unlike the symmetric case where it is always positive. For $D=4$ there is one direction which is transverse to the plane of rotation and one ``planar direction''. The eigenfrequencies of the planar mode were analyzed in \cite{Zahn:2013yma} but the determination of the corresponding contribution to the intercept has not been carried out yet.\footnote{This analysis will be part of a future publication \cite{InterceptPaper}.}
This is also the case about the Liouville mode in the presence of boundary conditions that follow from the existence of massive endpoints. Furthermore, as we have learned from fitting the spectra of mesons and baryons \cite{Sonnenschein:2014jwa,Sonnenschein:2014bia}, it turns out that (i) When considering trajectories of the orbital angular momentum (not the total angular momentum that includes also the spin), one finds that the intercept is always negative. As will be discussed in the next subsection this is very crucial for the hadronic string. (ii) The intercept is also a function of the spin and isospin of the corresponding hadron. At present we do not have a theoretical model that can fully account for these two properties. We therefore, in section \ref{sec:pheno}, will base  our phenomenology analysis on using the different experimental values of the intercept as were determined from the spectra of the hadrons \cite{Sonnenschein:2014jwa,Sonnenschein:2014bia}. 

Next we have to determine what is the relation between the quantum trajectories and the intercept. In \cite{PandoZayas:2003yb} the fluctuations  along all bosonic directions  were inserted into the  expressions of the Noether charges associated with the rotation in space and translation in time, namely $J$ and $E$ and then considering the difference $J-LE$ it was shown that 
\be\label{deltaa}
L \delta E = -2 \int d\sigma \langle H_{ws} \rangle = -2a  \rightarrow  \qquad \delta(E^2)= -\frac{a}{\alp}
\ee
This was done for the case of strings with no massive endpoints.  For that case these relations can be also put in the trajectories as  $\delta J = a$ and $\delta E = 0$. This is does not mean that there is no quantum correction to the energy. It means that the relation between $J$ and the full quantum corrected energy can be expressed in this form. 

For the massive case as discussed above the eigenfrequencies are different and in addition there are contributions to both the energy and angular momentum form the boundary terms. Moreover as will see in the next subsection the intercept enters the boundary equation of motion itself in the form  of a Casimir force. In our work on the spectra \cite{Sonnenschein:2014jwa,Sonnenschein:2014bia} we found that also for the stringy hadrons, which all have non-zero masses on their endpoints, using \ref{deltaa}, even though theoretically not fully justified, yielded very nice fits. We will therefore use this approach also for the decay (see section \ref{sec:model_intercept}).

\subsection{Repulsive Casimir force} \label{sec:casimir_force}
Let us now consider the quantum correction to the classical length of the string. From eq. \ref{deltaa} we can see that the intercept corrects the energy by adding the Casimir term \(E_C = -2a/L\). The modification of the length can be traced back to a corresponding Casimir force,
\be F_C = -\frac{\pa}{\pa L}E_C = -2a/L^2\,, \ee
which is added to the classical boundary condition of eq. \ref{eq:boundary_cl}. Thus we obtain the boundary equation
\be\label{repCasf} \frac{T}{\gamma} = \frac{2\gamma m \beta^2}{L} - \frac{2a}{L^2} \ee
It is worth mentioning that this is an approximated form of the full quantum corrected boundary equation of motion. 
One property that follows immediately from this equation is that for $a<0$, which as mentioned above must be the case for all hadrons, the endpoints of the string feel a repulsive Casimir force. This force will prevent the string from collapsing to zero size even for the case of vanishing orbital angular momentum.

Solving the quadratic equation, the length as a function of \(\beta\) and the parameters \((T,m,a)\) is
\be L=\frac{\gamma^2 m \beta^2}{T}\left(1+\sqrt{1-\frac{2aT}{\gamma^3m^2\beta^4}}\right) \ee
The string obtains a ``zero point length'' at \(\beta = 0\), given by
\be L_0^2 = -\frac{2a}{T}\,.\ee

In the non-symmetric case, with two different endpoint masses \(m_1\neq m_2\), we have two different equations at the two endpoints:
\be \frac{T}{\gamma_i} = \frac{\gamma_i m_i \beta_i^2}{\ell_i} - \frac{2a}{L^2} \,.\ee
It is important to note that the first term on the RHS is the centrifugal force, and it depends on the radius of rotation \(\ell_i\), while the second term comes from the whole string, whose length is \(L = \ell_1+\ell_2\). The velocities are related by \(\omega_1 = \omega_2\), or
\be \frac{\beta_1}{\ell_1} = \frac{\beta_2}{\ell_2}\,.\ee
The solution can be obtained by plugging in
\be \ell_2 = \frac{\beta_2}{\beta_1}\ell_1 \ee
into the boundary condition for \(m_1\):
\be \frac{T}{\gamma_1} = \frac{\gamma_1 m_1 \beta_1^2}{\ell_1} - \frac{2a}{\ell_1^2(1+\beta_2/\beta_1)^2} \,.\ee
Then solving a quadratic equation for \(\ell_1\) as function of \((\beta_1,\beta_2)\),
\be \frac{T}{\gamma_1}\ell_1^2 - \gamma_1 m_1 \beta_1^2\ell_1 + \frac{2a}{(1+\beta_2/\beta_1)^2} = 0\,.\ee
whose solution is given by
\be \ell_1 = \frac{\gamma_1}{2T}\left(\gamma_1 m_1 \beta_1^2+\sqrt{(\gamma_1 m_1 \beta_1^2)^2-\frac{8aT}{\gamma_1(1+\beta_2/\beta_1)^2}}\right)\,. \ee
The last equation to solve is the boundary condition for \(m_2\), which can be written as
\be \frac{T}{\gamma_2} = \frac{\gamma_2 m_2 \beta_1\beta_2}{\ell_1} - \frac{2a}{\ell_1^2(1+\beta_2/\beta_1)^2} \,.\ee
Using the expression above for \(\ell_1\), we get an equation we can solve numerically to obtain \(\beta_2\) as a function of \(\beta_1\) (and, of course \(T\), \(m_1\), \(m_2\), and \(a\)). Then \(L\), as well as \(E\) and \(J\), are all reduced to functions of the parameter \(\beta_1\), like in the symmetric case.

The quantum modification of the string length and the phenomenology behind it will be further discussed in section \ref{sec:model_intercept}.

\section{The decays of long holographic string} \label{sec:decaywidth}
As was described in the  section \ref{sec:HISH}, the  holographic hadronic string, especially for large angular momentum, has the structure of a flat string that stretches close to the ``wall'' and  is connected with   approximately vertical strings  to flavor branes. In principle the string can break apart at any point   both in the flat horizontal part as well as in the vertical segments. We consider  here first the former option and in subsection \ref{sec:decay_vertical} we will address also the latter. In the holographic setup the two endpoints that result from the breaking cannot be in thin air but rather must be attached to flavor branes. Therefore the breakup mechanism can take place only when the string fluctuates, reaches a flavor brane and there breaks  apart.  The  fluctuations and the probability to reach a flavor brane will be addressed in section \ref{sec:decay_exponent}.
 This type of holographic string  breakup can be mapped to  the HISH model, namely, to a string  with  massive particles at its ends    that resides in non-critical four dimensions. For that one has to postulate that the probability for a split of the HISH includes also the suppression factor which in the holographic picture is associated with the probability of reaching a flavor brane due to fluctuations.

We consider the mechanism of the split of a string into two strings in several setups that lead to the one of the hadronic string. In the next subsection we discuss the decay of an open string with no massive endpoints in flat critical dimensions. We then generalize to non-critical dimensions. Next we discuss the case of a string with massive endpoints, before finally discussing the decay of the hadronic string.
  
\subsection{The decay of an open string in flat  spacetime in  critical  dimensions}
\begin{figure}[ht!]
			\centering
				\includegraphics[width=0.66\textwidth]{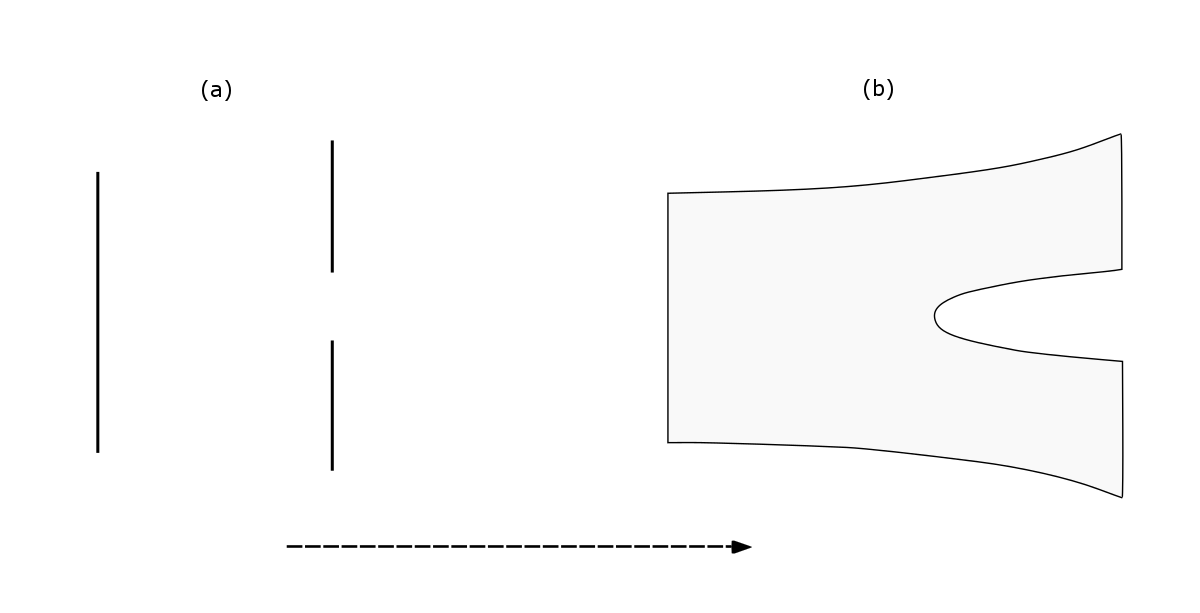}
				\caption{An open string breaks into two open strings. (a) Snapshots of before and after the split. (b) The worldsheet diagram of the split. The arrow represents the time direction. \label{basicsplit}}
	\end{figure}

We  discuss in this subsection the probability that an open string in flat critical dimension  will break apart. This process is depicted in  figure \ref{basicsplit}.

 Since   the string can split at any point along its spatial direction,  it is intuitively  expected  that the decay
rate $\Gamma$  will be proportional to the length of the string $L$, 
 \be \Gamma\sim L.\ee 
 This property, as will be shown next, is very non-trivial to derive explicitly and in fact it is only an approximated result valid for long strings. 

The  decay rate of an open string into two strings   was analyzed  by many  authors. The first treatments of the problem were given in   \cite{Dai:1989cp,Mitchell:1989uc,Wilkinson:1989tb,Mitchell:1988qe,Okada:1989sd}. Since then, it was further addressed in the context of the decay of mesons in  \cite{deVega:1995yj,Manes:2001cs,Iengo:2002tf,deVega:2003hh,Chialva:2003hg,Chialva:2004xm,Chen:2005ra,Cotrone:2005fr,Iengo:2006gm,Bigazzi:2006jt,Gutperle:2006nb,Bigazzi:2007qa,Sadeghi:2011cf,Sen:2016gqt}.

\begin{figure}[ht!]
			\centering
				\includegraphics[width=0.68\textwidth]{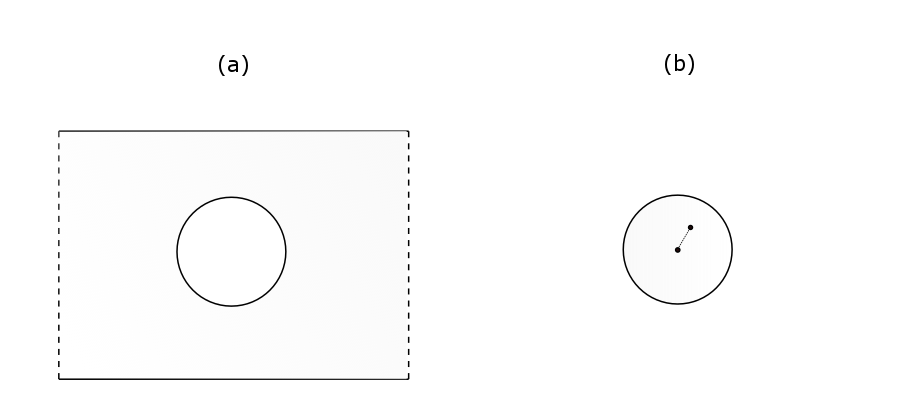}
				\caption{(a) Worldsheet of splitting and rejoining of a winding state. The dashed lines are periodically identified.  (b)  Conformal map of (a) into a disk. A branch cut in $x^\mu$ runs between the two vertex operators \cite{Dai:1989cp}. \label{decaydiagrama}}
	\end{figure}

We start with briefly reviewing the analysis of \cite{Dai:1989cp}.  
The idea of that paper is to compute the total decay rate, which is related via the optical theorem to the forward amplitude, by computing  the imaginary part of the self-energy diagram. A trick used in this paper is to
 compactify one space dimension and consider the case of incoming and outgoing strings that are stretched around this compact dimension.  This implies that the relevant vertex operator is the  simple  closed string one. 
The full process is  depicted in figure \ref{decaydiagrama}a
which shows a winding state that splits and rejoins.
When the initial and final states are represented
as vertex operators, the world-sheet is as shown
in figure \ref{decaydiagrama}b, a disk with two closed string vertex
operators.  

The corresponding amplitude takes the form
\be
\label{splitamplitude}
i{\cal A}_2 = \frac{iTN}{g^2} L \left [
\frac{\kappa}{2\pi\sqrt{L}}\right ]^2 \int_{|z|<1}\!{\rm d}^2z\, \langle:
e^{iP \cdot X(0)}::e^{-iP\cdot X(z)}:\rangle\,,
\ee
where $N$ relates to $SU(N)$, $SO(N)$ or $Sp (N)$ groups to which the type I open string is related. The
constant $g$ is the coefficient of the open string tachyon
vertex operator,  and $\kappa$ is the $D$-dimensional gravitational
coupling.
 We would like to determine the dependence of the  decay width on the length $L$.
The linear dependence on $L$  in front of the integral  follows from  the zero mode along the direction of the string. The $\frac{1}{\sqrt{L}}$ factor for each vertex operator follows from the normalization of the center of mass wave function of the string. 
A  further  dependence on $L$ follows from the $L$ dependence of the energy $E$ and the  left and right momenta, which are given by
\be\label{EP}
P_L = ( E,LT,0) \qquad P_R = ( E,-LT,0)\qquad E= \sqrt{(TL)^2-8\pi T}
\ee
The expression for the energy includes the classical $TL$ term and the correction due to quantum fluctuations which is the second term in the square root. For an open string the latter takes the form $a/\alp$ as in eq. \ref{Arvis}, where $a$ is the intercept. The open string intercept at critical dimensions is 
$a=\frac{D-2}{24}= 1$ and hence $\frac{a}{\alp}=2 \pi T$. For a closed string, which we discuss here, both the tension and the intercept  are  twice that of an open string and hence the result $8\pi T$. On the other hand the classical contribution for the open and closed strings are the same. 
Using that the vertex operator of a closed winding state is
\be
e^{iP\cdot X}= e^{i(P_L\cdot X_L + P_R\cdot X_R)}
\ee
and the standard OPE for $X_L$ and $X_R$ one finds that 
\bea\label{vvope}
\langle : e^{iP\cdot X(0)}::e^{iP\cdot X(z)}: \rangle&=& z^{-\frac{P_R^2}{4\pi T}}
\bar z^{-\frac{P_L^2}{4\pi T}}( 1-z\bar z )^{\frac{-P_R\cdot P_L}{4\pi T}}\CR
&=& |z \bar z|^{-2}(1-z\bar z )^{\tilde J}\CR
\eea
where $\tilde J \equiv \frac{L^2T}{2\pi}-2$.\,\footnote{ We use this notation since for a   rotating string the classical  angular momentum $J= \frac{L^2T}{2\pi}$.}
Upon substituting \ref{vvope} into \ref{splitamplitude} it was shown in \cite{Dai:1989cp} that the dependence of the amplitude on $L$ is given by
\bea
i{\cal A}_2 &=& \frac{iTN\kappa^2}{2\pi g^2}\lim_{t\rightarrow 0}\frac{\Gamma(t-1)\Gamma(1-\tilde J)}{\Gamma(t-\tilde J)} \CR 
&=& \frac{iTN\kappa^2}{2\pi g^2}\left (\tilde J\pa_{\tilde J} \ln[ \Gamma(-\tilde J)]+ \lim_{t\rightarrow 0}\frac{\tilde J}{t}\right)\CR
\eea
where $t$ was introduced as a regulator. The imaginary part of the term in the brackets is $-\pi \tilde J$
or more precisely $\sum_k \pi k \delta ( \tilde J - k )$ for $k= 1,...$.

  Thus we finally  get that 
\be
\text{Im}{\cal A}_2 = -\frac{iTN\kappa^2}{2 g^2}\tilde J
\ee
Since ${\cal A}_2$ gives the mass squared shift of the winding state, then the width is given by 
 \be\label{GammaL}
\Gamma=-\text{Im} \delta(m)=-\text{Im}\frac{1}{2m}\delta(m^2)=\frac{TN\kappa^2}{4 g^2}\frac{\tilde J}{E} 
\ee
The asymptotic expression was further simplified in \cite{Dai:1989cp} by relating $\kappa$ to $g^2$. The leading dependence on $L$ finally reads, in $D = 26$ dimensions,
\be\label{kappag}
\frac{\Gamma}{L}= \frac{g^2 T^{13} N}{4(4\pi)^{12}}
\ee

\begin{figure}[ht!]
			\centering
				\includegraphics[width=200bp]{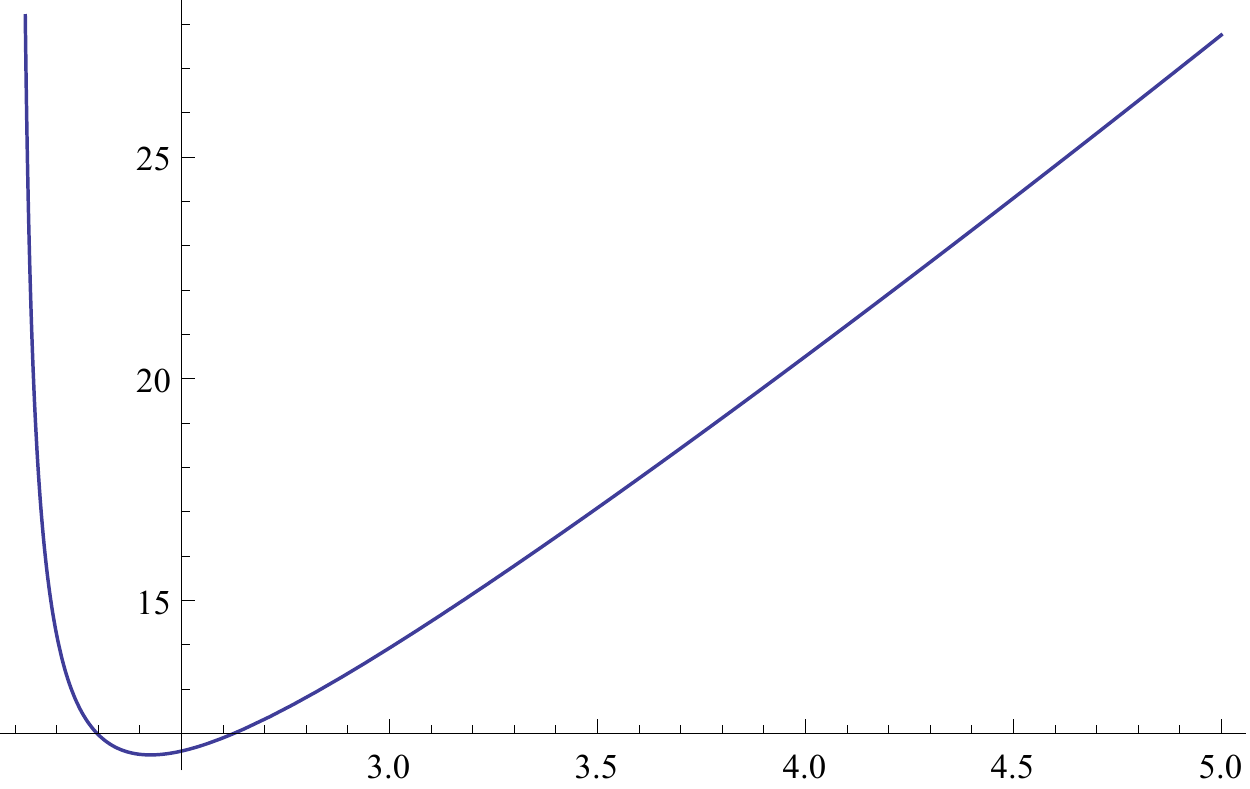}
				\caption{The dependence of the width on the length, per eq. \ref{lindep}.}
				\label{deviation}
	\end{figure}

The linearity dependence of $\Gamma$ on $L$ is a property of the asymptotic behavior at large $\tilde J$. However, for finite values of $\tilde J$ the expression  deviates  from exact  linearity.
 This follows from the fact that      the form of $\Gamma\sim \frac{\tilde J}{E}$ does not imply an exact linear dependence on $L$ since

\bea\label{lindep}
\Gamma&=& \frac{ T N\kappa^2}{4  g^2}\left[ \frac{L^2T - 4\pi}{\sqrt{(TL)^2 -8\pi T}}\right ]\sim \frac{N\kappa^2}{ g^2} \frac{TL}{4}\left[1 - \frac{4}{{\alp}^2(TL)^4}+ ... \right ]\CR
&=& \frac{ T N\kappa^2}{4  g^2}\left[ L_{tot} + \frac{4\pi}{T}\frac{1}{L_{tot}}\right]\CR
\eea
where we defined a ``total'' length $L_{tot}= \sqrt{L^2-\frac{8\pi}{T}}$, which includes a contribution from the intercept. In section \ref{sec:pheno} we argue that a similar definition of the length can be used in phenomenology.
  In figure \ref{deviation}  the decay width in the form of $\Gamma  \frac{4  g^2}{ T N\kappa^2} $  is drawn as a function of $TL$ for $\alp=0.9$ GeV$^{-2}$.
 \subsection{The decay width of  rotating and excited  strings} \label{sec:decaywidth_rot}
Hadrons in the HISH model are rotating and excited strings and thus we would like to use the width of a static string as determined above to infer that of a rotating string. This was also addressed in \cite{Dai:1989cp}. The basic idea is that the decay rate per unit length, which is a constant in the leading order approximation, is the same also for a rotating string, and for the rotating string one must take count of time dilation along the string.
 
For the rotating string, each point along the string moves at a different velocity, experiencing a different rate of time dilation. If we assume that the decay probability per unit length is constant along the string, then the total decay rate of a rotating string is given by integrating along the string
\be \Gamma_{\text{rot}} = \int \left(\frac\Gamma L\right)_{\text{stat}} \frac{d\sigma}{\gamma(\sigma)}\,,\ee
with the usual relativistic factor of time dilation, \(\gamma^{-1}(\sigma)\), which is position dependent in this case. 
The latter is  in fact a result of the transformation from the rest frame of the string to the laboratory frame in which the string rotates. Thus the total decay width of a long  string of length $L$ is
\be\label{gammarot}
\Gamma =\left ( \frac{\Gamma}{L}\right )_{\text{stat}} \int_{-L/2}^{L/2} d\sigma \sqrt{1-\omega^2\sigma^2} = \frac{\pi}{4} \left ( \frac{\Gamma}{L}\right )_{\text{stat}} L
\ee 
where we used the rotating solution of section \ref{sec:rotating_string} in the massless limit of \(\beta = \omega L / 2 = 1\).

Non-collapsed  open strings with finite length $L$ occur when there is a force  of  that can  balance the  string tension. For a rotating string this is naturally   the centrifugal force.    For  non-rotating strings   there can be a repulsive electrostatic force if the endpoint particles are charged with charges of the same sign. Of course, this force will not balance the tension for the case of opposite charges.  As  was discussed in section \ref{sec:casimir_force}, for all hadrons the intercept is negative, $a<0$, and there is  repulsive Casimir force that prevents the non-rotating string from collapsing. 
In such  a  case, similarly to the rotating string case, assuming that  for long  open strings  the decay width per unit length is a constant, the only thing left to do in order to determine the decay width is to multiply it with the string length. In section \ref{sec:rotating_string_cl} this length  was derived for a rotating string. For an excited  string with no massive endpoints, the $n^{\text{th}}$ excited level has a decay width of  
\be\label{excited}
\Gamma_{n} = \left(\frac{\Gamma}{L} \right ) \sqrt{\frac{2\pi ( n-a )}{T}} 
\ee
\subsection{The decay width of strings with massive endpoints} \label{sec:decaywidth_massive}
\begin{figure}[ht!]
			\centering
				\includegraphics[ width=0.48\textwidth]{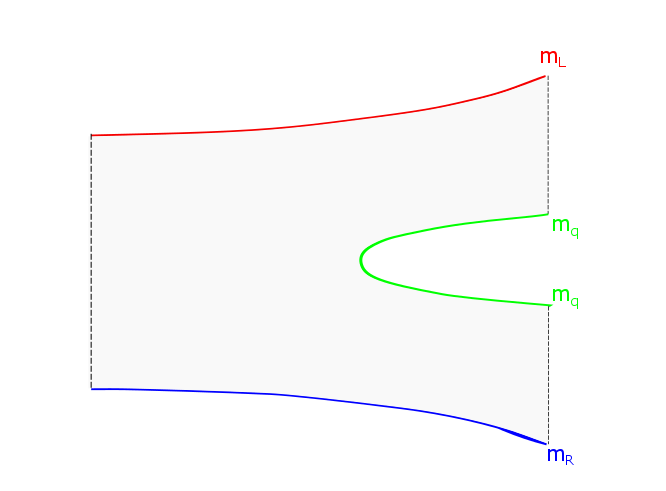}
				\caption{The split of a string with massive endpoints.}
				\label{massivesplit}
	\end{figure}

As was discussed in section \ref{sec:HISH}, the hadron in the HISH model is a string with massive particles on its ends. Massive quarks at the ends of the string are naturally also present in phenomenological models that follow directly from QCD. The string worldsheet diagram of \ref{basicsplit} is modified by the masses at the boundaries and is drawn in figure \ref{massivesplit}.

 The mechanism of the generation of the pair of massive particles created at the point of the split will be discussed in detail in section \ref{sec:decay_exponent}. For now we would like to examine 
 how is  the decay width affected by the properties of the endpoint particles and in particular their masses.
The computation of the decay width is affected by the endpoint masses in three ways: (i) The dependence of the decay width on the masses enters  via the dependence of the length of the string on the mass, i.e. by modifying the relation between the total mass of the hadron and the length, and for a rotating string also via the time dilation. (ii) The different boundary condition. (iii) The dependence of the intercept of the endpoint masses as discussed in section \ref{sec:quantum_massive}. As for (ii), whereas the computation of \cite{Dai:1989cp} is based on assuming Neumann boundary conditions, for the case of a string with massive endpoints this is not be the case. We would like to argue that for string with light endpoint particles, that is a string for which the main contribution to the total mass of the hadron comes from the string and not from the endpoint particles, the result should not be dramatically affected. However, for the case where the string is ``short'' and a large fraction of the total mass comes from the endpoint particles the modification is more meaningful. In section \ref{sec:model_phase_space} we address this issue and argue that, for the purposes of phenomenology, inserting a phase space factor can partially compensate for the deviation.

As for (i), the result for the time dilation factor in a rotating string with no particles on its ends \ref{gammarot} can be generalized also to the case of massive endpoints. For the case of the same mass on both ends, using eq. \ref{repCasf} we get
\be \label{eq:gamma_wl}
\Gamma= \frac12\left(\frac{\Gamma}{L}\right )_{\text{stat}}L\left [ \sqrt{1- \left(\frac{ \omega L}{2}\right)^2} + \frac{2}{\omega L} \arcsin\left[\frac{\omega L}{2}\right]\right ]
\ee
 where now the time dilation is a function of the velocity $\beta = \omega L/2$, which is given by
\be
\beta^2=  \left [1 - \left(\frac{\frac{-a}{L} + \sqrt{(\frac{a}{L})^2 + 2m (TL + 2m)}}{TL+2m}\right)^2\right ]
\ee
if one solves the force equation \ref{repCasf} for \(\beta\).

\begin{figure}[ht!]
			\centering
				\includegraphics[ width=200bp]{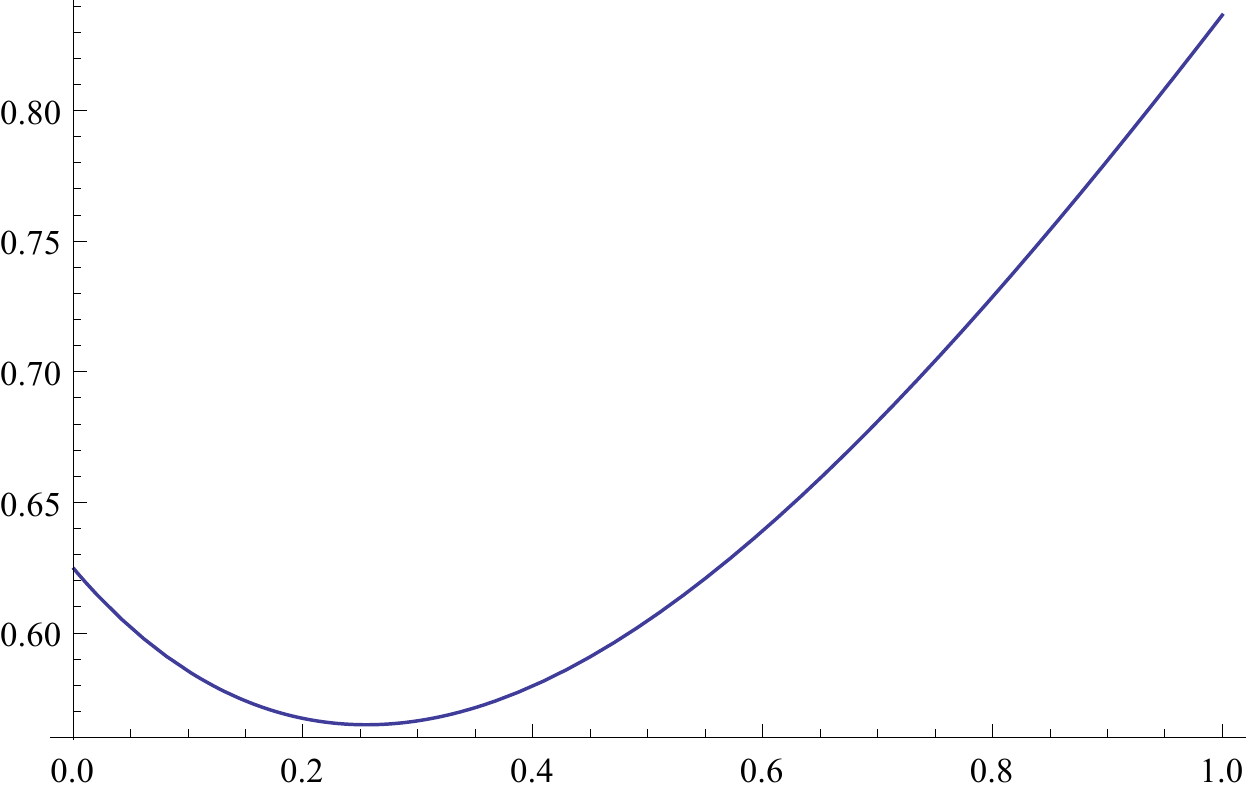}
				\caption{ The dependence of the decay width on the length for typical values of the parameters: $a=-0.5$, $m=100$ MeV  and $T=0.17$ GeV\(^2\), following eq. \ref{eq:gamma_wl}.}
				\label{decaywidthmass}
	\end{figure}

The outcome of this is that for a rotating string with massive endpoints the decay width deviates from linearity as can be seen from figure \ref{decaywidthmass}. At the high velocity limit, \(\beta= \to1\), we can write the expansion
\be \Gamma = \left(\frac{\Gamma}{L}\right )_{\text{stat}}\left(\frac{\pi}{4} L + \frac{\pi}{4}\frac{m}{T} -\frac{4}{3\sqrt2}\left(\frac{m}{T}\right)^{3/2}L^{-1/2}+\mathcal O(L^{-3/2})\right) \ee

Finally, in the asymmetric case, we can generalize to
\be \Gamma \propto \sum_iT\int_{0}^{\ell_i} d\sigma\sqrt{1-\omega^2\sigma^2} = \sum_i\frac{T\ell_i}{2}\left(\sqrt{1-\beta_i^2}+\frac{\arcsin\beta_i}{\beta_i}\right)\,. \ee

For excited string state for some given angular momenta we have argued that the decay width behaves as in eq. \ref{excited}. For a string with massive endpoints, as was shown in 
 section \ref{sec:quantum_massive}, the eigenfrequencies $n$ should be replaced by $\omega_n/\pi$ where the new eigenfrequencies are solutions of eq. \ref{transcendental1}. Thus the decay width reads 
\be\label{excited2}
\Gamma_{n} = \left(\frac{\Gamma}{L} \right )_{\text{stat}} \sqrt{\frac{2\pi ( \frac{\omega_n}{\pi}-a(q) )}{T}} 
\ee
where the intercept is a function of $q=\frac{m}{TL}$.
\subsection{ The decay width in non-critical dimensions}
So far following \cite{Dai:1989cp} we discussed the decay of  strings in the critical dimension  that wrap a   compactified  direction.
To address the decay width in any dimension  $D$ below the critical dimension we can check what modifications should be made to the analysis of 
\cite{Dai:1989cp} and we can also follow the analysis of \cite{Mitchell:1989uc,Wilkinson:1989tb,Mitchell:1988qe} that considered the case of general $D$ dimensions.  These latter papers  did not use the trick of compactifying a space dimension but rather analysed   a split of an open string  into two open strings.

Lets follow first the former approach. Inspection of the treatment of \cite{Dai:1989cp} reveals two points where the use of $D = 26$ was made: (i) Expressing $\kappa$ in terms of $g^2$ in \ref{kappag}. This of course does not affect the leading order linearity in $L$. (ii) The expression of the energy of the wrapped string \ref{EP}.
This expression is based on using an intercept $a=\frac{D-2}{24} =1$ which is well known to be the case for $D = 26$.
 The computation of the intercept of non-critical bosonic string theories in $D$ dimensions should include the contributions from  the transverse directions as well as  the contribution of the Liouville term,  or in a different formulation the Polchinski-Strominger term. This calculation  was performed in \cite{Hellerman:2013kba}. For the rotating string this term diverges. In \cite{Hellerman:2013kba} a regularization and renormalization of the PS term was performed and the outcome was that 
\be\label{intercept}
a= a_{cr}+ a_{PS}= \frac{(D-2)}{24} + \frac{(26-D)}{24} =1
\ee

If we incorporate this into  the expression for the energy  at any $D$ dimension we get
\bea
E &=& \sqrt{ ( TL)^2 - 8\pi T }= \sqrt{ ( TL)^2 - 8\pi T(a=\frac{26-2}{24}) }\ \ \   \rightarrow \CR
E &=&\sqrt{ ( TL)^2 - 8\pi T(a_{tr}+ a_{PS} )}= \sqrt{ ( TL)^2 - 8\pi T(\frac{D-2}{24}+ \frac{26-D}{24}) }= \sqrt{ ( TL)^2 - 8\pi T}\CR
\eea
Thus we conclude that the leading order linearity in $L$ seems, following \cite{Dai:1989cp}, to hold also for a general dimension
 $D$.
Next we want to check whether  this property holds also following the analysis of \cite{Mitchell:1989uc,Wilkinson:1989tb,Mitchell:1988qe}.
The authors of these papers also computed the imaginary part of the self-energy diagram but unlike \cite{Dai:1989cp} they did not use a string wound over a  compactified dimension   but rather computed the diagrams that contribute to the self energy  of an open string. The calculations were done taking that the string resides  in  $D$ dimensions. However,   they  included only the transverse modes in the final
decay products and not also the ``Liouville'' mode for $D\neq 26$.  
The result of their calculations is  that the dependence of the width on the length $L$ takes the following form
\be
\Gamma\sim L^{\frac{D-14}{12}}= L^{\frac{D-2}{12}-1}
\ee

Note that for the critical dimension $D = 26$ the width is linearly dependent on $L$ as was the asymptotic result of \cite{Dai:1989cp}. However, for $D = 4$, it implies that $\Gamma\sim  L^{-\frac{5}{6}}$ which means the width decreases with $L$.
 
It is well known that a consistent string theory in non-critical flat $D$ dimensions has to take into account also the contribution of the Liouville mode. Thus, the computations of \cite{Mitchell:1989uc,Wilkinson:1989tb,Mitchell:1988qe} are incomplete. Note that according to their result the imaginary part of the self energy (without dividing by $E$) is
\be
\text{Im}[\delta(m^2)]\sim  t^{\frac{D-2}{24}}=t^{a}
\ee
where in the last expression we have noted that $\frac{D-2}{24}$ is the intercept $a$ of a bosonic string in $D$ dimensions.  As was explained above, the intercept, if one includes the Liouville mode, is $a=1$ at any $D$ dimension. 
Thus if such a transformation between the critical and non-critical theories occurs also in the case of the self energy the result is that asymptotically for large $L$  
\be
\Gamma\sim \frac{t^a}{E} \sim L^1
\ee
namely,  the decay width is linear in $L$  in $D$ dimensions just as it is for $D = 26$.
As was emphasized above for a string in the critical dimension, it is true also for any $D$ dimensions and in particular in $D = 4$ that the linearity  holds only for very long strings with $LT \gg \sqrt{ 8\pi T}$ and for smaller length of the string there are deviations from linearity.

\subsection{The decay of a stringy hadron}
A stringy hadron as described  in section \ref{sec:HISH} is a string with charged massive spinors on its endpoints. The full quantum treatment of such a string is not fully understood. For the basic string with no endpoints the quantum effects are encapsulated in (i)  the intercept that modifies the whole spectrum  including the ground state and (ii)  a tower of quantum excited states.  The intercept associated with the string with endpoints is affected by the endpoint masses, charges and spins. The dependence on all these factors has not yet been determined. 
 Therefore   rather than computing it we take the value extracted from the measured spectrum. We denote this intercept by $a_{exp}$. 
In the previous subsection we argued that the Linear dependence in $L$ in non-critical dimensions  is heavily related to the fact that $a=1$. Since now we advocate using $a_{exp}$  it may seem that we argue that the linearity is lost. We want to emphasize that this is not the case. The deviation of $a$ from unity is due to non string effects like the mass charge and spin of the endpoints. Thus, though the latter affect the overall value of $a$, it is not clear if and how  they  affect the incorporation of the ``longitudinal''  Liouville mode on top of the transverse mode which are modes of the string itself. This issue will be further investigated as part of an attempt to find a theoretical description that is in accordance with the phenomenological intercept.   An important property of $a_{exp}$ is that it is negative $a_{exp}= -|a_{exp}|$ (as was discussed  in \ref{sec:casimir_force}), and is in charge of the repulsive Casimir force.

For the case of a string with no  massive endpoints,  we replace the expression for $E$ in \ref{EP} by 
\be
  E\rightarrow  \sqrt{ ( TL)^2 + 8\pi T |a_{exp}|}\,, \qquad \tilde J \rightarrow \frac{L^2 T}{2\pi} + 2 |a_{exp}|\,.
\ee
Correspondingly,  the leading behavior of the decay rate \ref{lindep} is replaced by 
\be
\Gamma \sim \frac{N\kappa^2}{4 g^2} TL \left[1 + \frac{4 |a_{exp}|^2}{{\alp}^2(TL)^4}+ ... \right ]
\ee
For the case of a string with massive endpoints we take the same expression for  $\frac{\Gamma}{L}$ and then multiply it with the expression of the length that takes into account the massive endpoints as was derived in section \ref{sec:rotating_string}.

\section{Exponential suppression of pair creation} \label{sec:decay_exponent}
Now that we have discussed the decay width associated with a breakup of a open string
without or with massive endpoints in critical or non-critical dimensions, we would like to analyze the probability of such a process taking place for  hadrons.
Prior to presenting the analysis for holographic stringy hadrons we first briefly review the determination of the probability in standard non-stingy description. We review the   mechanism in a stringy holographic setup following  \cite{Peeters:2005fq}, and discuss corrections due to spacetime curvature and massive endpoints of the string. We conclude this section by briefly discussing the multi-breaking of the string.
\subsection{The decay as a Schwinger mechanism} \label{sec:decay_Schwinger}
In  QCD the breaking mechanism is that  of creating a quark-antiquark pair at the point where the chromoelectric flux tube is torn apart and turns into two flux tubes. A similar process of creating an electron positron pair in a constant electric filed was analyzed in \cite{Schwinger:1951nm} and is referred to as the Schwinger mechanism. The analog of this  mechanism in QCD was determined using a WKB approach in \cite{Casher:1978wy,Gurvich:1979nq} and an exact path integral in  \cite{Nayak:2005pf}.  
The model  relating the decay of a meson with  these computations of the probability of a quark antiquark is based on  assumptions: (i) That at the hadronic energy
scale of 1~GeV the quarks can be treated as Dirac particles with
mass $m_q$. The quark masses are not universally defined: in \cite{Gurvich:1979nq} the masses are taken to be the QCD masses, in  \cite{Casher:1978wy} the constituent quark masses and in \cite{Nayak:2005pf} the masses are not specified at all. (ii) That there is a chromoelectric flux tube of
universal thickness which is being created in a timescale that is
short compared to the hadronic timescale. (iii) In the WKB approach  the chromoelectric field is
treated as a classical longitudinal Abelian field.  The flux tube is parametrised by
the radius of the tube $r_t$, the gauge coupling $g$ (which is also the
charge of the quark), and the electric field~${\mathcal E}_t$. The
energy per unit length stored in the tube is equal to the string
tension,
\begin{equation}
T_{\text{eff}}=\frac{1}{2} {\cal E}_t^2\pi\, r_t^2=   \frac{1}{4} g\, {\cal E}_t
\end{equation}
where in the last part of the equation the Gauss law was used.  It is
easy to verify that $g^2=4r_t^2/\alp$. When the radius of the flux
tube is smaller than the size of the tube but larger than the distance
scale relevant to pair production, i.e.~when it is of the order of
$r_t\sim 2.5~\text{GeV}^{-1}$, the coupling constant is indeed weak,
$g^2/8\pi<1$. In \cite{Nayak:2005pf} the expression of the  effective tension was found to be a more complicated function of the two Casimir operators of  the $SU(3)$ 
color group $E^a E^a$ and $d_{abc}E^a E^b E^c$.
The probability of a single
pair-creation event to occur is given by
\begin{equation}
\label{e:singledecay}
{\cal P}_{\text{pair prod.}} = \exp\left( - \frac{2\pi m_q^2}{g{\cal E}}\right).
\end{equation}
Generalizing this to multiple pair creation one derives~\cite{Casher:1978wy} the decay probability per unit time
and per unit volume,
\begin{equation}\label{decaypro}
{\cal P}= \frac{g^2 {\cal E}^2 }{16\pi^3} 
  \sum_{n=1}^\infty \frac{1}{n^2} \exp\left(-\frac{2\pi m_q^2 n }{g{\cal E}}\right)
  =  
  \frac{T_{\text{eff}}^2}{\pi^3} \sum_{n=1}^\infty \frac{1}{n^2} \exp\left(-\frac{\pi m_q^2 n }{2 T_{\text{eff}}}\right)\,.
\end{equation}   
This probability was then used to determine the vacuum persistent probability, which was found to be $1- e^{-V_4(t_M){\cal P}}$ where $t_M$ is the
meson lifetime measured in its rest frame. For $V_4(t_M)= V t_M$ the corresponding decay width is given by  $\Gamma= V {\cal P}$. Thus taking that ${\cal P}$ is given in \ref{decaypro} what is left to do is to compute the volume.  In \cite{Casher:1978wy} the volume of the system
was determined  for two cases: (i)  a rotating flux tube (ii)  a one dimensional
oscillator. For the first case we use $M =\pi T_{\text{eff}} L$ which
implies that $V_4(t_M)= \pi r_t^2 L t_M$ and hence the decay width is
$\Gamma= \pi r_t^2 L {\cal P}$ and finally
\begin{equation}
\left(\frac{\Gamma}{M}\right)_{\text{rot}} = \frac{2 r_t^2}{T_{\text{eff}}}{\cal P} .
\end{equation}

For the case of the oscillator, the relation between the length and
the mass is given by $M= T_{\text{eff}} L$ and therefore on average
$V_4(t_M)= \frac{1}{2} \pi r_t^2 L t_M$ and hence $\Gamma=\frac{1}{2}
\pi r_t^2 L {\cal P}$ which means that
\begin{equation}
\left(\frac{\Gamma}{M}\right)_{\text{osc}}=\frac{\pi}{4} \left(\frac{\Gamma}{M}\right)_{\text{rot}} .
\end{equation} 
 From eq. \ref{decaypro} two properties are immediately clear: the exponential suppression does not depend on the length of the string, while the total probability scales linearly in the length. This linearity property is clearly in accordance with the stringy calculation presented in section \ref{sec:decaywidth}. We would like to check whether the exponential suppression factor shows up also in the holographic stringy picture. Furthermore, if it does, an interesting question is the exact form of the exponential factor. From the discussion above we have that ${\cal P}_{\text{pair prod.}}\sim   \exp\left( - C \frac{m_q^2}{T_{\text{eff}}}\right)$ where $C$ is a numerical factor.  In fact, assuming an exponential suppression factor that does not depend on the length of the string, one gets this form just from dimensional arguments. The exact form depends on what is the value of $C$ and what are the values of the masses and tension. The tension used in both \cite{Casher:1978wy} and \cite{Gurvich:1979nq} is the same and in fact is the one determined from the hadron spectrum \cite{Sonnenschein:2014jwa} and \cite{Sonnenschein:2014bia}. However, as mentioned above there is a big difference between the values of the light quark masses taken in \cite{Casher:1978wy} and \cite{Gurvich:1979nq}. The former advocates using the constituent quark masses whereas the latter the QCD masses. Naturally we expect that in  the stringy holographic setup  the string endpoint masses defined in section \ref{sec:HISH} will be the relevant masses. 
	
\subsection{The suppression factor for  stringy holographic hadrons} \label{sec:decay_holoexp}
As was  described in section \ref{sec:HISH} mesons are described in holography by flat horizontal strings that stretch in the vicinity of the ``wall'' and are connected with vertical segments to flavor branes. For baryons there is a similar string that on one side is connected  directly to a flavor brane like a meson but on the other side it is connected to a baryonic vertex that connects via two short strings to flavor branes. Since the key player for the decay is the horizontal string which is in common to both the mesons and baryons we first discuss them together. Later, in section \ref{sec:decay_baryons} we discuss the special features of the suppression factor for baryons.
 
\begin{figure}[ht!]
			\centering
				\includegraphics[width=0.48\textwidth]{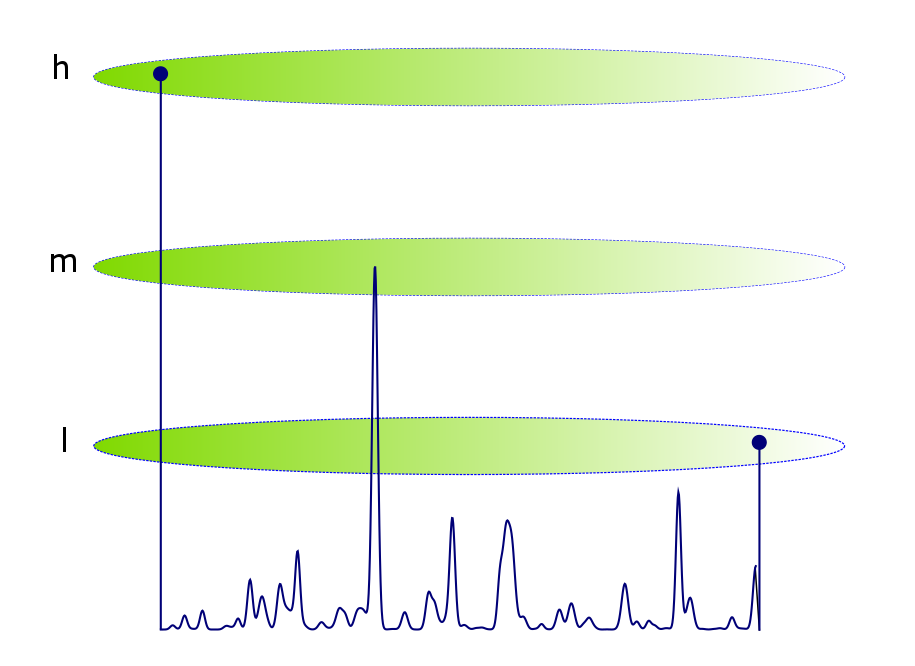}
				\caption{Quantum fluctuations of the horizontal segment of a heavy-light meson reach the medium flavor brane.}
				\label{quantumfluc}
	\end{figure}

Quantum fluctuations along the horizontal segment of the string can bring the string up to one of the flavor branes. When this happens the string may break up, and the two new endpoints connect to the flavor brane. This process is demonstrated in figure \ref{quantumfluc} where fluctuations of  a meson built from one heavy and one light endpoints reach a medium flavor brane. 
 
\begin{figure}[ht!]
			\centering
				\includegraphics[width=0.90\textwidth]{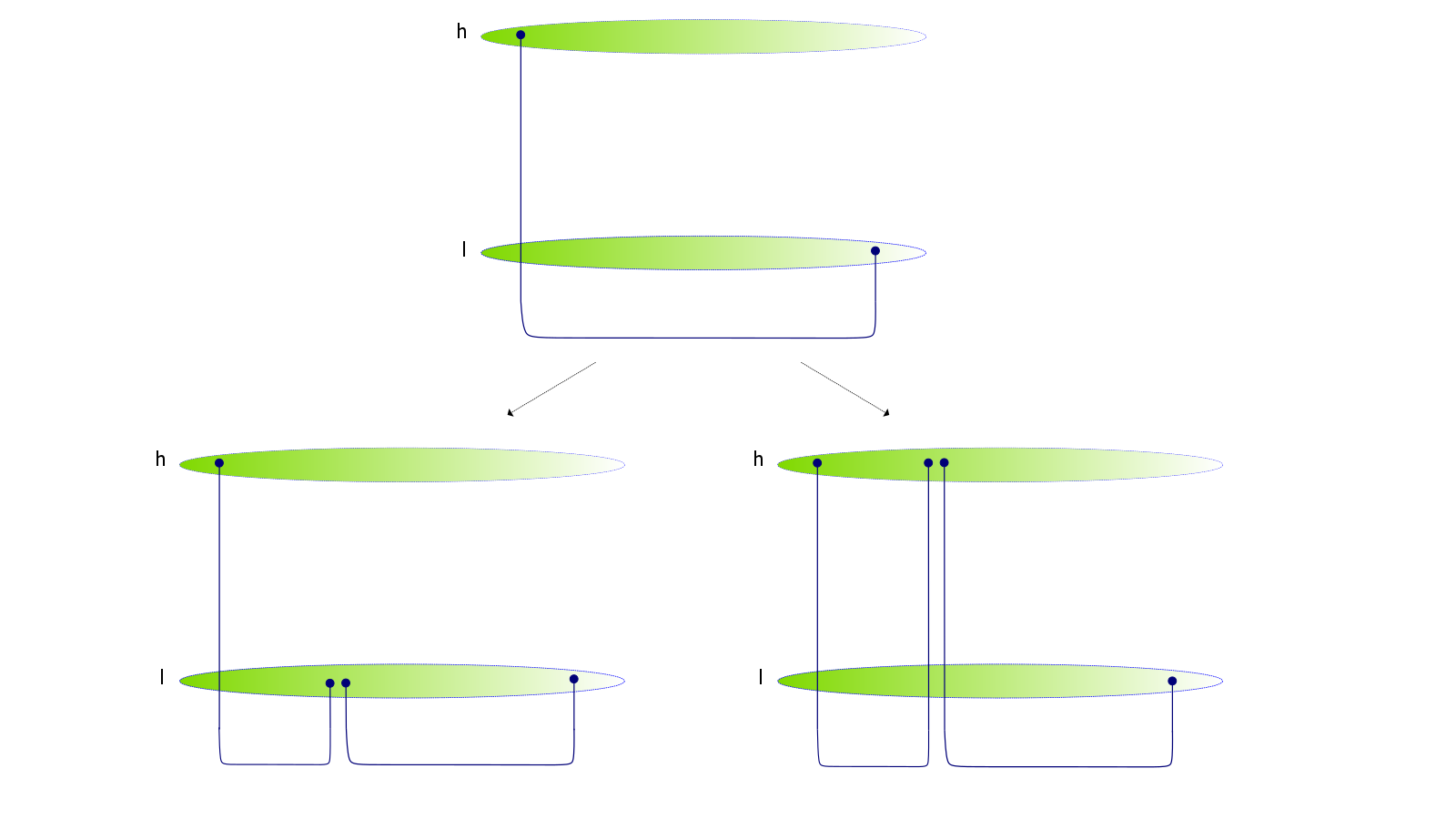}
				\caption{Possible decays of a heavy-light meson via light or heavy quark pair creation. The ratio between the two channels will include an exponential suppression term of the form \(e^{-C(m_h^2-m_l^2)/T}\).}
				\label{posholdecasy}
	\end{figure}

In fact as shown in figure \ref{posholdecasy} there is more than one  possibility for the decay of such a meson. Quantum fluctuations can reach the light or heavier flavor branes. Of course, there are also kinematical constraints on whether or not a hadron can decay in a channel where heavier quarks are created.

A calculation of the probability of the fluctuations of the  horizontal segment of a hadronic string to reach a flavor brane  was first performed  in \cite{Peeters:2005fq}. It was further discussed in \cite{Bigazzi:2007qa,Bigazzi:2006jt,Cotrone:2005fr}. In the following lines we  review the analysis of \cite{Peeters:2005fq} and then discuss  further  corrections to the basic computation of aspects of the probability. It should be noted here that the following calculations are for the most part made assuming non-rotating strings. We assume in the rest of the paper that the correction due to the rotation of the string is not very significant, so that the basic picture exponential suppression of pair creation, which is the probability of the string to fluctuate (in a direction that is of course transverse to the plane of rotation) and hit a flavor brane, is unaltered.

Basically one has to quantize the horizontal segment of the holographic stringy meson and determine the probability that the quantum  fluctuations of the string  will bring parts of it to a flavor brane.  For this purpose one has  to 
 construct  the string wave function.
 The starting point  is      the classical  holographic rotating  string  configuration   discussed in section \ref{sec:HISH}. We then have to  determine the spectrum of bosonic quantum 
fluctuations~$\delta X^M(\tau,\sigma)$ around this string
configuration.  The wave function has to be  expressed in terms of normal coordinates. Generically, the normal
coordinates~$\{{\mathcal N}_n (X^M)\}$ are nontrivial functions of all
target space coordinates $X^M$, due to the fact that the target space
metric is curved.  Each mode ${\mathcal N}_n$ is described by its own
wave function $\Psi_n[{\mathcal N}_n]$, and the total wave function is
just a product of wave functions for the individual modes,
\begin{equation}
\label{simple}
\Psi\big[\{ {\mathcal N}_n \} \big] = \prod_n \Psi_n\big[{\mathcal N}_n (X^M)\big] \, .
\end{equation}
 
 Once the wave functions are known   one  has 
to extract  the probability that, due to quantum fluctuations, the
string touches the brane at one or more points. We will call this
probability~${\cal P}_{\text{fluct}}$ and it is formally given by
\begin{equation}
\label{fluct-prob}
{\cal P}_{\text{fluct}} = 
\int'_{\{ {\mathcal N}_n\}} \big|\, \Psi\big[\{{\mathcal N}_n\}\big] \, \big|^2\,,
\end{equation}
where the prime indicates that the integral is taken only over those
string configurations $\{ {\mathcal N}_n \}$ which satisfy the
condition
\begin{equation}
\label{e:abovebrane}
\max\big(u(\sigma)\big) \geq u_f\,.
\end{equation}
This is a complicated condition to take into account, because
$u(\sigma)$ is a linear combination of an infinite number of
modes. While the constraint is simple in terms of~$u(\sigma)$, it thus
becomes highly complicated in terms of the modes~${\cal N}_n$. The
probability~\eqref{fluct-prob} only measures how likely it is that the
string touches the brane, independent of the number of points that
touch the brane. Note that this is a dimensionless probability, not a
dimensionful decay width. The passage between ${\cal P}_{\text{fluct}}$ 
and the width $\Gamma$ was discussed in detail in \cite{Peeters:2005fq}. The final result 
is that the two quantities can be related in the following way
\begin{equation}
\label{el}
\Gamma_{\text{approx}} 
 = A \Big( T_{\text{eff}}\,{\mathcal P}_{\text{split}}\,\times\,
 L\, \Big)\,\times\, {\cal
   P}_{\text{fluct}}\, ,
\end{equation}
where $A$ is certain numerical factor independent of the string tension, the masses of the parent meson,  its endpoints, or the pair created.

We start with  reviewing  the computation in the context of a string bit model in flat space time. Next we discuss the case of a continuous string in flat spacetime.   We check the impact of the curvature of the holographic spacetime and of the masses of the HISH  endpoint particles and finally we summarize the holographic determination of the suppression factor.  


\subsubsection{String bit model in flat space}\label{s:stringbits}
\begin{figure}[t]
\begin{center}
\includegraphics[ width=0.60\textwidth]{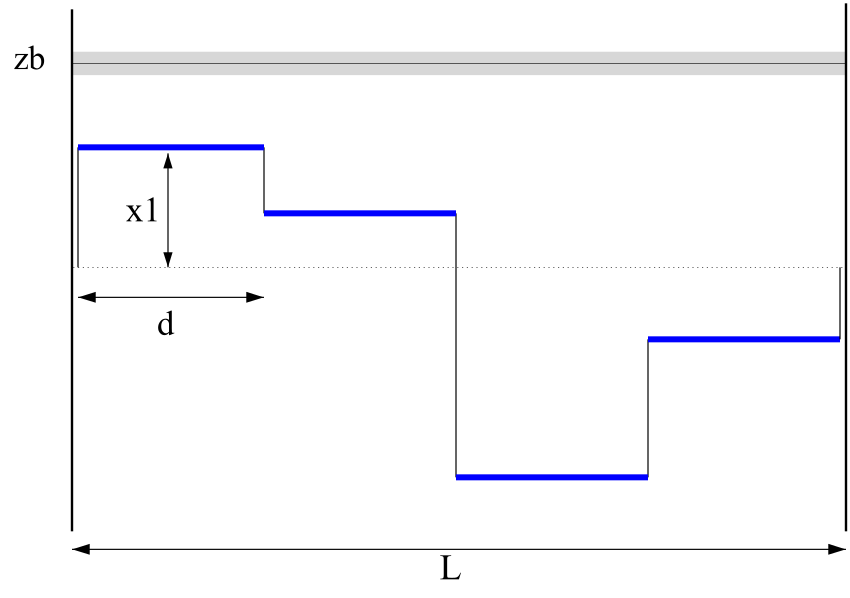} 
\vspace{-2ex}
\end{center}
\caption{The discretised string \cite{Peeters:2005fq}, consisting of a number of horizontal
  rigid rod segments connected by vertical springs. Also depicted is the flavor brane at $z=z_B$. The flavor brane has a finite width. \label{f:beadbox}}
\end{figure}

 A simple  way to calculate ${\cal P}_{\text{fluct}}$ is in the context of a toy model  where instead of a continuous string one uses
a discrete set of beads and springs (see figure \ref{f:beadbox}), whose number is then taken to be large. This of course introduces a certain
approximation, but it has the advantage that the integration over the
right subset of configuration space becomes much more manageable. The calculation of the probability  to reach the flavor brane in the context of this toy model was computed in  \cite{Peeters:2005fq}. For the convenience of the reader we review briefly it in the following lines.

The goal is to  compute the probability that, when the system is in the
ground state, one or more beads are at the brane at~$z=z_B$.
In order to write down the wave function as a function of the
positions~$z_i$ of the beads, we have to go to normal coordinates in
which the equations of motion decouple. Denote the string tension
by~$T_{\text{eff}}$, the number of beads by~$N$, their individual
masses by~$M_b$ and the length of the system by~$L$, which satisfies~$L
= d\,(N+1)$, where \(d\) is the distance between neighboring beads. The action is given by
\begin{equation}
S = \frac{1}{2}\int\!{\rm d}t\, \left(\sum_{n=1}^N M_b\,\dot{z}_n^2
- \frac{T_{\text{eff}}}{d}\,\sum_{n=1}^{N+1} \big(z_n - z_{n-1}\big)^2\right)\,,
\end{equation}
where $z_{N+1}\equiv 0$. This action corresponds to taking Dirichlet
boundary conditions at the endpoints, i.e.~infinitely massive quarks
at the endpoints of the string. As was discussed in section \ref{sec:HISH}, the horizontal string segment can be viewed as having massive endpoints. The corresponding boundary conditions are neither Dirichlet nor Neumann. This of course affects the eigenfrequencies of the normal modes. The analysis of \cite{Peeters:2005fq} is based on Dirichlet boundary condition which we continue to follow. In section \ref{sec:decay_exp_mass} we comment on the possible modification due to massive endpoints. 

The normal modes and their frequencies associated with this action
are given by
\begin{equation}
\label{e:normalmodes}
y_m = \frac{1}{N+1}\sum_{n=1}^{N} \sin\left(\frac{m n \pi}{N+1}\right) z_n\,,\qquad
\omega_m^2 = \frac{4\,T_{\text{eff}}\,N(N+1)}{M \, L}\, \sin^2\left( \frac{m \pi}{2 (N+1)}\right)\,,
\end{equation}
 
These expressions have been written in such a way that it is easy to take the continuum limit~$N\rightarrow\infty$ while keeping~$L$, $T_{\text{eff}}$
and the total mass of the whole string~$M=N\,M_b$ fixed. In particular,
\begin{equation}
\lim_{N\rightarrow\infty} \omega_m^2 
 = \frac{m^2 \pi^2\, T_{\text{eff}}}{M \,L}\,= \frac{m^2\pi^2\,}{L^2}
\end{equation}
  The action then reads
\begin{equation}
\lim_{N\rightarrow\infty} S(M =T_{\text{eff}}L) = 
T_{\text{eff}} \int\!{\rm d}t\int_0^L\!{\rm d}\sigma\,\Big[ \dot{z}(\sigma)^2 - {z'}(\sigma)^2\Big]\,.
\end{equation}
The system is now decoupled and the action for the normal coordinates
is
\begin{equation}
S = (N+1) M_b\, \int\!{\rm d}t \sum_{m=1}^N \big( \dot{y}_m^2 - \omega_m^2 y_m^2\big)
\end{equation}
The wave function is a product of wave functions for the normal modes,
\begin{equation}
\Psi\big(\{y_1,y_2,\ldots\}\big) =
  \prod_{m=1}^N  \left(\frac{2(N+1)M_b \omega_m}{\pi}\right)^{1/4}
                 \exp\left( -(N+1)M_b \omega_m\,y_m^2 \right)\,.
\end{equation}
The wave function~$\Psi(\{z_1,z_2,\ldots\} )$ is now obtained simply
by inserting the normal modes \eqref{e:normalmodes}, which of course
results in a complicated exponential in terms of the~$z_n$. Note that
the width of the Gaussian behaves as
\begin{equation}
\lim_{N\rightarrow\infty} (N+1) M_b \omega_{m}
 = \lim_{N\rightarrow\infty} (N+1) \frac{T_{\text{eff}}L}{N}\frac{m\pi}{L}
 = T_{\text{eff}} \pi m\,.
\end{equation}
This expression depends linearly on~$T_{\text{eff}}$ and is independent of~$L$, in
agreement with  what will be found out for the continuum analysis of section \ref{sec:decay_exp_continuum}.

  For each bead
position, we define the integration intervals corresponding to being
``at the brane'' and being ``elsewhere in space'' by
\begin{equation}
\label{e:bdydefs}
\begin{aligned}
I_{\text{brane}} &: \big[ -z_B - \Delta,\, -z_B \big]\; \cup \;
                    \big[ z_B,\, z_B + \Delta \big]\,,\\[1ex]
I_{\text{space}}  &: \big\langle -\infty,\, -z_B - \Delta \big]\; \cup\;
                    \big[-z_B,\, z_B\big]\; \cup\;
                    \big[ z_B+ \Delta,\, \infty\big\rangle\,.
\end{aligned}
\end{equation}
Here $\Delta$ is the width of the flavor brane, which of course has
to be taken equal to a finite value in order to be left with a finite
probability. The probability of finding a configuration which has,
e.g., one bead at the brane and all others away from it, is then given
by
\begin{multline}
\label{e:onebeadexample}
{\cal P}(\text{one bead at brane}) = \\[1ex]
 \sum_{i=1}^N  \int_{I_{\text{brane}}}\! {\rm d}z_i \;
         \prod_{k\not= i} \int_{I_{\text{space}}}\! {\rm d}z_k\;\;
         J\big(\{y_1,y_2,\ldots,y_N\},\{z_1,x_2,\ldots,z_N\}\big)\\[1ex]
         \times\Big|\Psi\big(\{z_1,z_2,\ldots,z_N\}\big)\Big|^2\,.
\end{multline}
The factor~$J$ is a Jacobian arising from the change of normal
coordinates to the original positions of the beads.  
The total decay width  of  one-meson into
 two-mesons. is a sum of decay widths
labeled by the number of beads which are at the brane,
\begin{equation}
\Gamma_{\text{meson} \rightarrow \text{2 mesons}}
  = \sum_{p} \Gamma^{(p)}_{\text{meson} \rightarrow \text{2 mesons}}\,.
\end{equation}
where the partial width~$\Gamma^{(p)}$ is given by
\begin{equation}
\label{e:bits_decay_width}
\Gamma^{(p)}_{\text{meson} \rightarrow \text{2 mesons}}
 = \sum_{\substack{\text{all configurations}\\\text{with $p$ beads at brane}}} p
 \,\cdot\, {\mathcal P}_{\text{configuration}}\,
 \cdot T_{\text{eff}}\,\, {\mathcal P}_{\text{split}}\,
 \cdot\, \text{length per bead}
\,.
\end{equation}
The factor~$p$ occurs because a configuration with~$p$ beads at the
brane can decay in~$p$ different ways into a two-string configuration.

The basic property of the decay, derived in section \ref{sec:decaywidth}, the linearity of the decay width with the length of the string is easy to get 
in the discrete picture.  Namely, consider the system
with~$T_{\text{eff}}$ fixed and $N$ fixed (and large in the continuum limit). The
total length (and thus the total mass) is now changed by varying the
spacing~$d$.   In fact, as long as
$\sum_i {\cal P}_{\text{bead $i$ at brane}}$ scales linearly in~$N$,
one obtains a linear dependence of the decay width on~$L$. For the
partial width~$\Gamma^{(N)}$ the proportionality with~$L$ is in
actually trivial,
\bea
\Gamma^{(N)}_{\text{meson} \rightarrow \text{2 mesons}} &=&
      N
\cdot {\cal P}_{\text{all beads at brane}}
\cdot T_{\text{eff}}\, {\cal P}_{\text{split}}
\cdot d \frac{N+1}{N}\  \CR
&=& {\cal P}_{\text{all beads at brane}}
\cdot T_{\text{eff}}\, {\cal P}_{\text{split}}
\cdot L\,   .
\eea

\begin{figure}[t]
\begin{center}
\includegraphics*[trim=100px 50px 100px 60px,width=.8\textwidth]{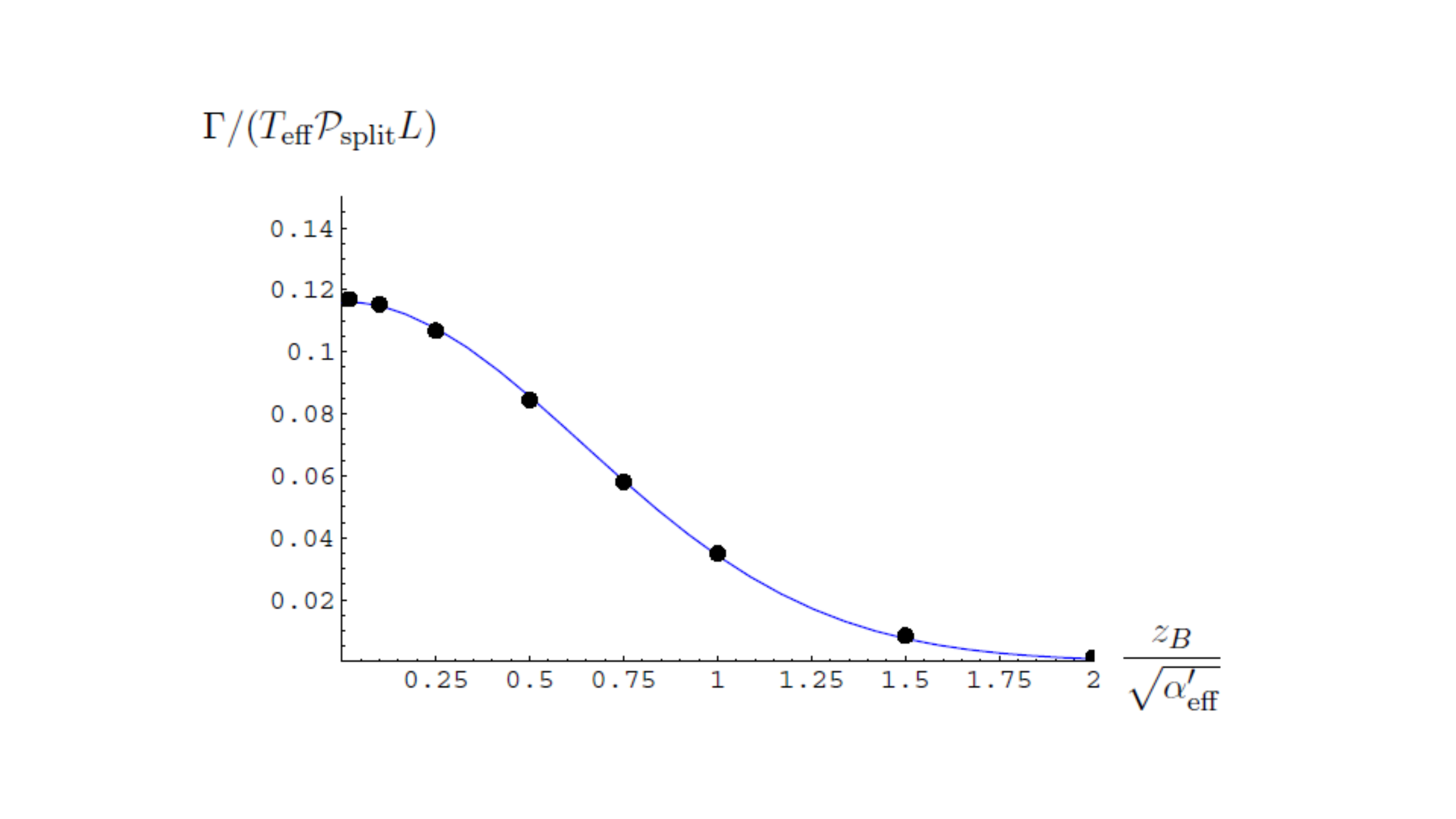}
\end{center}
\caption{Total decay width (divided by~$T_{\text{eff}}{\cal
  P}_{\text{split}}\,L$) for a six-bead system, with a brane width
  of~$0.1\,\sqrt{\alp_\text{eff}}$, as a function of the
  distance~$z_B$ of the IR~``wall'' to the flavor brane. Superposed is a
  best-fit Gaussian, with parameters $\Gamma = 0.12\,\exp( - 1.22\,
  z_B^2/\alp_{\text{eff}} )$. The string is allowed to split if a
  bead is at the brane, while the other beads are allowed to be anywhere
  (both below and above the brane). Figure from \cite{Peeters:2005fq}.\label{f:sixblobgauss}}
\end{figure}
\begin{figure}[t]
\begin{center}
 
\vspace{3ex}
\includegraphics*[width=.6\textwidth]{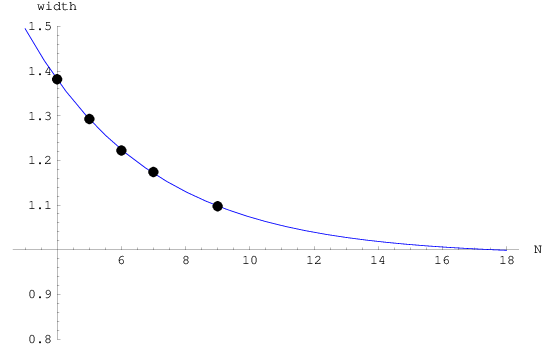}
\end{center}
\caption{The behaviour of the width of the Gaussian (numerical scaling factor in the LHS of eq. \ref{e:bitsresult}) shape of the decay
  width, given as a function of the number of beads~$N$. A best-fit
  exponential curve has been superposed, which suggest that the scale
  approaches one in the large-$N$ limit. Figure from \cite{Peeters:2005fq}. \label{f:widthfit}}
\end{figure}

To determine the decay width one has to perform the complicated integrations. This was done in  \cite{Peeters:2005fq}   numerically  using Monte-Carlo integration.
  An example of the decay width of a six-bead
system is given in figure~\ref{f:sixblobgauss}. By computing the decay
width for various values of~$N$ and extrapolating to large-$N$ (figure \ref{f:widthfit}), one
finds that the decay width is well approximated by
\begin{equation}
\label{e:bitsresult}
\Gamma_{\text{beads}} = \text{const.} \cdot \exp\left( - 1.0\,
\frac{z_B^2}{\alp_{\text{eff}}}\right) \cdot T_{\text{eff}}\,{\cal P}_{\text{split}}
 \cdot L\,.
\end{equation}
This extrapolation includes not just an extrapolation to large-$N$,
but also an extrapolation to small value of the brane
width.  The exponential suppression factor when translated to the quark antiquark masses takes the form
\be
\exp\left( - 1.0\,
\frac{z_B^2}{\alp_{\text{eff}}}\right)=\exp\left( - 2\pi\,
\frac{m_{sep}^2}{T_{\text{eff}}}\right)
\ee

\subsubsection{Continuous string in flat spacetime} \label{sec:decay_exp_continuum}

Next, following \cite{Peeters:2005fq}, we review the computation of the probability to reach a flavor brane  for the case of a continuous string in flat spacetime. The holographic background is non-flat but for mesons with light endpoint masses and for the creation of a light quark pair the fluctuations of the strings are close to the vicinity of the wall and hence a flat-space time is a reasonable approximation. In the next subsection will discuss the effects of the curvature. 

The starting point is the classical solution \ref{eq:rotating_solution} which we now rewrite as \begin{equation}
\label{e:openstringrod}
X^0 = L \tau\,, \quad
X^1 = L \sin\tau\sin\sigma\,,\quad
X^2 = L \cos\tau\sin\sigma\,,\quad u=u_\Lambda\,,
\end{equation}
where $\sigma\in [ -\pi/2, \pi/2 ]$
Next we consider the quantum fluctuations around this classical configuration along the holographic radial direction.  We define the coordinate $z$ as follows
\be
z\equiv C\sqrt{\frac{u-u_\Lambda}{u_\Lambda}}
\ee
where $C$ is a constant which depends on the particular holographic model used.  The action that controls these fluctuations is given by 
\begin{equation}
\label{Sfluct}
S_{\text{fluct}} = \frac{L}{2 \pi \alp_{\text{eff}}} \int\!{\rm d} \tau {\rm d} \sigma\,  
\left[ - (\partial_\tau z)^2 + \frac{1}{L^2} (\partial_\sigma z)^2 + \ldots \right] \, .
\end{equation}
Here the dots refer to fluctuations along other  directions. The boundary conditions for these fluctuations depend in the HISH picture on the endpoint masses. For nearly massless case we should have used Neumann boundary conditions and for heavy masses a Dirichlet boundary condition.  Here following 
\cite{Peeters:2005fq} we use the latter boundary condition but we claim that the final result will not be sensitive to this choice. 
  The
fluctuations~$z(\sigma, \tau)$ can be written as
\begin{equation}
\label{ansatz}
z(\sigma,\tau) = \sum_{n>0} z_n \cos(n\sigma)\,.
\end{equation}
Using this expression in the action and integrating over the~$\sigma$
coordinate, the action for the fluctuations in the $z$-direction
reduces to
\begin{equation}
\label{lotLHO}
S_{\text{fluct}} = \frac{L}{2 \alp_{\text{eff}}} \int\!{\rm d} \tau\,
    \left[  \sum_{n>0}\left( - (\partial_\tau z_n)^2 +
		\frac{n^2}{L^2} 
  z_n^2 + \ldots \right)\right]\,.
\end{equation}

The relevant, transverse part of the wave function is now given by
\begin{equation}
\label{simple2}
\Psi[\{z_n \}] = \prod_{n=1}^\infty \Psi_{0}(z_n)\,,
\end{equation}
where the wave functions for the individual modes are given by
\begin{equation}
\label{wavefnf}
\Psi_0(z_n) = \left(\frac{n}{\pi \alp_{\text{eff}} } \right)^{1/4} \exp\left(-
  \frac{n}{2\alp_{\text{eff}}} \, z_n^2\right) \, , 
\end{equation}
where all coordinates $z_n$ are unconstrained (i.e.~run from
$(-\infty, +\infty)$).

From this, one
obtains a \emph{lower} estimate for the probability that the string
touches the brane,
\begin{equation}
\label{Pmin}
{\mathcal P}^{\text{min}}_{\text{fluct}} = 1- \lim_{N \rightarrow \infty} \,
\int_0^{z_B} \! {\rm d}z_1 \int_0^{z_B} \! {\rm d}z_2 \cdots
\int_0^{z_B} \! {\rm d}z_N \, \,  \big| \Psi( \{ z_n \}) \big|^2  \ .
\end{equation}
The integral~\eqref{Pmin} was  evaluated numerically in \cite{Peeters:2005fq}. The final expression for the decay
width in the flat space approximation takes the form
\begin{equation}
\label{gammacont}
\Gamma_{\text{flat}} =  \Big( \text{const}.\times 
T_{\text{eff}}\,{\mathcal P}_{\text{split}}\,\times\,
 L \Big) \,\times\, \exp\left( - 1.3\frac{z_B^2}{\alp_{\text{eff}}}\right)  \, .
\end{equation}

\subsubsection{Curved spacetime corrections} \label{sec:decay_exp_curved}
The holographic spacetime discussed in section \ref{sec:HISH}  is characterized by 
 higher than five  dimensional curved  spacetime  with a boundary.  The coordinates of this spacetime include  the coordinates of the boundary spacetime,    a radial coordinate, and additional  
 coordinates transverse to the boundary and to  the radial direction. 
We  assume that the corresponding  metric depends only on the radial
coordinate such that its general form is \begin{equation}\label{metric}
ds^{2}=-g_{tt}\left(u\right)dt^{2}+g_{uu}\left(u\right)du^{2}+g_{x_{||}x_{||}}\left(u\right)dx_{||}^{2}+g_{x_{T}x_{T}}\left(u\right)dx_{T}^{2}\end{equation}
 where $t$ is the time direction, $u$ is the radial coordinate,
$x_{||}$ are the space  coordinates on the boundary and $x_{T}$ are the
transverse coordinates. We adopt the notation in which the radial
coordinate is positive defined and the boundary is located at $u=\infty$.
In addition, a ``horizon''  exists at $u=u_{\Lambda}$, such
that the spacetime is defined in the region $u_{\Lambda}<u<\infty$, instead
of $0<u<\infty$ as in the case where no horizon is present.

It is useful to define
\begin{equation}\label{fg}
\begin{aligned}f(u)^2 & \equiv g_{tt}\left(u\right)g_{x_{||}x_{||}}\left(u\right)\,,\\
g(u)^2 & \equiv g_{tt}\left(u\right)g_{uu}\left(u\right)\,.\end{aligned}
\end{equation}
It was found out in \cite{Kinar:1998vq} that the condition for a confining Wilson loop takes the form:  
\begin{enumerate}
\item $f$ has a minimum at $u_{min}$ and $f\left(u_{min}\right)\neq0$, \textbf{or}
\item $g$ diverges at $u_{div}$ and $f\left(u_{div}\right)\neq0$
\end{enumerate}
For long Wilson lines 
$ u_{min}\rightarrow u_{\Lambda}$ or $u_{div}\rightarrow u_{\Lambda}$,
and correspondingly the effective string tension   $T_{\text{eff}}$ is given by 
 \begin{equation}
T_{\text{eff}}=  T f\left(u_{\Lambda}\right)\label{eq:string tension}\end{equation}

\begin{figure}[t]
\begin{center}
\includegraphics[width=.48\textwidth]{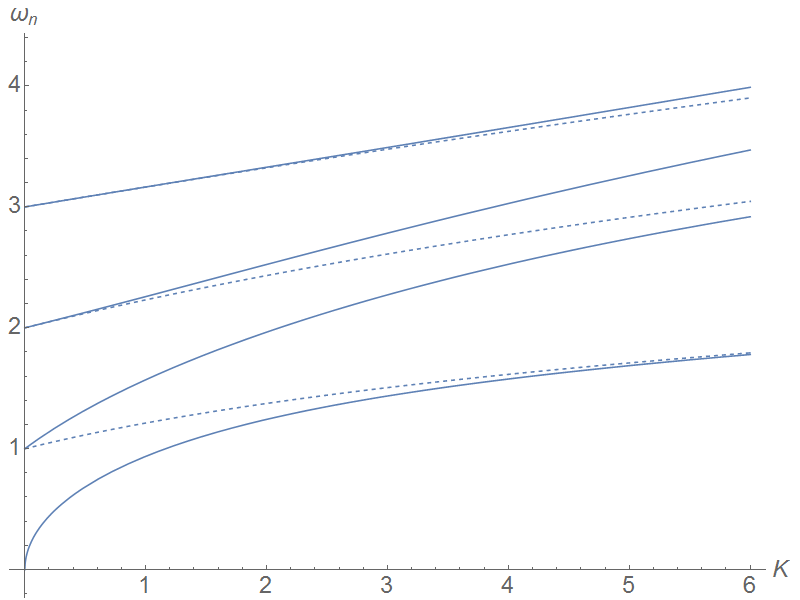}
\end{center}
\caption{The first few eigenfrequencies $\omega_n$, given
  in eq. \ref{e:allfrequencies}, as a function of the
  parameter~$K$. These frequencies correspond to modes satisfying the
  equation of motion~\protect\eqref{e:mathieu} with Dirichlet boundary
  conditions. Solid lines correspond
  to~$\omega^2_n = a_n(K/2)+K$ while the dashed lines correspond
  to~$\omega^2_n = b_n(K/2)+K$.  For $K=0$ and for
  $K\rightarrow\infty$, the spectrum is degenerate. For intermediate
  values of~$K$ there is level splitting.\label{f:frequencies}}
\end{figure}

As was discussed in section \ref{sec:HISH} the classical configuration of a hadron has a U-shaped profile (see figure \ref{fig:mapholflat}).
The classical string undergoes bosonic quantum fluctuations along various directions as well as fermionic fluctuations. These fluctuations in holographic backgrounds were discussed in \cite{PandoZayas:2003yb,Bigazzi:2004ze}. Assuming that the fluctuations along the holographic radial direction can be decoupled from the fluctuations along  other directions,  the action that controls them is given by 
 
\begin{equation}
\label{e:etaction}
S = 
T_{\text{eff}}\int\!{\rm d}\tau{\rm d}\sigma\,\left[
   \big( \dot z^2 - {z'}^2\big)
 -K\cos^2(\sigma)   
 z^2\right]\,.
\end{equation}
Thus the difference between action for fluctuations in the flat background given by \ref{Sfluct} and the action in curved spacetime is the $\sigma$-dependent worldsheet mass term. This mass term depends on $L^2$. In \cite{PandoZayas:2003yb} the coefficient of $L^2$ was referred to as $m_0^2$. Here we parametrize slightly differently, using
\be
 K = c_K(g_s N) T_{\text{eff}} L^2  \,,
\ee
 where $c_K$ is a dimensionless  factor which depends on the particular model chosen. The values of $cT_{\text{eff}}$  for several holographic models were determined in \cite{PandoZayas:2003yb,Bigazzi:2004ze}. For the models of  Klebanov-Strassler, Maldacena-Nuñez, and Witten it was found that $c_K\sim \frac{0.19}{g_s N}$, $c_K\sim \frac{1.97}{(g_s N)^{3/2}}$, and $c_K\sim 1.27\frac{1}{\sqrt{g_s N}}$ respectively.
Hence, the dependence of this term on $g_s N$ is not universal and varies between the different models.  
The equation  of motion for the fluctuations is  
\begin{equation}
\label{e:mathieu}
\left[- \frac{{\rm d}^2}{{\rm d}\tau^2} +  \frac{{\rm d}^2}{{\rm d}\sigma^2} 
-  2 K\big( 1 + \cos(2\sigma)\big) \right]
z(\tau,\sigma) = 0\,.
\end{equation}
 Imposing, 
 as for the flat spacetime background, Dirichlet  boundary conditions, 
$z(\tau, -\frac{\pi}{2}) = z (\tau, \frac{\pi}{2}) = 0\, $
and factorizing the solution according to $
\label{e:factorise}
\eta(\tau,\sigma) = e^{i\omega \tau} \, f(\sigma)\,, $
with a real frequency~$\omega$ the resulting equation in~$\sigma$ is
the Mathieu equation~\cite{PandoZayas:2003yb,Bigazzi:2004ze}. The
solution which satisfies the boundary condition at the \emph{left} end
(i.e.~$\sigma=-\pi/2$) is given, up to an overall multiplicative
constant, by
\begin{multline}
\label{e:etasolD}
f(\sigma) = 
C( \omega^2 -K , \frac{K}{2}, -\frac{\pi}{2})
      S(\omega^2 - K , \frac{K}{2} ,\sigma)
     -S( \omega^2 -K  , \frac{K}{2}, -\frac{\pi}{2} )
      C(\omega^2 -K , \frac{K}{2} ,\sigma)\,,
\end{multline}
where $C$ and $S$ are the Mathieu functions. We now need to tune $\omega^2$ such that the
boundary condition at the \emph{right} end (i.e. $\sigma=\pi/2$) is
satisfied.
This boundary condition at~$\sigma=\pi/2$ can be satisfied by making
use of the Mathieu characteristic functions~$a_n(q)$ and~$b_n(q)$,
which give the value of the first parameter of the even and odd
Mathieu functions respectively, such that they are periodic with
period~$2\pi n$.  We use the following properties of the Mathieu
functions,
\begin{equation}
\begin{aligned}
S( a_n(q), q, \pm\pi/2) = S( b_n(q), q, \pm\pi/2) &= 0  &&\quad \text{for even $n$,}\\[1ex]
C( a_n(q), q, \pm\pi/2) = C( b_n(q), q, \pm\pi/2) &= 0  &&\quad \text{for odd $n$.}
\end{aligned}
\end{equation}
These properties imply that for even~$n$, the second term
of~\eqref{e:etasolD} vanishes and the first one satisfies both
boundary conditions. For odd~$n$, the situation is reversed, and the
first term in~\eqref{e:etasolD} vanishes altogether while the second
term satisfies both boundary conditions. We thus see that the boundary
condition at~$\sigma=\pi/2$ is satisfied for any of the frequencies
\begin{equation}
\label{e:allfrequencies}
\begin{aligned}
\omega^2_n = a_n(K/2) + K\quad & \text{for $n\geq 0$}\,,\\[1ex]
\omega^2_n = b_n(K/2) + K \quad & \text{for $n> 0$}\,.
\end{aligned}
\end{equation} 
This spectrum has been plotted in figure~\ref{f:frequencies}. At
leading order these $w^2$ behave like~$n^2$ but there
are~$K$-dependent (and thus~$L$-dependent) corrections.

Thus the action of the harmonic oscillators is that of \ref{lotLHO} but with the frequencies $\omega_n$ instead of $n$.  From the expressions given above for $c_K$ for the various different confining models it is clear that for large 't Hooft parameter $g_s N$ the corrections are suppressed by negative powers of $g_s N$. However,  for real hadrons this is not the case anymore . Correspondingly  
the wave
function for the ground state of this harmonic oscillator behaves like
\begin{equation}
\label{effpsi}
\Psi[z_n]=\left(\frac{\sqrt{ \big(a_n(K/2) + K\big)}}{\pi \alp_{\text{eff}} } \right)^{1/4}\exp\left[ - \frac{1}{2\alp_{\text{eff}}}   
   \sqrt{ \big(a_n(K/2) + K\big)}\, z_n^2 \right]\,.
\end{equation}

\begin{figure}[t]
\begin{center}
\includegraphics[width=.6\textwidth]{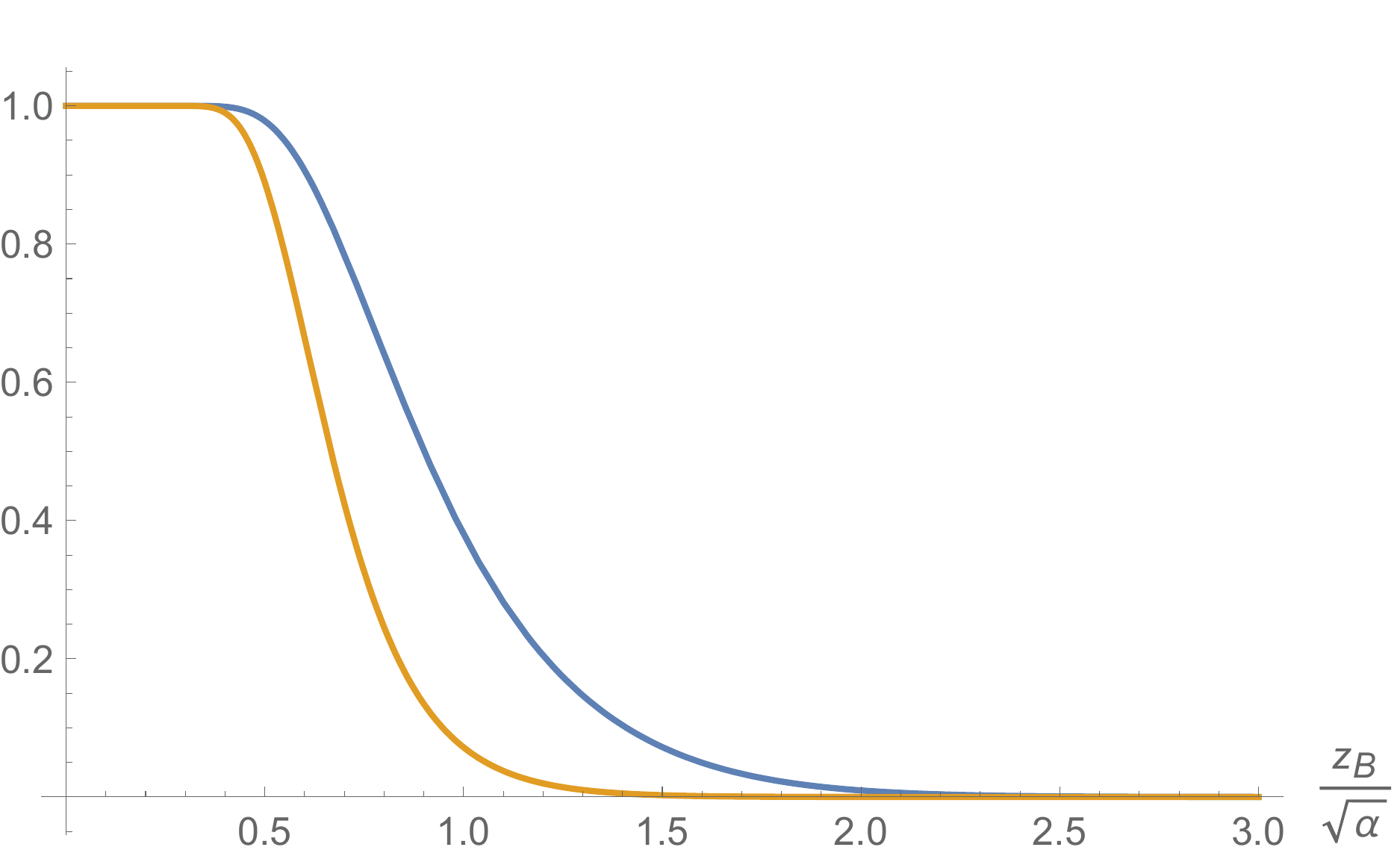}
\end{center}
\caption{The probability to fluctuate (\({\mathcal P}^{\text{min}}_{\text{fluct}}\) of eq. \ref{Pmin}) for the flat spacetime $K = 0$ (blue line) and for curved spacetime with $K = 10$ (yellow), drawn as a function of \(\frac{z_B}{\sqrt{\alp_{\text{eff}}}}\). \label{curvedsuppression}}
\end{figure}

In \cite{Bigazzi:2004ze} it was shown that for small values of the parameter $K$, namely for hadrons of small mass,  one can treat the mass term as a perturbation of a Schrödinger equation and in this way approximate the frequencies as follows
\be\label{wn}
\omega_n \sim \sqrt{n^2+ \frac{K}{2}}
\ee 
In figure \ref{curvedsuppression} we draw the suppression factor as a function of ${z_B}/{\sqrt{\alp_{\text{eff}}}}$ for the flat case $K=0$ and for the case of $K=10$. It is clear from the comparison between the two graphs that the effect of the curvature of spacetime is to enhance the suppression factor. 

\subsubsection{ String with massive endpoints} \label{sec:decay_exp_mass}
\begin{figure}[ht!]
\begin{center}
\includegraphics[width=.6\textwidth]{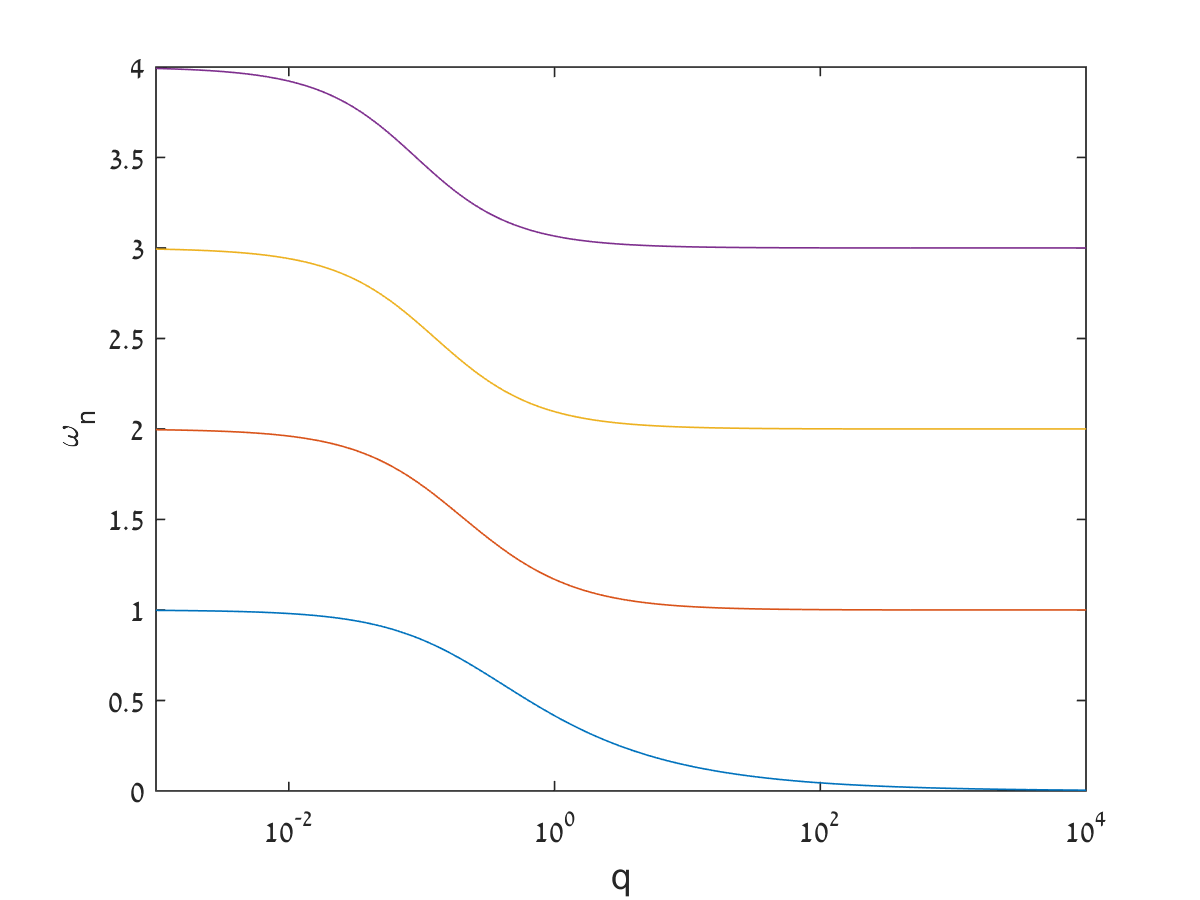}
\end{center}
\caption{The frequencies of the first excited states as a function of $q = m/TL$. \label{massiveendpointsupp}}
\end{figure}

So far we have discussed the suppression factor due to the fluctuations that are subject to Dirichlet boundary conditions. This corresponds to having infinitely heavy endpoints. However, as was described in section \ref{sec:HISH} the holographic stringy hadron can be approximated  by a string in flat spacetime  with endpoints with finite mass. The latter corresponds to the energy of the  vertical segments of the holographic string. Thus the spectrum of fluctuations discussed above both for flat and curved backgrounds will be further corrected due to the boundary conditions imposed by the finite massive endpoints. Here following the spirit of the HISH model we take that the massive endpoints are of a string residing in flat spacetime. The  structure of the fluctuations in the plane of rotation and those transverse to the plane is different. For our purpose only the latter are relevant. To determine boundary condition we add to the action of the string an action of a relativistic massive particle on its ends. The latter can be written as
\be
S_p = - m \int d\tau \sqrt{-(\dot X)^2}
\ee  
In the orthogonal gauge (for details see for instance \cite{Sonnenschein:2016pim}) the boundary equations of motion for the fluctuations $z(\tau,\sigma)$ takes the form
\be
 \left [ T z^\prime \pm m \pa_\tau\left (\frac {\dot z}{\sqrt{-(\dot X)^2}} \right )\right]_{\sigma=\pm\sigma_b}=0 
\ee
where $\pm\sigma_b$ are the boundary values of $\sigma$.
The corresponding frequencies $\omega_n$ are determined by the following transcendental equation
discussed in section \ref{sec:quantum_massive}
\be\label{transcendental}
\tan( \omega_n) =\frac{ 2 q \omega_n}{q^2 \omega_n^2 -1}
\ee
where $q= \frac{m}{TL}$.
The effect of the mass on the frequency of the first excited states is drawn in figure
\ref{massiveendpointsupp}. If $q=0$ or $q\rightarrow\infty$ on both ends, then $\omega_n=n$.
The frequency as a function of $q$ interpolates between a state $n$ for the Dirichlet boundary condition $q=0$ and a state $n+1$ for the Neumann boundary condition $q\rightarrow\infty$. If the string is rotating then eq. \ref{transcendental} may be further modified.

Since the starting point for the computations of section \ref{s:stringbits} was \(q\to\infty\), and it was done using Dirichlet boundary conditions, reducing the endpoint masses to a finite value increases the frequencies and hence enhances the exponential suppression factor to lower the probability for pair creation. On the other hand, there is the zero mode, which is not present at \(q\to\infty\), but shows up for finite \(q\). This is the lowest curve in figure \ref{massiveendpointsupp}. This adds a mode with small frequency and a wide wave function, which works the other way to reduce the suppression. Whether the endpoint masses enhance or reduce the suppression is then a function of the relative strengths of these competing effects.
\subsubsection{Summary holographic suppression factor} \label{sec:decay_exp_sum}
The outcome of the analysis of the previous subsections  is that the determination of the suppression factor  in holography involves several layers.
  If the decaying hadron is built from light quarks one can ignore the corrections due to massive endpoints given in section \ref{sec:decay_exp_mass}. If  the decay involves the creation of a light quark antiquark pair, then the corrections due to curved spacetime (section \ref{sec:decay_exp_curved}) are small. For the case of hadron with heavy endpoint particles and for the creation of a heavy pair one has to take into account the correction to the basic exponential suppression factor. Whereas the latter depends only on the masses of the quark antiquark pair and does not depend on the total mass of the hadron and not on the masses of the endpoint particles, the full expression of the suppression does depend also on the total mass and the endpoint masses.
	One can parametrize the suppression factor as follows
	
\be
\exp\left(-\text{suppression factor}\right) =  \exp\left(-2\pi C(T_{\text{eff}},M, m_i ) \frac{ m_{sep}^2}{T_{\text{eff}}}\right)\,, \ee
with
\be
\qquad C(T_\text{eff},L) = C(T_{\text{eff}},M,m_i) \approx 1 + c_c\frac{M^2}{ T_{\text{eff}}}   +\sum_{i=1}^2 c_{m_i} \frac{m_i}{M}\,.
\ee
We have expanded  the dimensionless coefficient $C$ for the case that $c_c\frac{M^2}{ T_{\text{eff}}}\ll 1$ and  $c_{m_i} \frac{m_i}{M}\ll 1$ for both massive endpoints $m_1$ and $m_2$. The coefficients $c_c$ and $c_{m_i}$ measure the impact of the curvature and of the massive endpoints respectively. In section \ref{sec:decay_exp_curved} it was shown that generally \(c_C > 0\), enhancing the suppression. The coefficients \(c_{m_i}\) can be either positive or negative. Using the relation between the length of the hadron and its total mass derived in section \ref{sec:rotating_string} it is thus clear that the dependence on the length translates into a dependence on both the total mass of the hadron and the masses of the endpoints. 
 
As we will see in section \ref{sec:pheno_lambdas}, the experimental data from which we can cleanly extract the suppression factor is rather scarce, and therefore we do not find good reasons to attempt a more precise prediction of the dependence of the suppression factor on the total mass of the hadron and the masses of its endpoint particles. On the other hand, the suppression factor plays an essential role in describing fragmentation processes  that take place due to multi-breaking which is discussed in the next subsection. 
 
\subsection{Multi string breaking and string fragmentation} \label{sec:decay_multi}
\begin{figure}[ht!]	\centering
				\includegraphics[width=0.48\textwidth]{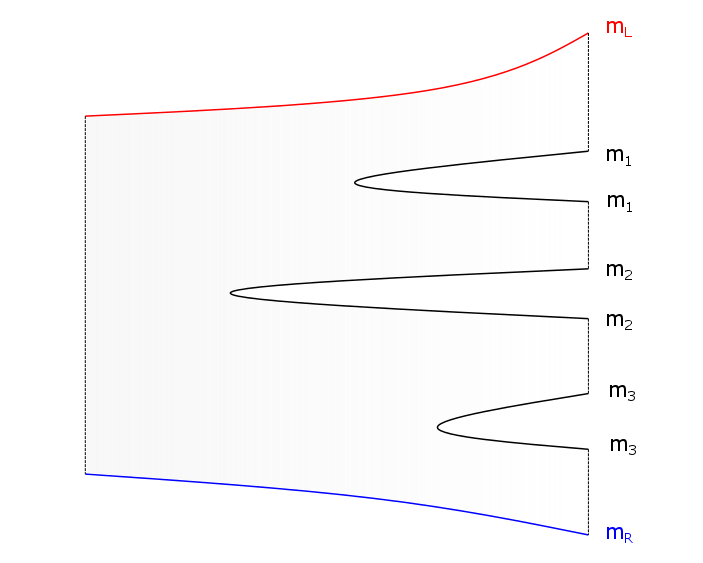}
				\caption{The multisplit of a string with massive endpoints into four strings.}
				\label{multimassivesplit}
	\end{figure}

The basic process of a string splitting into two strings can of course repeat itself and thus eventually describe a decay of a single string into $n$ strings.
This applies not only for a string with no particles on its ends but also to strings with massive endpoints. Thus the figure \ref{massivesplit} is generalized for the case of a  decay into multiple strings to the figure \ref{multimassivesplit}. 
 
The expression for the probability of the multiple decay in Schwinger mechanism 
\ref{decaypro} can be generalized into 
\begin{equation}\label{decaypro2}
{\cal P}=    
  \frac{T_{\text{eff}}^2}{\pi^3} \sum_f\sum_{n_f=1}^\infty \frac{1}{n^2} \exp\left(-2\pi n_f \tilde C  \frac{ m_{sep,f}^2}{ T_{\text{eff}}}\right)\,.
\end{equation}   
where $f$ runs over the possible flavors of the created pair, \(n\) is the number of pairs created, and  $m_{sep_i}$ is the mass of the quark and antiquark generated at the splitting point. The endpoint masses and the curvature affect the constant \(\tilde C\).

\begin{figure}[ht!]
			\centering
				\includegraphics[ width=200bp,trim={100bp 100bp 100bp 50bp}]{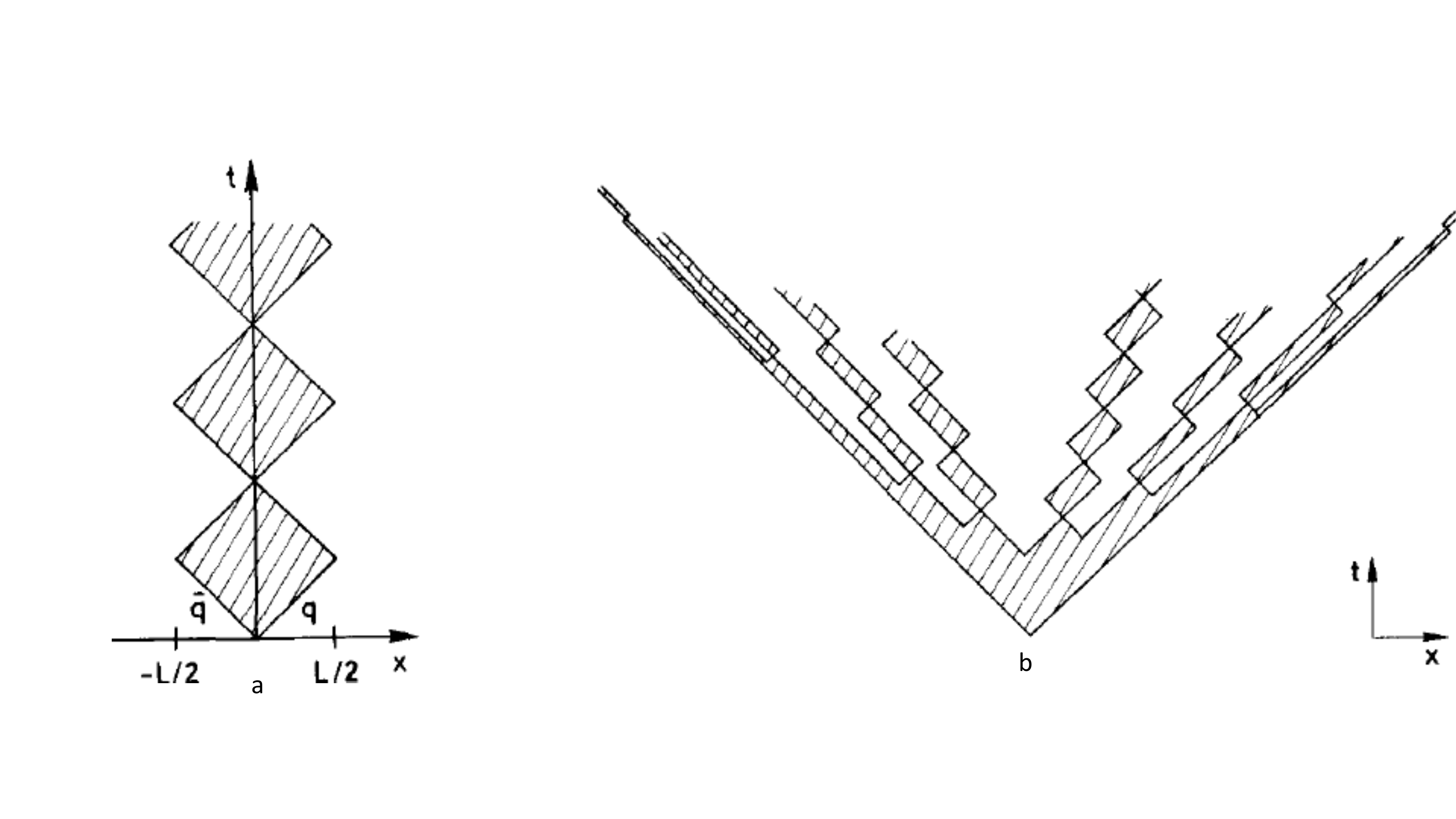}
				\caption{String fragmentation in the Lund model \cite{Andersson:1983ia}. (a)  The yo-yo oscillation of the flux tube between a $q$ and a $\bar q$ pair. (b) Multi-breaking of the original flux tube.}
				\label{Lund}
	\end{figure}

The process of multi-breaking is sometimes referred to as the ``string fragmentation mechanism'' (SFM)  which is the generator of jets that  emerge from hadron-hadron  and $e^+ e^-$ collisions. However, the multi-breaking described in figure \ref{multimassivesplit} is not the same as the one used in models of the SFM. In fact there are several different mechanisms depending on the angular momentum of the string, the momenta of the original string endpoints and the momenta of the endpoints (quark-antiquark pair) created. In our discussion of sections \ref{s:stringbits}--\ref{sec:decay_exp_sum} we have assumed that the string is at rest, and there are no momenta to the endpoints. Determining the decay width of a rotating string from the one at its rest frame was discussed in section \ref{sec:decaywidth_rot}. For the SFM the situation is different. Its underlying picture is based  on having non-trivial and opposite momenta to the two endpoints of the string. In fact the Lund model \cite{Andersson:1983ia} starts with a string with a longitudinal ``yo-yo'' mode, as in figure \ref{Lund}a. It is well known that there is no longitudinal modes in a string with massless endpoints but a string with massive ones can have such a mode. Furthermore, as can be seen from  figure \ref{Lund}b, in the Lund model  the generated quark and antiquark also carry equal but opposite momenta. We defer the analysis of such systems and in particular the structure of jets that follows from the HISH model to a future research project. What is common to the HISH model and the Lund model is that the basic probability of  breaking  the string and creating a pair has the same exponential suppression structure in both setups. This suppression is usually attributed to the Schwinger mechanism, but this can be replaced if necessary with the exponential factor rising from holography.

\section{Decay processes of hadrons} \label{sec:decay_hadrons}
The tearing of an open string as described in the previous sections is the main mechanism for strong decays of mesons. In this section, we build upon the simple picture of open string decays to discuss baryon decays, based on the stringy description of baryons found in \cite{Sonnenschein:2014bia}, and glueball decays following \cite{Sonnenschein:2015zaa}. We also discuss sub-leading decay processes of mesons, namely Zweig suppressed decays of quarkonia and the isospin violating decay \(D^*_s\to D_s+\pi\), both of which requiring that the string endpoints meet and annihilate. We then apply a similar mechanism for the decay of stringy exotic hadrons \cite{Sonnenschein:2016ibx}. Finally we discuss the decays associated with a breakdown of the vertical segments of the holographic string.

 \subsection{Decays of baryons} \label{sec:decay_baryons}
\begin{figure}[ht!] \centering
					\includegraphics[width=.66\textwidth]{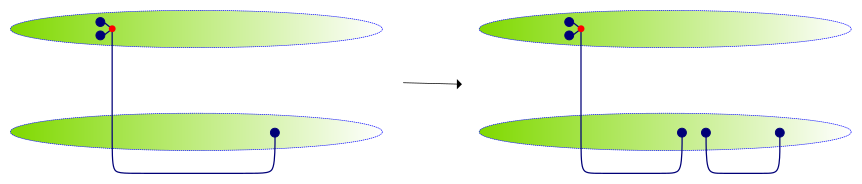}
					\caption{\label{fig:ssddecay}  A doubly strange baryon with an \(ss\) diquark decaying into a doubly strange baryon and a non-strange meson.} 
\end{figure}

As was described in section \ref{sec:HISH} and drawn in figure \ref{holtoflat2}, the string configuration that follows from the spectra of baryons is that of a single flat string that connects a quark on one end and a baryonic vertex (BV) plus a diquark in the other end. This is mainly based on the experimental fact that the slope $\alp$ of the baryon trajectories is within 5\% the same as that of the meson trajectories. It is thus clear that the leading holographic decay mechanism of a baryon is the same as that of a meson. Quantum  fluctuation can lift  the string up to a flavor brane and a breakup of the string will create two new vertical segments, namely a quark and antiquark pair. Therefore the same suppression factor that was found above for mesons occurs also for baryons. A decay process of a holographic baryon is depicted in figure \ref{fig:ssddecay}.

Naturally, a baryon decays into a baryon and a meson. In HISH a hadron is a baryon if it contains a baryonic vertex which is a junction that connects 3 strings, and in particular it connects to a diquark via two very short strings and to a quark with an ordinary long hadronic string. Thus the outgoing baryon that follows a decay of a baryon necessarily includes the diquark. For a baryon of given flavor content, there are a priori several options of a diquark. In holography in principle one should be able to determine the interaction between the string endpoints that reside on flavor branes. These are flavor and electromagnetic interactions mediated by the flavor gauge bosons on the probe branes. The diquark formed should have obviously the lowest energy in comparison with the energy of other possible diquarks. This type of calculation has not been done but instead in \cite{Sonnenschein:2014bia} the structure of certain baryons was determined from the spectrum. In particular it was shown that for $\Xi$ a better fit is achieved for a diquark of the form $(ds)$ over a diquarks of $(ss)$.

Another way to determine what is diquark pair and which is the stand-alone quark is by identifying the decay products. Suppose that a given baryon has the quark content of $q_1q_2q_3$. Denoting a diquark formed from  $q_i$ and  $q_j$ by  $(q_iq_j)$
we can have the following configurations and correspondingly the following decays

\bea
 & & [(q_1q_2)q_3]\qquad\qquad  \ \ \ \  [(q_1q_3)q_2]\qquad\qquad \ \ \ \  [(q_2q_3)q_1] \CR
& & \ \ \ \   \Downarrow \qquad\qquad  \ \ \ \ \ \ \ \ \ \ \ \ \ \ \ \ \Downarrow \qquad\qquad \ \ \ \  \ \ \ \  \ \ \ \ \  \ \ \ \  \Downarrow \CR
& & [(q_1q_2)Q_i] [\bar Q_iq_3]\qquad  [(q_1q_3)Q_i] [\bar Q_iq_2]\qquad [(q_2q_3)Q_i] [\bar Q_iq_1]  \CR
\eea
where $Q_i \bar Q_i$ is the pair produced at the split point, and $i$ runs over the various flavors that can be created. Suppose that $q_3$ stands for a charm quark and $q_1$ and $q_2$ are not charmed,  then it is clear that if the outcome of the decay is a charmed meson (left option) the diquark does not include the charmed quark whereas if there is a charmed baryon (middle and right options) then the diquark does include a charmed quark. This enables us to determine the diquark structure of a given baryon and in particular if it composed of only one structure or it can be a mixture  of various options which follows in this example by observing that there are decay channels with both charmed meson and others with charmed baryon. In section \ref{sec:baryon_structure} we analyze several decays of flavored baryons.
\subsubsection{ Decays of Y-shaped baryons} \label{sec:decay_Y}
\begin{figure}[ht!]
			\centering
				\includegraphics[ width=.60\textwidth]{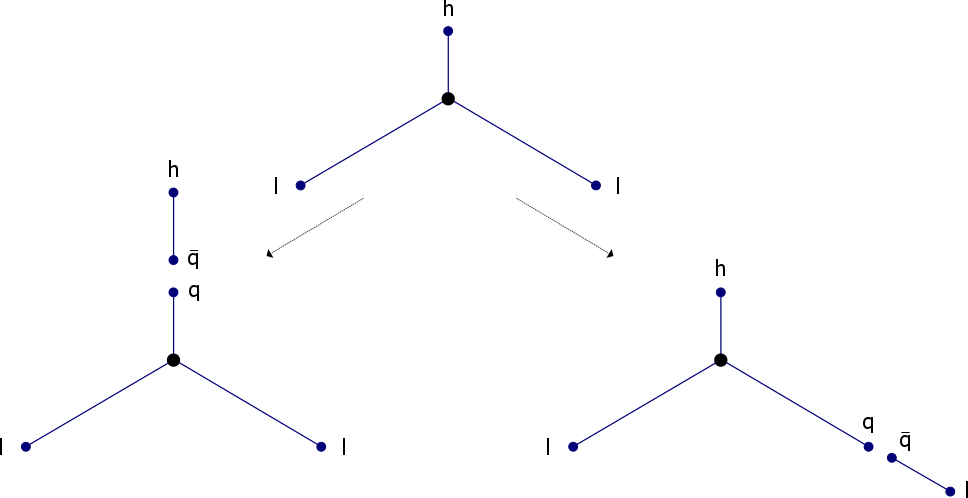}
				\caption{The HISH setup of  a decay of a Y-shaped baryon composed of two light quarks and one heavy quark.}
				\label{decayY}
	\end{figure}

We have argued above that baryons are found to furnish (modified) Regge trajectories with a slope close to that of mesons and hence the conclusion that also the structure of a baryon is based on a single string connecting a quark to a diquark. However, there are baryonic states that do not fit exactly the trajectories and some of them may have a structure different from a single string and in particular form a Y-shaped string.  As was discussed in \cite{Sonnenschein:2014bia} one can show that the Y-shape is unstable at the classical level as a string configuration. However, since we consider states that undergo strong decay anyway and hence have a very short life time, one may observe one of these ``metastable'' Y-shaped baryons. In this case there are three string arms that can break. If the string endpoints are all of the same flavor than there is not much difference in terms of the decay products  between a decay of Y-shape and a single string. On the other hand, if there are both light and heavier quarks on the  endpoints of a given baryon then the breaking of different arms will yield different decay products and one can estimate their relative width.
We demonstrate this in figure \ref{decayY} this time in the HISH and not the holographic setup.  Denoting by  $m_l$ and $m_h$ the light and heavy endpoint masses respectively, a breaking of the short arm connecting to the $m_h$ yields $ B+ M_h$, where $M_h$ stands for a meson that includes the heavy quark, or $B_h + M$ when the long arm connecting to $m_l$ is split. We assume that the pair created is of a light quark, which will always be the dominant mode.
	Assuming that  the linearity relation between the decay width and and the length we find that the ratio between the  decay width of the two possible decays drawn in the figure are 
		\be \frac{\Gamma_i}{\Gamma_j} = \frac{l_i}{l_j}
		\ee
		where $l_i$ are the length of the two different arms connecting the baryon vertex to the different type of quarks. We can determine the length of the different arms $l_i$ using the results of section \ref{sec:rotating_string_asym} and \ref{sec:casimir_force}. Notice that the arms are strings that have on one end a massive quark and on the other the baryonic vertex, which may also have a finite mass of its own. In \cite{Sonnenschein:2014bia}, based on the analysis of the spectra of baryons we have concluded that the best fit is achieved when the mass of the baryonic vertex can be neglected. However, in general we have to consider the arms as being suspended by two masses, one of the quark and another of the BV. The boundary equation \ref{repCasf} applies then only to the quark endpoint and on the vertex end there is a ``static stability'' caused by the three arms.
\subsection{Decays of glueballs} \label{sec:decay_glueball}
As was briefly explained in section \ref{sec:HISH} and was described in details in \cite{PandoZayas:2003yb,Sonnenschein:2015zaa}, glueballs are described in holographic and  HISH models as rotating folded closed strings. In the former case  they are localized in the vicinity of the holographic wall. Possible Regge trajectories associated with closed string glueballs were presented in \cite{Sonnenschein:2015zaa}. 

\begin{figure}[ht!]
			\centering
				\includegraphics[ width=0.95\textwidth]{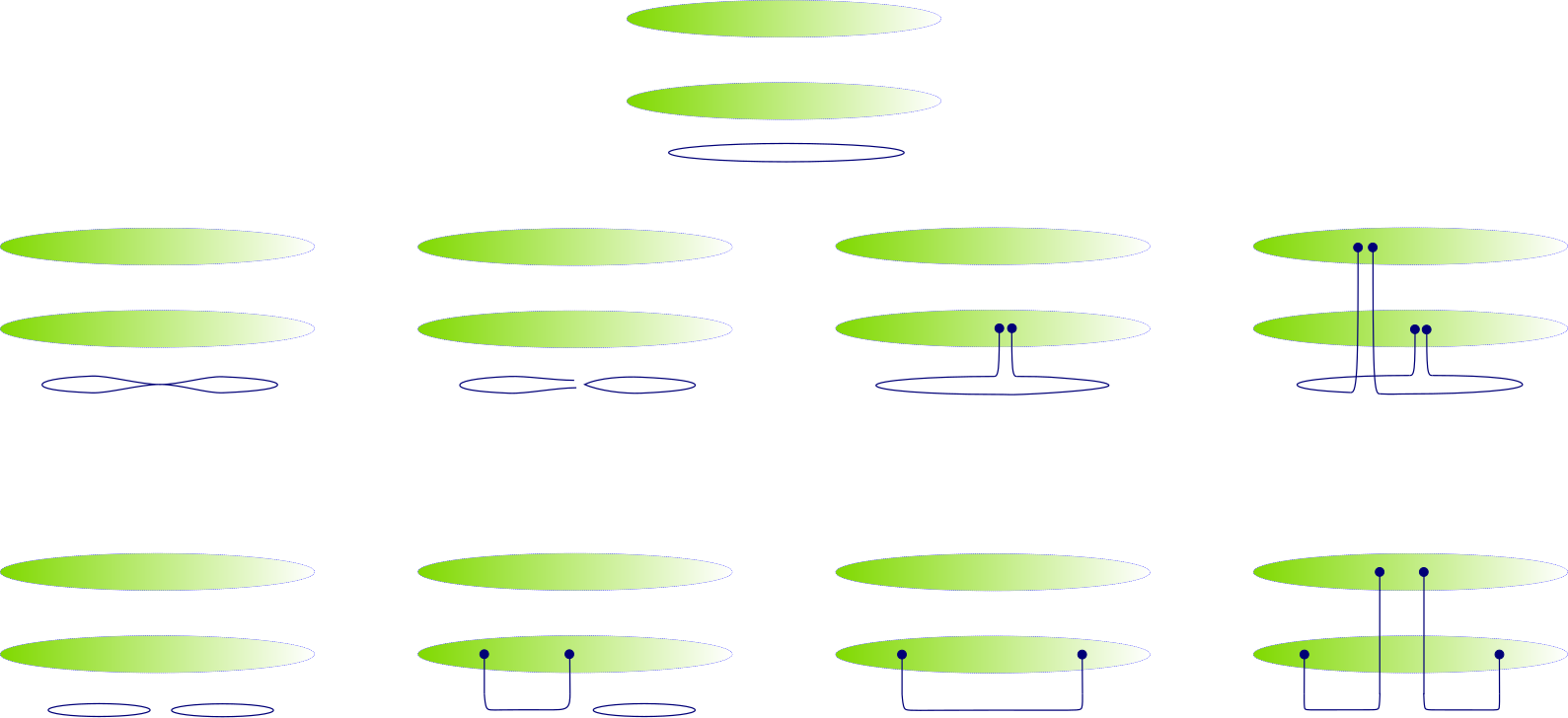}
				\caption{Four possible ways for a glueball to decay.}
				\label{decayglueball}
	\end{figure}

The decay processes of glueballs are also related to the breaking apart of the string, in this case the closed string. The simplest channels of the  decay of a glueball are:
(a) Decay into two glueballs, (b) decay into a glueball and a meson, (c) decay into a meson, and (d) decay into two mesons. These different channels are depicted in figure \ref{decayglueball}. We have assumed that the probability that the folded closed string will have a crossing point where it breaks apart as $\sim L \Gamma_{\text{cross}}$. Thus, the widths of the various decays behave as
\begin{align}
\Gamma_a &\sim L \Gamma_{\text{cross}}
\qquad &\Gamma_b &\sim L \Gamma_{\text{cross}} e^{-2\pi C m_{sep}^2/T} \nonumber \\ 
\Gamma_c &\sim L  e^{-2\pi C m_{sep}^2/T}\qquad 
&\Gamma_d &\sim L^2  e^{-2\pi C m_{sep}^2/T} e^{-2\pi C m_{sep}^{\prime2}/T}\,.\end{align}
In  figure \ref{decayglueball} we took that in (b) and (c)  a light quark meson is been created and in (d) two mesons with one light and one medium quarks. Obviously there are more options for these channels.
As will be discussed in section \ref{sec:pheno}, phenomenologically  the most relevant decay is of type (d) with two light mesons. Option (a) is less likely since the lightest glueball is in fact heavy and the mass of all known candidates is smaller than twice the mass of the lightest glueball. Because of this fact the phase space of option (b) is also smaller than that of (d). For option (c) to take place there should be a meson and a glueball with the same mass (with mixing between them). In such a case it will eventually decay into two mesons like in (d). 
For the channel of decaying into two mesons (d)  there is a
 distinguishing feature of closed strings, that unlike open strings,  can decay to any of the three channels of two light mesons made of only \(u\) and \(d\), two strange mesons, or two \(\ssb\) mesons. Each creation of a pair of \(\ssb\) will be exponentially suppressed compared to light quark creation with a factor of $ \exp\left(-2\pi C (m_s^2-m_{u/d}^2)/T\right)$. This ratio is the strangeness suppression factor, and we denote it by \(\lambda_s\). This means that we can expect a glueball to decay to either of the three type of channels, with a defined hierarchy between them. Normalizing each mode by its available phase space, we will obtain the ratio
\be \frac{1}{P.S.}\Gamma(GB\to\text{2 light}) : \frac{1}{P.S.}\Gamma(GB\to\text{2 strange}) : \frac{1}{P.S.}\Gamma(GB\to\text{2 }\ssb) = 1 : \lambda_s : \lambda_s^2\,\ee
This result is in contrast to the usual approach to glueball decays, which assumes that glueball decays are flavor blind \cite{Mathieu:2008me}, and in particular that there is no suppression of \(\ssb\) creation beyond phase space differences. In holography flavor blindness can be found when describing glueballs as fields rather than strings \cite{Brunner:2014lya,Brunner:2015oqa}. The existence of strangeness suppression in glueball decays is then a test of the stringy picture.

The strangeness suppression factor can be measured in hadronic decays, as we will do in section \ref{sec:pheno_lambdas}. A numerical example for the expected branching ratios in glueball decays is given in section \ref{sec:pheno_glueball}.
\subsection{Zweig suppressed decay channels} \label{sec:decay_Zweig}
\begin{figure*}[t!]
			\centering
				\includegraphics[ width=0.95\textwidth]{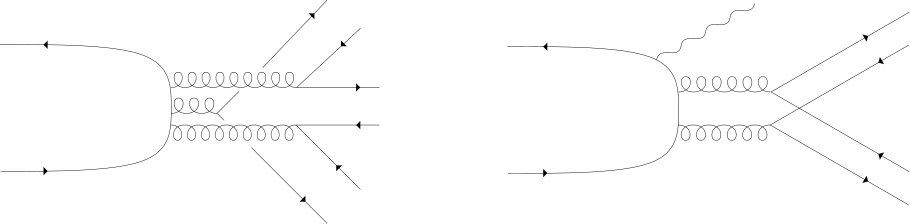}
				\caption{Zweig suppressed decay of quarkonia in QCD. \textbf{Left:} Decay via 3 gluons. \textbf{Right:} Radiative decay via 2 gluons and a photon.}
				\label{Zweig}
	\end{figure*}

Certain heavy quarkonia, namely $c\bar c$ or $b\bar b$ mesons, cannot decay via the mechanism of breaking apart of the horizontal string. The reason for that is simply that their mass is smaller than twice the mass of the lightest meson that carries the flavor of the quarks that constitute the quarkonium. The only way for these states to decay is via an annihilation of the $q\bar q $ pair. This process is referred to as the Zweig suppressed decay. In ordinary perturbative QCD this annihilation involves the creation of three gluons, or two gluon and a photon as depicted in figure \ref{Zweig}. 

The questions that this situation raises for the HISH model are the following: (i) What is the process of the $\bar  q q $ pair annihilation in the stringy picture. (ii) Is it indeed suppressed in comparison to the ordinary split of the horizontal string. (iii) What is the prediction for the decay width of the Zweig decay channel for a given meson.

\begin{figure}[ht!]
			\centering
				\includegraphics[ width=0.95\textwidth]{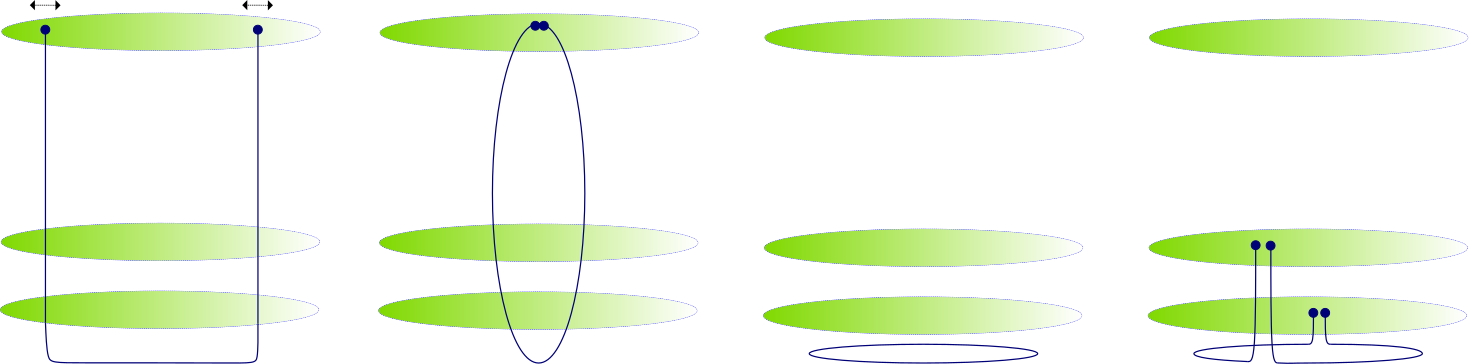}
				\caption{Holographic Zweig suppressed decay of a quarkonium meson via annihilation of the endpoints. The endpoints of the strings fluctuate and eventually meet and annihilate. The resulting closed string decays to two mesons by tearing at two different points.}
				\label{Zweigstringa}
	\end{figure}

\begin{figure}[ht!]
			\centering
				\includegraphics[ width=0.95\textwidth]{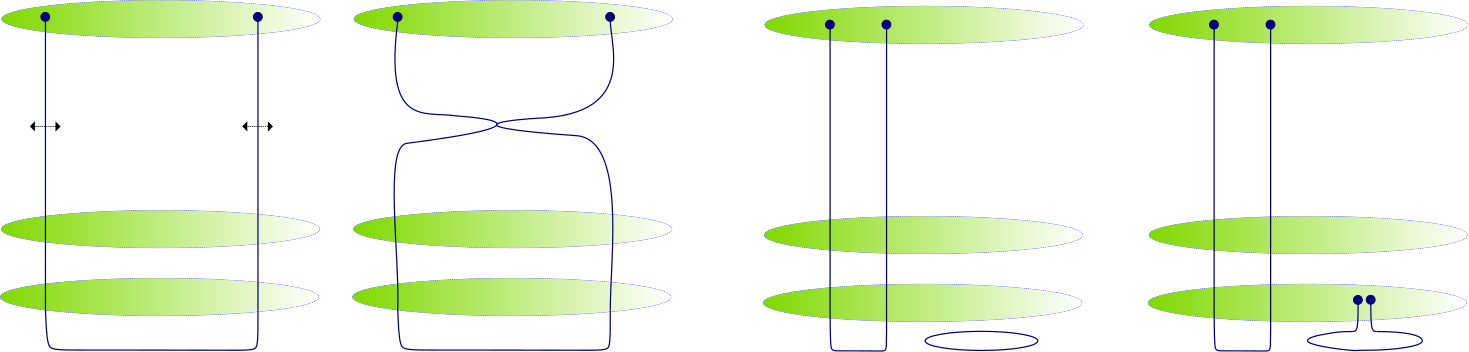}
				\caption{Holographic Zweig suppressed decay of excited quarkonium to a lower state. The vertical segments of the string fluctuate and meet. The string tears into an open string (the lower quarkonium state) and a closed string. The closed string tears to form a light meson.}
				\label{Zweigstringb}
	\end{figure}

The decay process when breaking of the horizontal string is not permitted can in fact take place in two ways. One is that due to fluctuations the two endpoints of the string that reside on the flavor brane (charm or bottom) hit each other and annihilate. This process, which is depicted in figure \ref{Zweigstringa}, transforms open string meson into a closed string glueball which then splits into  mesons as was discussed in section \ref{sec:decay_glueball}. An alternative process is that the vertical segments of the original quarkonium, again due to fluctuations, will collide and cause a  glueball closed string to be emitted from the original meson while still leaving a string with endpoints on the same flavor brane as before (see figure \ref{Zweigstringb}). Note that the difference in mass between the original quarkonium state and the lower state it decays into is due to the different length of the horizontal string.

 This process cannot take place for the lightest quarkonium state but only for an excited state, or a higher point on the corresponding Regge trajectory, and only when there is a phase space for decay into a lower point on the trajectory plus additional mesons. These two processes provide an answer to (i). As for (ii), since the total probability for a Zweig suppressed process involves the product of the probability of the annihilation of the endpoint pair times the probabilities of the break up of the produced glueball it is obvious that it is smaller than that of only breaking a meson or a glueball, although this is only a qualitative statement and it is unclear whether this mechanism can account for the large suppression of these processes. The detailed calculation of the width (iii) will involve an analysis of the fluctuations of the endpoints for (a) and of the vertical string for (b).  We leave this for a future research. Here we use a ``hand waving'' argument for the determination of the decay width.

For process (b) we need to describe the probability that due to the fluctuations the two segments will meet. This is analogous to the probability that the horizontal string segment hit a flavor brane. There are however two major differences: (i) In this case it is a fluctuations of two strings and not one, and more importantly, unlike the horizontal case where the tension is constant, along the holographic radial direction the string tension varies. A crude approximation will be to use an averaged tension. Then the probability of the two vertical segments to meet is crudely given by   
\bea
{\cal P}&=& \int_{-\infty}^{\infty} dx\, \psi(x-L/2) \psi( x+ L/2)= \CR
&=& \int_{-\infty}^{\infty} dx\, \exp[-T_{av}( x-L/2)^2] \exp[-T_{av} (x+L/2)^2] = \CR &=& \sqrt{\frac{\pi}{2T_{av}}}e^{-T_{av}L^2/2} = \sqrt{\frac{\pi}{2T_{av}}}e^{\frac{4(M-2m)^{2}}{9 T_{av}}}\CR
& &
\eea
where we have used the (classical) approximated length of eq. \ref{non-rel} in the last expression. In section \ref{sec:pheno_Zweig} we will confront these theoretical approximations with experimental data. 
\subsubsection{Virtual pair creation combined with a Zweig suppressed decay} \label{sec:virtualpair}

\begin{figure}[ht!]
			\centering
				\includegraphics[width=0.95\textwidth]{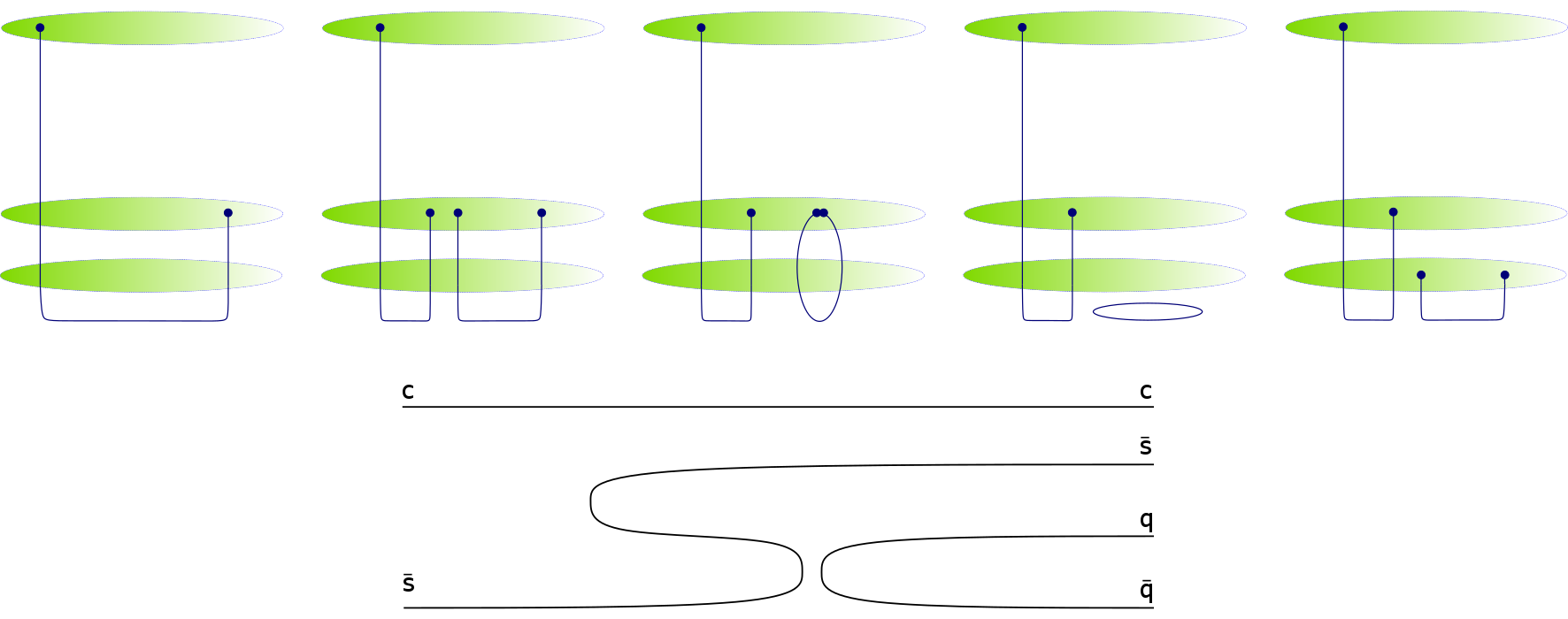}
				\caption{The stringy picture of the isospin violating decay \(D^*_s\to D_s\pi\). A virtual \(\ssb\) is created, which then decays via annihilation of the string endpoints and creation of a light quark pair. The bottom figure is an equivalent quark flow diagram of the same process.}
				\label{virtualpair}
	\end{figure}

So far we have discussed processes of string splitting where the pair of quark antiquark created are allowed by energy-momentum conservation laws. In principle we have to consider also cases where a virtual pair   is  been created in such a way that energy conservation is assured only for the full process. In figure \ref{virtualpair} we show an example of a pair of $s\bar s$ created at the splitting point of a hadron that has an $s$ quark on one of its ends. A Zweig suppressed decay mechanism of the meson built from one of the string segments with an $s$ and an $\bar s$ on its ends turns this meson into a glueball. The latter then fluctuates, reaches the $u/d$ flavor brane and breaks apart thus creating a neutral light meson. This mechanism can explain the observed isospin violating decay \(D^*_s\to D_s\pi\).

	The decay width for such a process includes two exponential suppression factors, one from the pair creation and another from the following annihilation, so that 
	\be
	\frac{\Gamma_i}{\Gamma}\sim \exp[{-2\pi C\frac{m_{sep}^2}{T}}]\exp[(-T_{Z}L^2/2]
	\ee
	where here $m_{sep}= m_s$, and $L$ is the length associated with the created \(\ssb\) meson.

\subsection{Decays of exotic hadrons} \label{sec:decayexotic}
Since the basic idea of the HISH models is that hadrons are strings, it seems natural to assume in this framework that also exotic hadrons, if such object ``exist'' are strings.  In fact, as we have argued in \cite{Sonnenschein:2016ibx} the decay process of such strings can be an important guiding line in their identification. 
As a prototype of the exotics we can consider the tetraquark which is described as 
 a single string with a massive diquark at one end and a massive anti-diquark at the other end. Therefore, its main decay mechanism will be like that of the mesons and baryons, via a breakup  of the long string connecting the two endpoints. The split 
 of the  string is accompanied with  a generation of a quark-antiquark pair at the two ends of the torn apart string. In this way we get a BV connected to a quark and a diquark on one side of the split string - which gives a baryon, while on the other side we get an antibaryon.

\begin{figure}[ht!] \centering
	\includegraphics[width=0.95\textwidth]{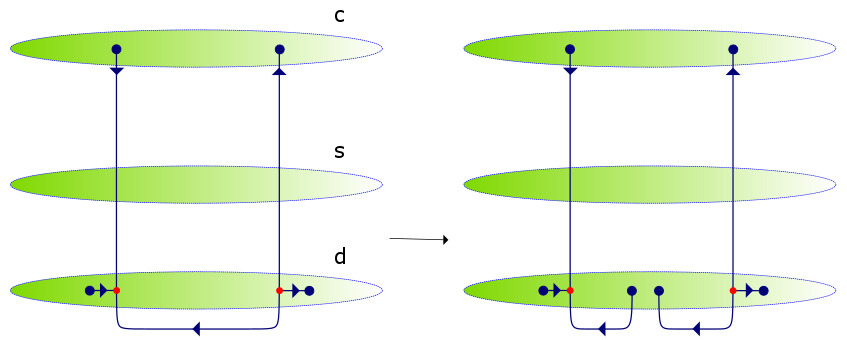}
				\caption{\label{holdecay} The decay  of a holographic tetraquark composed of a diquark of $c$ and $u$ and an anti-diquark of $\overline c$ and $\overline u$ into $\Lambda_c \bar\Lambda_c$. Outgoing (incoming) arrows signify quarks (antiquarks).}
			\end{figure}

\begin{figure}[ht!] \centering
	\includegraphics[width=0.55\textwidth]{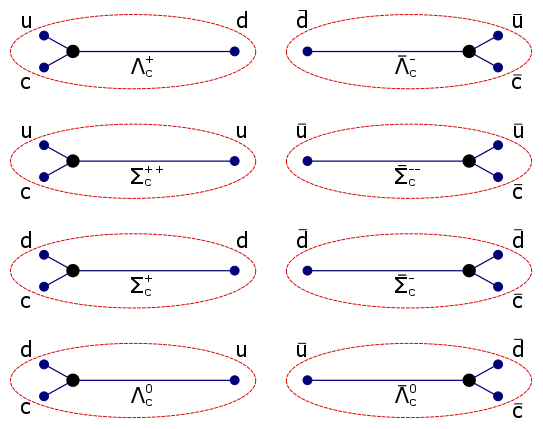}
				\caption{\label{HISHdecay} Different charmonium-like stringy tetraquarks and their decay modes.}
			\end{figure}

In figure \ref{holdecay} we draw a sketch of the holographic decay of the $Y_{(cd)(\overline{cd})}/Y_{(cu)(\overline{cu})}$ tetraquark into a $\Lambda_c \bar\Lambda_c$ pair. Figure \ref{HISHdecay} shows the same process in the HISH description including various other flavor combinations and their outcomes.

It is clear from these figures that the tetraquarks cannot decay through this mechanism of breaking the string into two mesons like $D$ and $\overline D$, but rather only to a baryon anti-baryon pair. On the other hand it is also clear that a stringy charmonium state decays via a string breaking and a creation of quark antiquark pair into a pair of charmed mesons, $D$ and $\overline D$.

The decay width of a tetraquark into a baryon and antibaryon via this mechanism 	has the same structure as that of a meson and a baryon, namely as for that latter it is   proportional to the length of the string \(L\) times the exponential suppression factor,
\be
\Gamma\sim L e^{-\frac{{m^q_{sep}}^2}{T}} \,,
\ee

In the holographic stringy model, we can construct additional stringy configurations \cite{Sonnenschein:2016ibx} forming exotic hadrons of higher orders (pentaquarks, hexaquarks, and so on). Also for such configurations the decay mechanism  based on  splitting  of a string  into two strings will take place. The results of the decay may be non-exotic hadrons like the baryon anti-baryon of above or a combination of exotic and non-exotic hadrons.
\subsection{Decays via breaking of the vertical segments} \label{sec:decay_vertical}
\begin{figure}[ht!]
			\centering
				\includegraphics[ width=200bp]{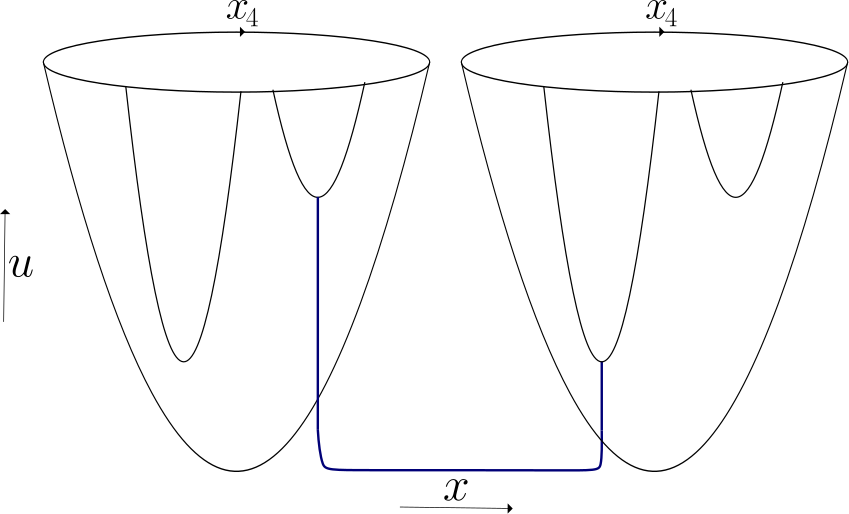}
				\caption{ The holographic setup of the light and heavy flavor branes with a heavy-light stringy hadron.}
				\label{flavorsetup}
	\end{figure}

So far in this section we have discussed only the probabilities that the horizontal segment of the  holographic hadron string will break apart (see figure \ref{quantumfluc}). However, there is nothing that prevents the vertical segments from  breaking apart so we should study the implications of such processes on possible decays of stringy hadrons.  Before doing that we would like first to clarify a point about the holographic setup.  Consider the case of a heavy light meson depicted in figure \ref{quantumfluc}. The left vertical arm that stretches from the heavy brane down to the ``wall'' seems to cross on its way both the ``medium'' and ``light'' flavor branes. Had this been indeed the picture we would have  anticipated that  this vertical string will disconnect  from the rest of the stringy hadron in either of these contact points with the flavor branes.  In fact we want to argue that the holographic setup is different as can be seen in figure \ref{flavorsetup}. In this setup the flavor branes associated with the light $u$ and $d$ quarks are separated along the compact $x_4$ direction from the $s$ and $c$ flavor branes. In this way, as can be seen in this picture, vertical segments do not intersect flavor branes. In  figure \ref{flavorsetup} this is shown for a heavy-light  stringy meson. Note that the meaning of this setup is that there are additional 
parameters of the setup which are the asymptotic separation distances along $x_4$ between the centers of the U shape probe branes.  
	
\begin{figure}[ht!]
			\centering
				\includegraphics[ width=200bp]{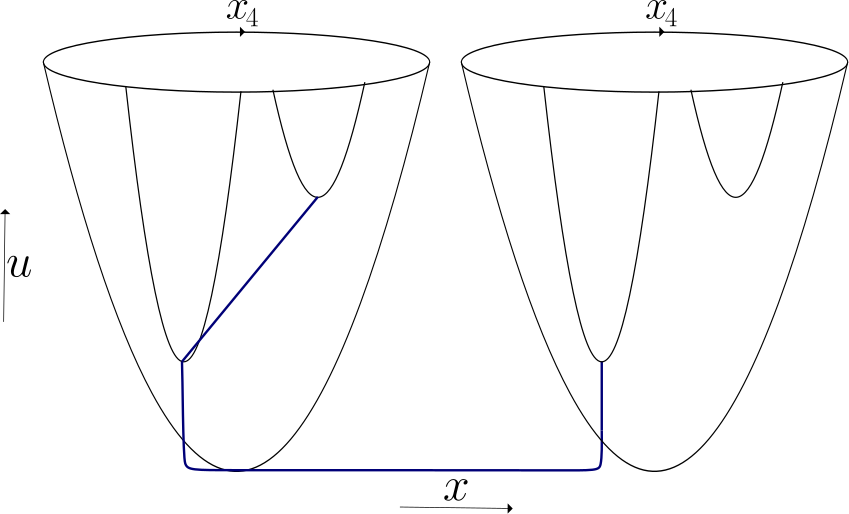}
				\caption{A vertical break. The vertical arm breaks by connecting to the light flavor brane thus creating a light stringy meson and a ``particle''.}
				\label{verticalbreak}
	\end{figure}

Now that we have clarified the flavor brane setup, we would like to discuss a process whereby a vertical segment breaks apart. Such a process  depicted in figure \ref{verticalbreak}	 can occur  for instance  if due to quantum fluctuations the left  vertical segment hits the medium  or light flavor branes. The result of the breakup is a stringy meson with a medium instead of heavy string endpoint particle on the left plus an additional string that connects  the heavy and medium flavor branes. Since this latter object stretches along the holographic direction $u$ and the compact coordinate $x_4$ but not along any space coordinate $x_i,\ i=1,2,3$, it is not a string from the point of view of the ordinary four dimensional spacetime but rather a point-like particle. To better recognize this string in the transverse directions, imagine a situation where all the four flavor branes $u$, $d$, $s$, and $c$ have been located in  the same radial coordinate. In that case there is  a $U(4)$ flavor gauge symmetry on the stack of the four flavor branes. A string between these flavor brane is  one of the 16 strings that constitute this symmetry. Now assuming that the $u$ and $d$ flavor branes are still at the same radial location, there will be four massless (zero length) strings and 10 massive ones associated with the massive gauge bosons related to the breaking of $U(4)$ to $U(2)\times U(1)\times U(1)$.

The process depicted in figure  \ref{verticalbreak} can be interpreted as a decay of $ D^0(c\bar u) 	\rightarrow  K^-(s\bar u) + (c\bar s)_v $ where the particle $(c\bar s)_v$ is a string the stretches along the $u$ and $x_4$ directions.. 

In a similar way to the computation of the width associated with the breaking of the horizontal segment, the width associated with the breaking of a vertical segment should be of a form like 
\be
\Gamma_{\text{vertical}} \sim \int_{u_\Lambda}^{u_c} du \exp\left(-C_v\frac{(\Delta_{x_4}(u))^2}{\alp(u)} \right)\,.
\ee
where we integrate over the radial coordinate \(u\) from the wall at \(u_\Lambda\) to the heavy flavor brane at \(u_c\), and assume each point's probability to fluctuate is exponentially suppressed with its distance (\(\Delta_{x_4}\)) from the lower flavor brane. The string tension is \(u\)-dependent, and so is the distance $\Delta_{x_4}$. The dependence of both on \(u\) is of course model dependent.

The question now is what is the interpretation of the decay processes associated with the vertical breaking and in particular what is the meaning of the ``vertical string'' created that connects the two flavor branes. We do not have a full answer to this question and we leave it for future investigation. Nevertheless let us mention some thoughts  about a possible scenario that may explain the role of  the vertical segments. 

Consider first the case of only light, $u$ and $d$,  quarks. For that matter we can still consider figure \ref{verticalbreak} but now with the right (left) vertical segment associated with a $d$ ($u$)  quark respectively. In terms of the quark content the decay process is therefore $d\bar u\rightarrow u\bar u+ (d \bar u)_v$ where as above we denote by $(d \bar u)_v$ the vertical string which is viewed as a particle from the ordinary spacetime point of view. In this way we have two kinds of mesons with light quarks: (i)  our ordinary stringy mesons that have been considered so far and (ii)  the vertical type mesons.  
 The question is, therefore,  with what type of  experimentally observed  mesons can the latter type  be affiliated with? One possibility is that the latter type is associated with the Goldstone bosons related to the spontaneous breaking of the chiral symmetry $SU(2)_L\times SU(2)_R\rightarrow SU(2)_D$. The vertical strings $(u\bar d)_v$ and $ (d\bar u)_v$ can be associated with $\pi^+$ and  $\pi^-$ respectively.  Recall, \cite{Sonnenschein:2014jwa} that the  ground state pion  does not fit well its corresponding Regge trajectory but rather it is lighter than what would follow from the trajectory structure.  However, vertical strings that connect the $d$ and $u$ flavor branes obviously cannot correspond to $\pi^0$ whose quark assignment is of the for $\frac{1}{\sqrt{2}}(u\bar u+ d\bar d)$. The same phenomenon can occur with the vertical segments that connect the strange flavor brane to that of a $d$ or $u$ flavor brane. These vertical strings that carry the $u\bar s, d\bar s, \bar d s,\bar u s$ may correspond to $K^+, K^0, \bar K^0, K^-$ respectively. Thus, the vertical segments have the quantum numbers of only six out of the eight Goldstone bosons. In addition to the missing $\pi^0$ also the $\eta$ with the quark structure of $\frac{1}{\sqrt{6}}( u\bar u+ d\bar d-2s\bar s)$ does not seem to relate to any of the vertical strings. As stated above we leave these issues for a future study.  

\subsection{Kinematic constraints for the decays of states on Regge trajectories} \label{sec:decay_Regge}
From the mass formulae of hadrons that lie on Regge trajectories, we can derive some simple results regarding their allowed decay modes, and relate kinematical constraints to the value of the Regge trajectory intercept.

\subsubsection{Light quarks}
We start with light quark (\(u/d\)) hadrons, whose Regge trajectories are linear, or very close to linear with small corrections from the string endpoint masses. States on a linear Regge trajectory obey the relation
\be N = \alp M_N^2 + a\,,\ee
where \(N = 0,1,2,\ldots\) is the sum of the orbital angular momentum \(L\) and the radial excitation number \(n\). Since \(N\) starts from zero, the intercept \(a\) defined by the above equation must always be negative. We denote the \(N-\)th hadron on a trajectory as \(H_N\).

For the linear trajectory, one can easily prove the inequality
\be M_N < M_{N-K} + M_K\,, \ee
which translates to
\be \sqrt{N-a} < \sqrt{N-K-a} + \sqrt{K-a}\,,\ee
or, after squaring both sides
\be a < 2\sqrt{(N-K-a)(K-a)}\,.\ee
For \(a<0\) this obviously holds for any \(N\) and \(K\). Therefore the decay \(H_N\to H_{N-K}+H_K\) is always forbidden.

Since \(M^2\) rises linearly with \(J\), at high angular momentum the mass difference between succeeding states decreases, as
\be \Delta M^2 = \frac{1}{\alp} \Rightarrow \Delta M = \frac1{2M}\Delta M^2 = \frac{1}{2\alp M}\,.\ee
If we look at the value above which a vector meson cannot be emitted in the transition from \(J\) to \(J-1\),
\be H_{N} \to H_{N-1} + \text{vector}\,, \ee
checking the inequality 
\be \Delta M_N = \frac{1}{2\alp M_N} < M_{\rho}\,, \ee
with \(\rho\) being the lightest vector meson. We can also use the isoscalar \(\omega\) meson, with a nearly identical mass. Using a typical value for the slope, \(\alp = 0.9\) GeV\(^{-2}\), we get the inequality
\be M_N > 700\text{ MeV}\,. \ee
This holds for any meson with \(J \geq 1\) so we can conclude that the \(S\)-wave decay between consecutive states in a Regge trajectory does not take place for any type of meson.

We can reframe the above as a statement about the intercept in a trajectory. The difference between the two lowest levels in a trajectory, \(N = 0\) and \(N = 1\), is the largest compared to the differences between higher consecutive levels, and it is
\be \frac{1}{\sqrt{\alp}}(\sqrt{1-a}-\sqrt{-a}) \ee
As a function of \(a\), this has a maximum at \(a = 0\) (we do not consider \(a>0\)) and decreases as \(a\) gets more negative. We can get an inequality on the intercept for when this difference is smaller than a given mass \(M_X\),
\be \sqrt{1-a}-\sqrt{-a} < \sqrt{\alp} M_X \ee
when
\be \sqrt{-a} < \frac{1-\alp M_X^2}{2\sqrt{\alp}M_X} \ee
For \(M_X = M_\rho = 776\) MeV, the lightest vector meson's mass, and assuming \(\alp = 0.9\) GeV\(^{-2}\) as before, the critical value of the intercept is
\be a = -0.10 \ee
So for any trajectory with \(a < -0.10\), there will never be the transition \(H_N \to H_{N-1} + \text{vector}\). This includes all light meson trajectories.

At some point, the mass difference is smaller than even the mass of pion and transitions between consecutive states in a trajectory (from \(J = N+1\) to \(J=N\)) are completely forbidden by kinematics. This happens when
\be \frac12 < \alp M_{N} M_\pi\,, \ee
or
\be M_N > \frac{1}{2\alp M_\pi} \approx 4\text{ GeV} \,,\ee
Since this occurs at such a high mass (for mesons made up of light mesons), it is not of much concern.

To summarize, we see that the mass formula for states in linear Regge trajectories, and with a negative intercept as found in nature, prevents decays of the form \(H_N\to H_{N-k}+H_k\). When the intercept is less than \(-0.11\), there are no decays of the form \(H_N\to H_{N-1}+\text{ vector}\).

At this point it is worth noting that the lightest mesons, namely the pions and kaons do not lie on Regge trajectories. Their excited partners form trajectories, but the ground states are lighter than what one would expect from the Regge trajectories. If the pseudoscalar pions and kaons were to lie on Regge trajectories, then they would have been heavier and their first excited partners would not have the possibility to decay strongly. In this way it is important that the pions and kaons are light, not only relative to \(\omega\), \(\rho\), \(\phi\), or other mesons, but also relative to the mass one would expect them to have based on the masses of the excited \(\pi\) and \(K\) mesons and the Regge trajectories they form.

\paragraph{Quark mass corrections}
In the last subsection we assumed linear Regge trajectories. With light quark masses, the first correction to the mass of the hadron, due to the masses is of the order \(J^{-1/4}\), and we find that
\be
M^2 = \frac{J}{\alp}\left(1+\frac{8\sqrt{\pi } }{3}\left(\frac{\alpha  m^2}{J}\right)^{3/4}\right)\,.
\ee
In the mass squared the first correction is of order \(J^{1/4}\). This is the classical formula. We can add an intercept \(J\) in order to obtain the quantum trajectories, as well as adding the dependence on the radial number. The two are done by replacing \(J\to N - a = J+n-a\) in the above equation.

With the quark mass corrections we can verify the inequality \( M_N < M_{N-k} + M_{k}\) numerically, and see that it still holds when adding masses, regardless of the value of the intercept (as long as it is negative).

The inequality \(a < -0.10\) guarantees that no decays of the form \(H_N \to H_{N-1} + \text{vector}\) exist in the massless case,
as \(H_{N} - H_{N-1} < M_\rho\) for any value of \(N\). This is for linear trajectories with the slope \(\alp = 0.9\) GeV\(^{-2}\). If we add masses and keep the Regge slope parameter constant, then the critical value of the intercept on the right hand side of the inequality is an increasing function of the masses. For example, if we take \(m_1 = m_2 = 100\) MeV, then with \(a < -0.06\) we are guaranteed there are no \(H_N \to H_{N-1} + \text{vector}\) transitions. When the endpoint masses are larger than \(\approx 300\) MeV then the inequality holds for any value of the intercept.

\subsubsection{Heavy quarks}
For hadrons made up only of massive quarks (\(s\), \(c\), \(b\)), the leading decay mechanism is, as always, via a tear in the string which produces a pair of light quarks.

These states will decay to states outside of their respective trajectories, so the discussion above does not apply to them.

We know that the lightest \(\ccb\) and \(b\bar b\) mesons cannot decay via a tearing in the string, simply for lack of available phase space: they are too light to be able to decay into a pair of charmed or bottom mesons as needed.

Schematically, we write the masses of the \(J/\Psi\) and \(D\) mesons as
\be M_{J/\Psi} = 2m_c + TL_0(a_{\Psi})\,,\qquad M_D = m_c + m_{u/d} + TL_0(a_D)\,. \ee
The quantity \(TL_0\) is a zero-point energy contributed by the string. The difference between the mass of the \(J/\Psi\) and the threshold for decays into \(D\bar{D}\) is
\be M_{J/\Psi} - 2m_D = -2m_{u/d} + T(L_0(a_\Psi)-2L_0(a_D))\,.\ee
The difference experimentally is of about 640 MeV. Now the mass of a \(u/d\) quark as a string endpoint is not nearly large enough to account for this difference, and in our formulation most of the difference is attributed to the quantum intercept, and the energy it adds to each system. This is in contrast to the approach where the extra mass is attributed to a constituent mass of the \(u/d\) quarks.

\begin{figure}[ht!]
			\centering
				\includegraphics[width=0.48\textwidth]{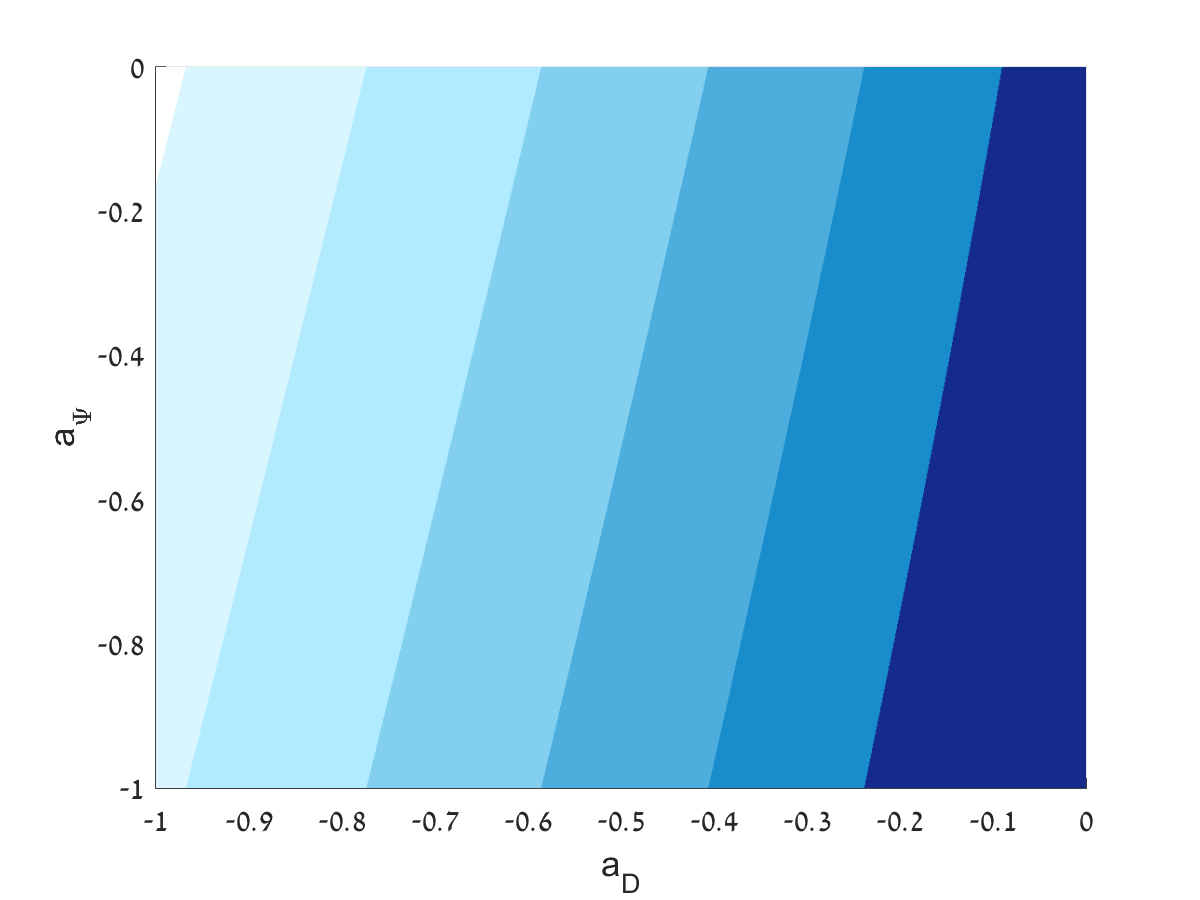}
				\includegraphics[width=0.48\textwidth]{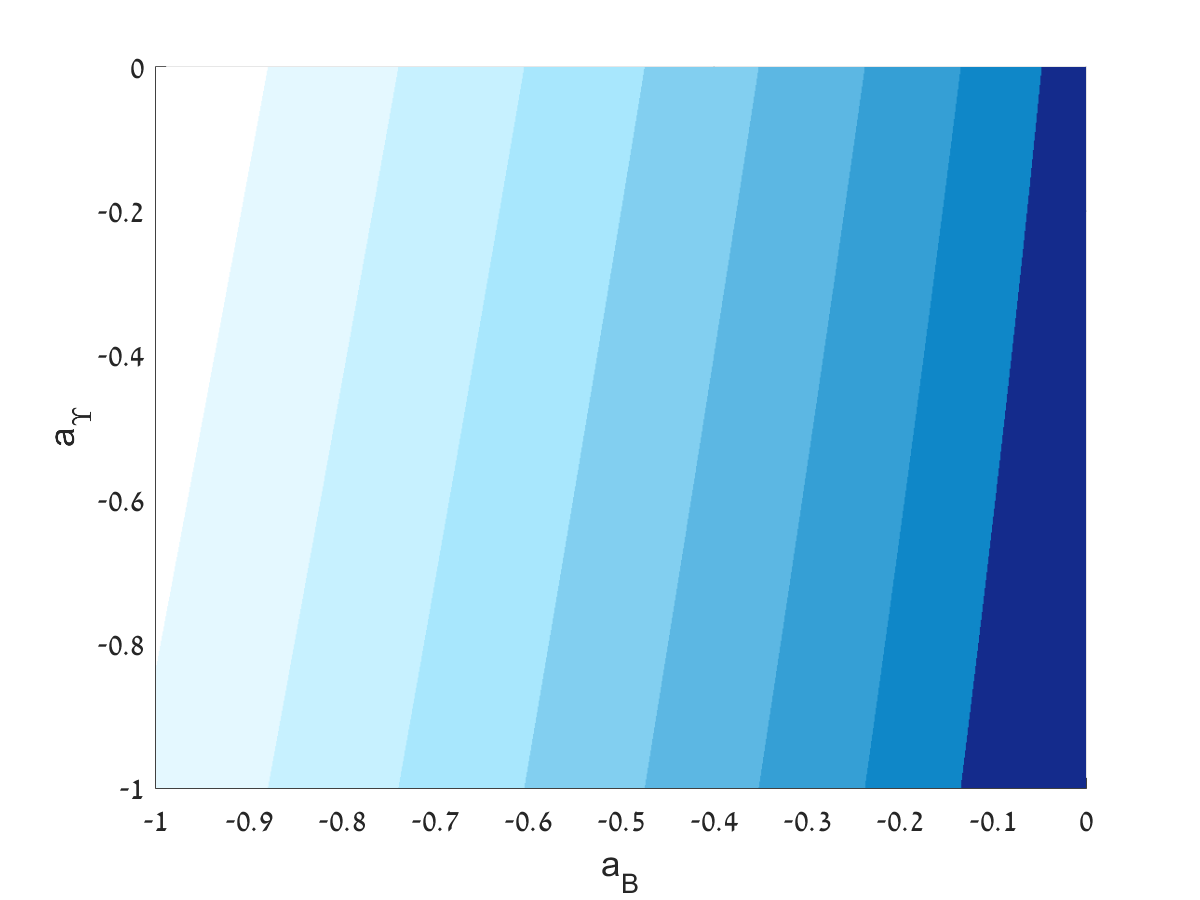}
				\caption{\textbf{Left:} The number of states below the open flavor threshold on the \(\Psi\) trajectory as a function of the intercepts of the \(D\) and \(\Psi\) trajectories. In the darkest area only the ground state is below threshold, in the next darkest the first excited state is also below threshold, and so on. It is assumed that the Regge slope is \(0.9\) GeV\(^{-2}\) and the quark masses \(m_c = 1500\) MeV, \(m_{u/d} = 60\) MeV. \textbf{Right:} The same for the analogous \(\Upsilon\) and \(B\) mesons. There we take \(\alp = 0.64\) GeV\(^{-2}\) (the slope of the orbital \(\Upsilon\) trajectory found in \cite{Sonnenschein:2014jwa}) and the quark masses \(m_c = 4730\) MeV, \(m_{u/d} = 60\) MeV.}
				\label{fig:ozi_intercept}
	\end{figure}

We can write the condition that open flavor decays be kinematically forbidden as an inequality of the intercepts of the Regge trajectories. This will give us a constraint on the intercepts of heavy-light and heavy-heavy mesons which we should obtain from the theory, if we want it to be consistent with experiment, even if only qualitatively. In particular, the number of states on the \(\Psi\) trajectory whose decays are suppressed depends on the intercepts. We plot this dependence in figure \ref{fig:ozi_intercept} (left side). The intercepts one obtains from the observed orbital trajectories of the \(D\) and \(\Psi\) are \(-0.19\) and \(-0.06\) respectively \cite{Sonnenschein:2014jwa}.

The second plot in figure \ref{fig:ozi_intercept} is for the analogous \(b\bar b\) mesons, \(\Upsilon\) and \(D\). The number of narrow peaks below the \(B\bar B\) threshold is a more sensitive test of the intercepts - and the Regge trajectories in general - in this case. The phenomenological intercept of the \(\Upsilon\) is close to zero. If the \(B\) meson sits on a trajectory with the same slope, then its intercept should be (based on the \(B\) meson mass) \(\approx-0.3\).

\section {Decay modes, spin, and flavor symmetry} \label{sec:decay_spin}
It is well known that considerations of the spin and  isospin or more generally flavor symmetry  of the initial  and final states are very important in determining which  decays are forbidden  and the relative decay width of the allowed modes. A natural question is how are such considerations realized in the holographic decay mechanism  of stringy hadrons. 

In the study of the spectra of the HISH models \cite{Sonnenschein:2014jwa,Sonnenschein:2014bia}, the spin and flavor symmetry have played a minor role. It was noted that the intercept depends on these degrees of freedom but this has not yet been fully understood theoretically. 
For decay spin and flavor symmetry are an important factor, and their effect cannot be simply absorbed into a single parameter such as the intercept.

In the following subsections will address the dependence of decays of hadrons on spin, of mesons and baryons on the isospin and $SU(3)$ flavor symmetry, and also on the Bose and Fermi symmetry of the hadron wave functions.

So far we have not dealt with the spin degrees of freedom of the stringy hadrons. In fact this is a topic that has not yet been fully studied in the context of the HISH models. In particular, one would like to see the dependence of the intercept on these degrees of freedom. Nevertheless, assuming that the spin of the hadrons is attached to the string endpoint particles, we can deduce certain restrictions on the decays processes that follow from spin considerations. This will be discussed in the following subsection.

\subsection{The spin structure of the stringy decays} \label{sec:decay_spinc}
In this section we intend to discuss the dependence of the decay width on the spins of the decaying and outgoing hadrons. There are two basic questions to address: (i) The role of the spin in the structure of the stringy hadrons and their decay. (ii) Is the dependence on the spin for stringy hadrons  the same or not in comparison to the dependence of non-stringy  hadrons.	

An  ordinary bosonic string in flat spacetime  with no particles on its ends can have angular momentum due to classical rotation.  In this case the total angular momentum is identical to the orbital one and there is no spin in the system.  Also quantum mechanically the open string  state of spin $n$, namely the $n^\text{th}$ rank  symmetric traceless tensor representation, takes the form 
\be
\prod_{j=1}^n\alpha_{-1}^{i_j} \vert0,p\rangle + \ldots
\ee 
where $\ldots$ stands for additional $n^{\text{th}}$ order creation operators acting on the vacuum that will be needed to complete the representation of the rotation group. Also in this case the quantum value is that of the total angular momentum.

The question then is whether this situation holds also for the holographic hadron string and in particular for the HISH string. 
A holographic stringy meson as depicted in figure \ref{fig:mapholflat} is composed of a ``straight  horizontal string'' stretching along the wall with two  vertical segments ending  on flavor D-branes. A baryon (figure \ref{holtoflat2}) has a structure of one similar vertical segment, and a second vertical segment that ends on a baryonic vertex, which is in turn connected with two short strings to flavor branes.

For a bosonic string we cannot assign spin to the horizontal segment so that the spin degrees of freedom could be associated with the vertical segments or the endpoints of the strings on the flavor branes. Had the vertical segments carried spin they should have been fermionic strings but it is not known how to connect a bosonic and fermionic strings. A possible scenario in this respect can be of worldsheet fermions which may be light on the vertical segments, and are connected to bosonic string on the horizontal segment, on which the fermions are heavier. In addition, for the baryons there are only two vertical segments but it is well known from the quark model  that the spin structure of a baryon should be composed from three spins. Thus a more reasonable scenario is to assign a spin to the connection points of the strings with the flavor branes. Thus in the HISH model the ansatz of the stringy hadron includes a particle at the endpoint of the string which is massive due to the energy of the vertical segment and carries a spin due to the contact with the flavor brane. For such a model one has to replace the endpoint particle action given in \ref{particle} with that of a fermionic particle. The analysis of such a model is left for future investigation. It is possible that the fact that the intercept of stringy hadrons is negative and correspondingly the Casimir force is repulsive, as was discussed in section  \ref{sec:casimir_force}, will follow from the bosonic string action combined with such an endpoint particle action. 

The decay of a stringy hadron can be affected by the spin degrees of freedom in two ways: (i) Conservation of the total angular momentum dictates what are the allowed outgoing states. (ii) If the suppression factor for pair creation depends on the spin. We did not find in discussions of the suppression factor in the literature an analysis of its dependence on the spin state of the created quark antiquark pair. We will assume here that if there is such a dependence it is weak in comparison with the exponential dependence on the mass of the created pair.

\begin{figure}[ht!]
			\centering
				\includegraphics[width=0.68\textwidth]{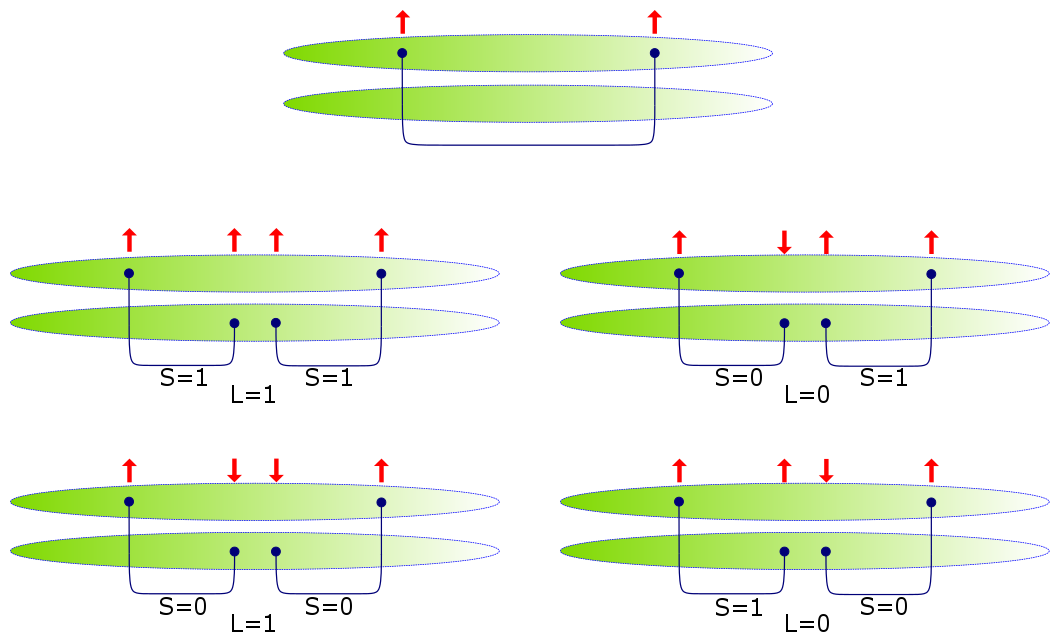}
				\caption{\label{spindependence} The spin structure of allowed decays of a neutral meson $M^0$ with spin $S=1$ into  $M^- + M^+$ mesons. The arrows indicate the values of $s_z$.}
	\end{figure}

The conservation law constraints from the stringy hadrons are the same as those of the particle picture. It is a reasonable assumption to make that the spins of the endpoints are unaffected by the creation of the pair even though this is not dictated by angular momentum conservation. The conservation law dictates then that if the spin of the created pair is one, then it must be accompanied by additional orbital angular momentum \(L_z = 1\) in the outgoing state. It remains to be seen how this affects the probability of pair creation (beyond the different kinematics). In figure \ref{spindependence} we demonstrate the spin structure of the allowed decays of a neutral meson of spin $S=1$. The relative orbital angular momentum in the outgoing state is equal to the spin of the created pair. Of course, the existence of each of these decays also depends on there being enough phase space for them.

\begin{figure}[ht!]
			\centering
				\includegraphics[ width=0.68\textwidth]{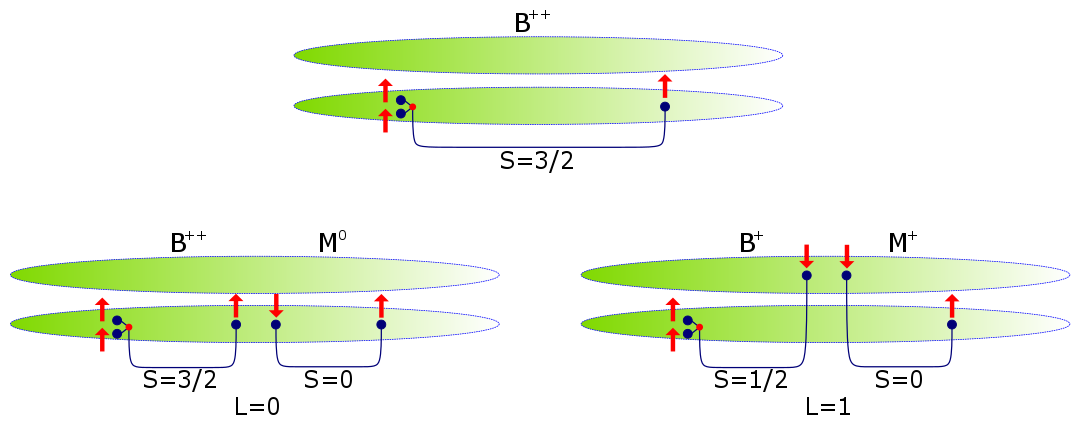}
				\caption{The spin structure of allowed decays of doubly charged baryon $B^{++}$ with spin $S=3/2$ into  a baryon and a meson.}
\label{baryonspindependence}
	\end{figure}

The spin structure of the decay of baryons is demonstrated in figure \ref{baryonspindependence}. A doubly charged baryon can decay (if kinematically allowed) to a a doubly charged baryon plus a neutral meson or to single charged baryon plus a charged meson.
 
\subsection{Isospin constraints  on decays of  holographic   stringy mesons}
Isospin approximate symmetry is realized in holography by the fact that the $u$ and $d$ flavor branes are located at roughly the same coordinate along the  holographic radial direction. It is well known that  the world volume of a stack of $N_f$ coincident flavor  branes  is characterized by a $U(N_f)$ flavor gauge symmetry. In fact in models like the Sakai-Sugimoto model where asymptotically there is a stack of $N_f$ flavor branes separated from a similar stack of flavor anti-branes the symmetry in the UV is a chiral flavor gauge symmetry of the form $U_L(N_f)\times U_R(N_f)$ that is geometrically  spontaneously broken in the IR to  $U_D(N_f)$ via the merger of the branes and anti-branes.  We would like to emphasize, especially since we are describing the hadrons as strings, that the realization of the flavor gauge symmetries is in terms of $N_f^2$ strings that connect the flavor branes. One of the rules in the dictionary of holography is that local symmetries in the bulk translate into global symmetry on the corresponding  boundary field theory  thus for $N_f=2$ there is an isospin  and baryon number global symmetries in the dual field theory. 

In the HISH language the requirement that the two flavor branes stretch at roughly the same radial coordinate  translates into
\be
m^d_{sep}-m^u_{sep} \ll m^u_{sep}.
\ee 
 We therefore  take that  the suppression factor, that  behaves as $\sim \exp\left(-\tilde C\frac{m^2_{sep}}{T}\right)$, is to a good approximation the same for creating a $u\bar u$ pair or $d\bar d$.
 
\begin{figure}[ht!]
			\centering
				\includegraphics[width=0.48\textwidth]{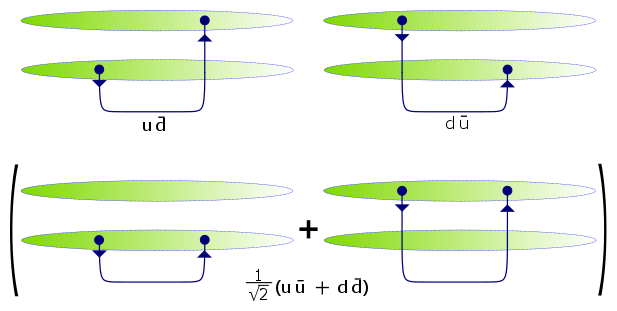}
				\caption{The stringy mesons of the isospin triplet. Top-left is the $M^+$ $\vert I=1,I_3=1\rangle$. Top-right is $M^-$ $\vert I=1,I_3=-1\rangle$. On the bottom row is the $M^0$ $\vert I=1,I_3=0\rangle$.}
				\label{mesonisospin}
	\end{figure}

We follow in this section the holographic, and not the HISH, description of hadrons and their decays. 
The endpoints of the hadronic strings, otherwise referred to as quarks,  reside on  the $u$ or $d$ flavor branes and hence  naturally  carry  isospin charges.  
Thus, the stringy mesons can be classified  
as  isospin eigenstates as is  depicted in figure \ref{mesonisospin}.
	We have denoted a quark by an arrow that comes out of a flavor brane and an antiquark by an arrow that comes into a flavor brane and we have used the convention that the isospin assignments are the following $(u\equiv  \uparrow),(d\equiv  \downarrow), (-\bar u\equiv  \downarrow), (\bar d \equiv  \uparrow)$. Note that the neutral meson $M^0$ is taken to be the linear combination of the strings that starts and ends on the $u$ brane or on the $d$ which corresponds in isospin terms  to the symmetric combination
	$\frac{1}{\sqrt{2}}( \vert\!\uparrow \downarrow\rangle+ \vert\!\downarrow\uparrow \rangle)$.
	
\begin{figure}[ht!]
			\centering
				\includegraphics[ width=0.9\textwidth]{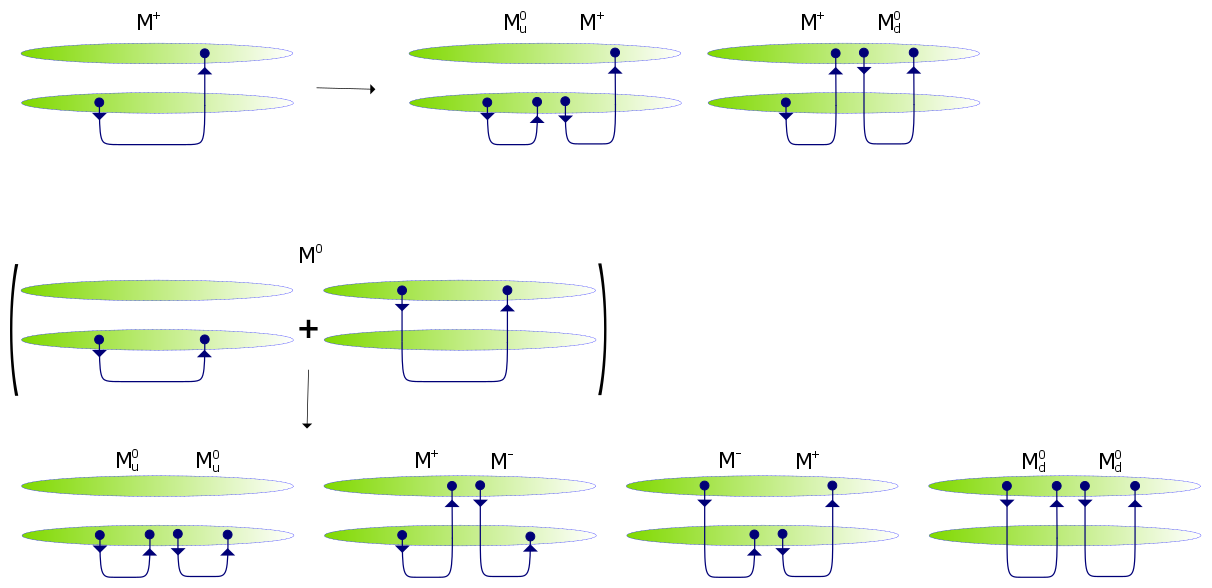}
				\caption{Possible decays processes of $M^+,M^0, M^-$.}
				\label{decaymesonisospin}
	\end{figure}

	The decay processes associated with the decay of these mesons involve obviously the breaking apart of the horizontal string and the attachment of its endpoints to either the $u$ or the $d$ flavor branes. These processes are drawn in figure \ref{decaymesonisospin}. 
 We see that the charged meson $M^+$  decays either to $M^+ + M^0_u$ or to $M^+ + M^0_d$, where $M^0_u$  and $M^0_d$ stand for a $\bar u u$ and $\bar d d$ respectively. Thus altogether we get of course the decay 
$M^+ + M^0$ and similarly for $M^-$. For the neutral meson $M^0$ we see that both components of it $M^0_u$ and $M^0_d$ decay into $M^+ + M^-$ and into $M^0_{u/d}+ M^0_{u/d}$ of their original flavor. So it seems that it decay  either to $M^+ + M^-$ or to $ 2 M^0$. However,  since by assumption the decay processes obey the isospin symmetry, the latter decay is forbidden since the corresponding Clebsch-Gordan coefficient vanishes.  Furthermore, as will  be discussed in subsection \ref{sec:BoseFermi} the decay process of 
$M^0 \rightarrow M^0+M^0$ is also forbidden because of arguments of the Bose symmetry of the wave function. 
\subsection{Isospin constraints  on decays of  holographic   stringy baryon} \label{sec:isospin}
\begin{figure}[ht!]
			\centering
				\includegraphics[ width=0.9\textwidth]{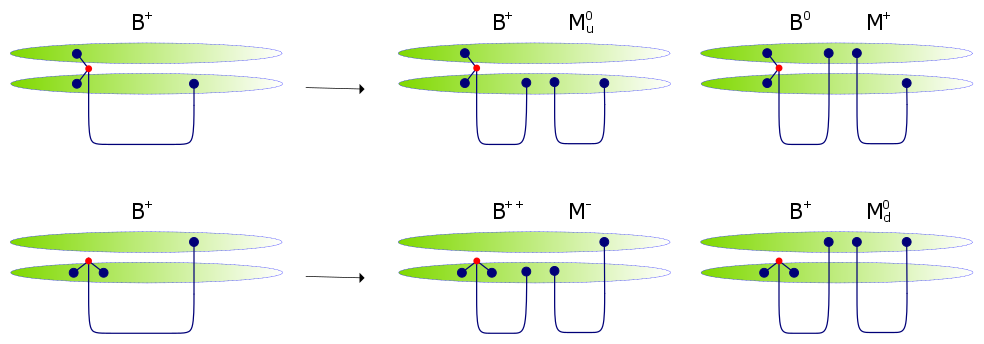}
				\caption{Possible decays processes of $B^+$ with different diquark configurations. The baryonic vertices are drawn out of flavor brane to make it easier to distinguish between the pictures.}
				\label{decaybaryonisospin}
	\end{figure}

Next we would like to analyze the isospin constraints on decays of baryons. As was found out in \cite{Sonnenschein:2014bia} the best stringy description of baryons is in terms of a string that connects a quark on one side and a baryonic vertex on the other side and the latter is connected to a diquark.
The baryonic vertex is immersed  on a flavor brane.\footnote{The baryonic vertex, which in holography is a wrapped D-brane,  is ``pulled up'' to the flavor brane by the strings and is pushed down by the interaction with the curved background. Thus, its location is determined by a static balance between these opposite forces. It was shown \cite{Seki:2008mu} that it is located on a flavor brane in the Sakai-Sugimoto model but  in other holographic models like in  \cite{Dymarsky:2010ci} it can be located also out of flavor branes.} For concreteness lets describe the decay of a positively charged baryon $B^+$. As is shown in figure \ref{decaybaryonisospin} the setup of this hadron can be with the baryonic vertex connected either to a $d$ and a $u$ quark (left) or to two $u$ quarks (right). The left $B^+$  can decay to $B^+ + M^0_u$ or  to $B^0 + M^+$, whereas the right one can decay to $B^{++} + M^-$ or  to $B^+ + M^0_d$.

The physical configuration corresponding to the \(B^+\) may be a mixture of the two possibilities, \((uu)d\) and \((ud)u\). In fact, this will be required by isospin. If we take a positively charged excited \(N\) baryon, with \(\vert I = 1/2; I_3 = 1/2\rangle\), then it should have decays to both \(\Delta^{++} \pi^-\) and \(\Delta^0\pi^+\). This ratio between these decays will be related to the mixing angle between the two different diquark configurations on one hand, and to the corresponding Clebsch-Gordan coefficients of isospin on the other. The mixing angle in this case cannot be zero since neither of the CG coefficients vanishes.
\subsection{Isospin breaking in strong decays}
So far we have discussed  the manifestation of isospin symmetry in the decays of meson and baryons. However, as was mentioned above also in holography isospin is an approximated and not exact symmetry. 
The locations of the $u$ and $d$ flavor branes  are not precisely the same. So far in holography the location are not determined from some underlying principle and the only way to determine $m^u_{sep}$ and $m^d_{sep}$ is by extracting it from experimental data. In \cite{BlechSirSon} the action of the string with particles on its ends was modified by adding charges to the endpoints and incorporating electromagnetic interactions. As a result of these interactions the structure of the massive  Regge trajectories have been modified. In particular one finds that even classically there is a non-trivial intercept. However, not surprisingly the changes of the trajectories are very small. There is one phenomenon that can be practically used  to determine the difference $m^d_{sep} -m^u_{sep}$ and that is the mass difference between a charged and a neutral hadron with otherwise the same properties. The landmark examples are the mass differences between the neutron and the proton and between the neutral and charged $\rho$ meson. From these two examples it was found out that the mass difference is of the order of $ \sim 2$ MeV. For hadrons with heavier ``quarks'' the difference can be some what different.
In principle  the  charges of the endpoint particles can affect the dependence of the decay width on the mass or the angular momentum of the hadron. The electromagnetic forces are now part of the boundary equation that determine the length of the string and hence the dependence of the width. We defer such a check to a future investigation. 

There are certain strong decays that violate isospin symmetry, an example being the decay $D^*_s \to D_s + \pi_0$. We would like to argue that such decays processes can follow from a virtual pair creation accompanied by a Zweig suppressed decay of a virtual \(\ssb\) state. This process described in section \ref{sec:virtualpair} provides the mechanism for an isospin breaking strong decay.

\subsection{Baryon number symmetry}
Whereas in QCD, or for that matter the quark model, the baryon number symmetry is identical to the quark number symmetry, in holography and hence also in HISH the situation is somewhat  different. The baryon number symmetry is the conservation of the number of baryonic vertices. The quark number symmetry is nothing but the fact that any open string has two ends and one associates an orientation to the string so  that one end on a flavor brane  is considered  a quark and the other antiquark. For a baryon the other end of the string is on a baryonic vertex (BV) which is a wrapped $D_p$ over a $p$-cycle (for instance in  it is a wrapped four brane over an $S^4$) and not on the flavor brane. The quark number symmetry is also part of the flavor branes symmetry. 
A setup of $N_f$  coinciding flavor branes has on its world-volume a $U(N_f)$ local symmetry realized by the zero length open strings that connect these branes.  The $U(1)\in U(N_f)$ is the quark number symmetry and according to the rules of holography this translates to a global $U(1)$ symmetry on the boundary field theory. Are the quark number and baryon number symmetries totally independent in holography? The answer is no and in fact they are related. As was shown in  \cite{Witten:1998xy} the action  on the world-volume of the baryonic vertex includes a CS term built from the RR form and the Abelian gauge field on the BV  brane. This term that dictates that there are $N_c$ strings attached to the baryonic vertex, relates the RR flux which when integrated over the p-cycle equals $N_c$ and the Abelian gauge field on the BV brane. A string that connects a flavor brane and the BV branes relates the $U(1)$  local symmetries on the BV and flavor branes. 

\begin{figure}[ht!]
			\centering
				\includegraphics[ width=0.9\textwidth]{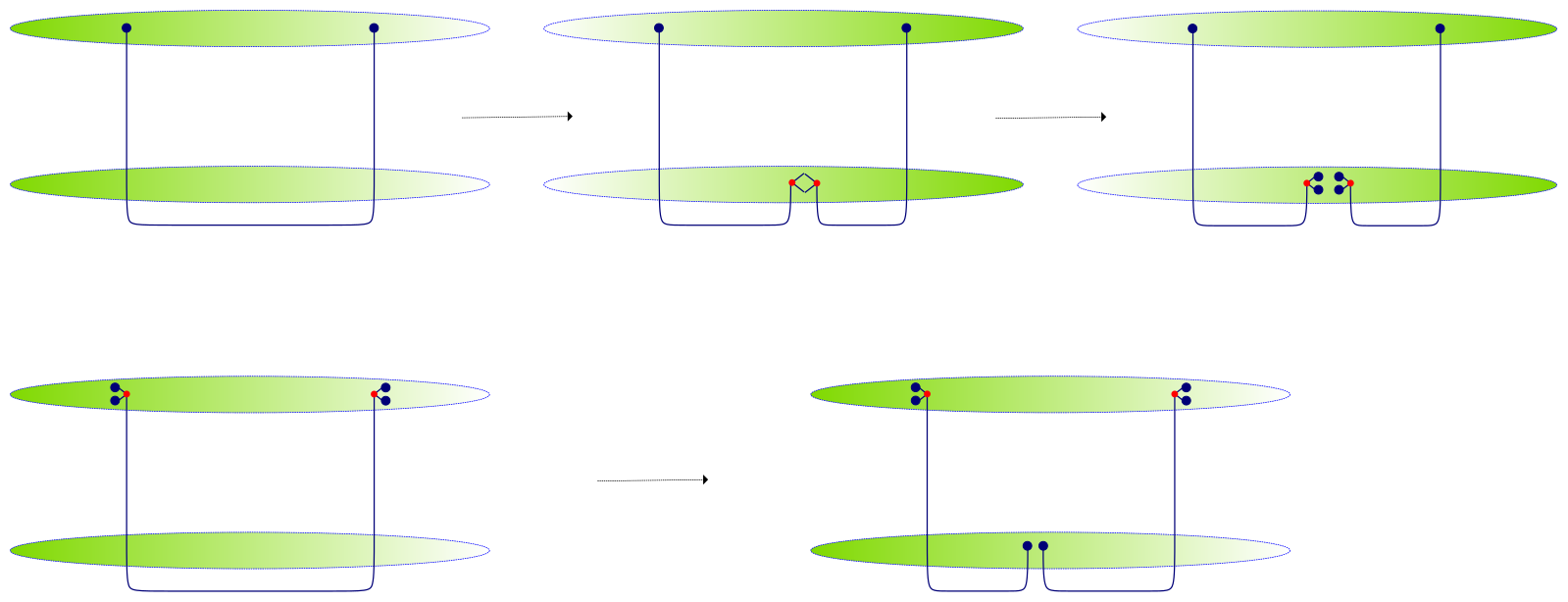}
				\caption{\textbf{Top:} Creation of a baryon anti-baryon pair via a string breakup and the creation of a BV $+$ diquark and an anti-BV $+$ anti-diquark. \textbf{Bottom:} Creation of the same via string split of an exotic hadron.}
				\label{BVantiBVpair}
	\end{figure}

The next question is if and how are the decay processes affected by this structure of the  holographic baryon number symmetry. We are by now very familiar with the mechanism of breaking of a stringy hadron which is accompanied by a creation of a quark antiquark pair. However we have not discussed processes in which a pair of baryon anti-baryon is being created, namely, a BV $+$ anti-BV configuration. In section \ref{sec:decayexotic} we have discussed the possible decay of an exotic hadron built from a string connecting a BV and anti-BV by a usual breaking of the string thus decaying to a baryon and anti-baryon. Such an outcome can in principle be a result of a breaking of an ordinary meson string breaking apart, but where the two string segments are connected directly to a flavor brane they can be connected via a BV and anti-BV as is depicted in figure \ref{BVantiBVpair}. The suppression factor for such a wrapped D-brane and anti-brane creation is not known to us and will be part of future investigation.  
 
\subsection{\texorpdfstring{$SU(3)$}{SU(3)} flavor  constraints  on decays of  holographic   stringy baryon}
As was mentioned in section \ref{sec:isospin}, in processes where the differences between the heights of the flavor branes can be neglected one can consider them to be coinciding and thus if there are $N_f$ of them then in the IR there is a $U(N_f)$ local symmetry on the flavor branes. This approximated symmetry can be useful in certain cases also for the first three flavors, namely $U(3)$. The symmetry can be applied on (i) the endpoint quarks of the original decaying hadron, and (ii) the endpoints of the outgoing hadrons. The latter are composed from strings with endpoints identical to the parent one on one side and two endpoints which  are the quark-antiquark pair generated by the split of the original string. In section \ref{sec:decay_exp_mass} we have discussed in length the exponential suppression factor and its dependence on the pair quark mass $m_{sep}$. The ratio between the creation of a strange pair of of $u$ or $d$ pairs manifestly breaks the $SU(3)$ flavor symmetry.
It is known since the early days of $SU(3)$ flavor symmetry \cite{Lipkin:1964zza} that it can be used to relate the amplitudes and hence the decay width of hadrons in a given $SU(3)$ representation to two hadrons in certain given representations.  

The remaining question then is can one relate decay width of hadrons with endpoint flavored quarks using the $SU(3)$. Comparing the decay width of  hadrons  with one identical quark and another that can be either $s$ quark or  a $u$ or $d$ one should anticipate, assuming $SU(3)$ symmetry that the ratios of the width should be just the ratios of the phase space factor. For instance the decays of vector mesons in the octet into two pseudoscalars also in octets or the decays of baryons in a decuplet into a baryon in an octet and a meson in an octet. In fact one can use U-spin and Isospin Clebsch-Gordan coefficients to relate these decay amplitudes. It seems that the same results can be obtained for the stringy mesons. For stringy baryons, due to the different possibilities of the structure of the diquark it is not obvious that the same relations follow. This is further discussed in section \ref{sec:baryon_structure}.

\subsection{Constraints from the symmetry properties of the overall  wave function} \label{sec:BoseFermi}
It is well known that in the particle picture certain  decays of hadrons, which are allowed in terms of the angular momentum and flavor conservation, are forbidden due to the symmetry of the overall wave function. This follows from the demand that the overall wave function for mesons be symmetric and for baryons antisymmetric. A typical example for these constraints is the decay of a neutral $S=1$ spin meson into two spin $S=0$ neutral mesons (i.e. \(\rho^0\to\pi^0\pi^0\)). Due to conservation of angular momentum the outgoing mesons have to be in P-wave and hence antisymmetric. Since the spin  and color wave functions are symmetric  the overall wave function is antisymmetric which is forbidden for a meson and hence this decay channel is forbidden. As was mentioned in section \ref{sec:isospin} it is also forbidden by using the isospin Clebsch-Gordan coefficients.

The question is therefore whether the stringy picture of hadrons admits the same type of overall wave function. We would like to argue that this is indeed the case. A substantial difference between the particle and the string picture is that the in the latter color degrees of freedom do not exist. For mesons this is not an issue and hence the example of  forbidden process of meson decay holds also in the stringy picture. The  question is what happens with the baryons. As was shown in \cite{Witten:1998xy} the wave function of the strings that connect the boundary and a baryonic vertex in the $AdS_5\times S^5$ string theory is indeed antisymmetric mimicking the contribution of the color wave function in the particle picture.  This follows from the fact that the string in that case are fermionic and not bosonic. The former has zero energy whereas the latter has positive energy. The ground state is non-degenerate because there are no fermion zero modes in the Ramond sector. The dual of this string configuration is an external baryon with infinitely heavy quarks.  A dynamical baryon corresponds to a system of $N_c$ strings  connecting a BV  wrapped D-brane to   flavor branes. For $AdS_5\times S^5$ these are $D3$ branes and  in this case again the lowest energy string is a fermionic one. 
 For confining backgrounds one has to check for each model whether the Ramond string (fermionic) has  a lower (or equal) energy to the Neveu-Schwarz one.  The string zero point energies for the NS and Ramond strings are as follows \cite{Polchinski:1998rr}
\be
NS: \ \ -\frac12 + \frac{\#_{ND}}{8}\,, \qquad\qquad R: \ \ 0 
\ee
where $\#_{ND}$ is the total number of directions for which the string has ND (Neumann-Dirichlet) or DN boundary conditions. In fact this is the condition for string connecting D-branes in flat spacetime and since the holographic confining models are all in curved spacetime there could be curvature corrections which are difficult to compute. 
It turns out that both in the Sakai-Sugimoto model \cite{Sakai:2004cn} and  in the model of \cite{Dymarsky:2009cm} the naive calculation shows that the fermionic string has the lowest energy and hence it seems that it has indeed the total antisymmetric nature that one anticipates from the color wave function in the particle description.
However, there is a severe flaw in this argument. Had the preferred configuration of the baryonic vertex  been that of a Y-shape in which is symmetric in terms of the three strings (for $N_c=3$ ) we could have argued based on the fact that these are fermionic strings that indeed it has  a totally antisymmetric wave function. However, as was explained in section \ref{sec:HISH} we know that the configuration is not of a Y-shape but rather that of an  asymmetric string configuration. The latter includes single long string which has a horizontal segment and a vertical one  connecting it to  a flavor brane (quark) and a diquark which is a system of  two strings  connecting  the BV and  flavor branes (see figure \ref{holtoflat2}) with no horizontal part. The latter are either two zero-length strings if the diquark is ``built'' from quarks with the same flavor  or one  of zero-length  and another  vertical string that connects the BV to a flavor brane on which it does not reside. 
In the holographic picture and in the HISH model we take that the horizontal string is a bosonic one and not a fermionic one. Moreover, it is clear that in this setup the three strings are different and   obviously the total wave function is not totally antisymmetric.

Thus, we have encountered an essential difference between the particle and stringy picture
of baryons (not of mesons). The question is whether there are known constraints on decay processes that follow from the antisymmetric property of the total wave function that cannot be explained in the stringy language.

\section{Phenomenology} \label{sec:pheno}
We now turn to the application of the HISH model to experiment. There are two main components to the stringy behavior of the decay width of a hadron. One is the overall factor of the string length, to which the decay width should be proportional. For the string without massive endpoints the length is proportional to the mass, and so is the decay width. For strings with massive endpoints, this is corrected. Much of this section will be dedicated to testing this linearity of the decay width of hadrons in their calculated string length, for hadrons lying on the Regge trajectories found in \cite{Sonnenschein:2014jwa,Sonnenschein:2014bia}.

The second factor is the exponential suppression of pair creation, which should be seen when comparing branching ratios of channels where different flavor quarks are produced. In practice this means we will compare channels where a light (\(u/d\)) quark-antiquark pair is produced to channels involving \(\ssb\) creation, and see if we can measure the ratio between them.

Unless explicitly stated otherwise, the data used throughout this section is taken from the Review of Particle Physics published by the PDG \cite{PDG:2016}.

\subsection{Fitting model} \label{sec:fitting_model}
In the following we construct a phenomenological model to be used in fitting the decay widths of various hadrons, based on the core assumption that the decay is proportional to the string length. The issues discussed in sections \ref{sec:model_intercept} and \ref{sec:model_phase_space} are the introduction of a ``zero-point'' quantum length to the string in the former, and phase space factors in the latter. We also discuss the time dilation factor in section \ref{sec:model_dilation}. After taking these steps, we write down the fitting formula of section \ref{sec:model_parameters}.

\subsubsection{String length and the phenomenological intercept} \label{sec:model_intercept}
In section \ref{sec:rotating_string_cl} we wrote the classical expressions for the energy, angular momentum, and length of the rotating string with massive endpoints. The quantization of the string and the resulting addition of the quantum intercept to the Regge trajectories were discussed in section \ref{sec:quantum_string}. A method to introduce quantum corrections to the string length was discussed in section \ref{sec:casimir_force}. Here we propose alternative prescriptions for defining the ``quantum length'' of the string, which will be necessary when confronting the model with data.

The intercept is most needed when we want to describe zero orbital angular momentum states as strings of finite length, which can then tear and decay like high spin strings. We are motivated to include these ground states in our model by the fact of their inclusion in their respective Regge trajectories, as well as the observation that their decay modes and widths are similar to those of their excited partners. The latter point may not seem very surprising, but the former is an important one. Placing a meson on a Regge trajectory is acknowledging it has some stringy behavior. It is a very non-trivial result that low spin hadrons display this behavior, and that even the ground states, with zero orbital angular momentum, do not break away from their respective Regge trajectories.

The question of how to insert the intercept into the equations describing the string length is a also question of the extrapolation of the model from long strings to short ones. Since we have no definitive answer as to why the model remains valid for the lowest spin hadrons when describing the Regge trajectories, but observe it is as an experimental fact, we may allow ourselves to look for an ansatz here that would be usable from a phenomenological perspective.

To state the problem in its simplest form, without rotation there is no classical string length. The finite decay width of the spin-zero ground state, if attributed to a finite ``zero point length'' of the string, must be due to quantum corrections and the Regge intercept. We wish to know how the intercept is related to this zero point length, and how the zero point length is manifested at finite, but still low, spins.

Classically, we have a simple relation between \(L\), \(E\), and \(J\), given by the formulae of the rotating string. Quantum mechanically, we need to add the intercept. The same situation appears when describing the Regge trajectories: we have a classical relation between \(J\) and \(E\), and add the intercept to reconcile the model with real world data. Once we have added the intercept, it is determined by the data. 

In section \ref{sec:quantum_massive} we discussed the introduction of the intercept in the case of a string with massive endpoints. The simplest way to introduce the intercept is by replacing \(J\to J-a\) in the classical expressions, and making no other modifications. This is the way the intercept was introduced for the Regge trajectories of \cite{Sonnenschein:2014jwa,Sonnenschein:2014bia}. Its disadvantage is that it is somewhat counter-intuitive: in particular, the zero spin of the ground state is obtained as a sum of some non-zero rotational angular momentum, which gives that state a finite length, and a negative intercept.

Another method, with the inclusion of a repulsive Casimir force, of the form \(F_C = -2a/L^2\), explicitly affects the length of the string, as it modifies the boundary equation through which determine the length of the string.

A third, purely phenomenological way of including the intercept is assuming that the quantum length, to which the decay width is proportional, is related to the classical length by
\be L^2 = L_{cl}^2 + L_0^2\,. \ee
This is motivated by the fact that, the intercept can be added linearly to \(E^2\), which is in turn proportional to \(L^2\). So we might write,
\be E^2 = E_{cl}^2 - \frac{a}{\alp}\,, \ee
on one hand, while on the other hand,
\be E^2 = (\frac\pi2 TL)^2 = \frac{\pi^2}{4}T^2(L_{cl}^2+L_0^2)\,. \ee
We separate \(L^2\) into a classical part, proportional to \(E_{cl}^2\), and attribute the contribution from the intercept to the zero length
\be L_0^2 = -\frac{8}{\pi} \frac{a}{T}\,.\ee
Previously we had another, conflicting value for the zero length. When we added the Casimir force to the boundary equation (section \ref{sec:casimir_force}), the result was that \(L_0^2=-2a/T\). The discrepancy is that between different methods of extrapolating the long string result, \(J = \alp E^2 + a\) to short string lengths. We can write, generically, that the zero point length is
\be L_0^2 = -C\frac{a}{T}\,,\ee
and leave the constant \(C\) as a parameter to be fitted.

To summarize, there were four options explored:
\begin{enumerate}
\item Do not include intercept in the string length. Calculate the classical length for states of given \(J\), and leave out the \(J = 0\) ground state. This gives good results but leaves out states whose decays our model might be able to describe.
\item Include the intercept by shifting \(J\to J-a\). This is how we fitted the Regge trajectories in previous works, and works well for the decay widths as well.
\item Include the intercept by taking \(L^2 = L_{cl}^2-Ca/T\). This is simply an ansatz, motivated by theory but not a result of it. The motivation came from looking at high energy, long strings, and we simply assume we can extrapolate the relation to short strings as it stands. The optimal value of the constant \(C\) is obtained from the fits, and it is in the range 1.6--2.0.
\item Calculate the quantum length from the modified boundary equation (eq. \ref{repCasf}) including the repulsive Casimir force. This prescription gives an explicit calculation of a quantum length of the string, but it can be complicated, and, in the case of the Regge trajectories, assuming a correction of this form does not match the data as well, possibly because the ansatz of eq. \ref{repCasf} is not valid for short strings.
\end{enumerate}

When fitting the decay widths, we chiefly used the second method, \(J\to J-a\). A comparison between the different methods is offered in section \ref{sec:pheno_test}.

\subsubsection{Phase space} \label{sec:model_phase_space}
To fit the decay widths of hadrons, we may need to take account of the available phase space for each decay channel. This is especially true of low spin, unexcited hadrons, who have little available phase space for their decays. To illustrate the necessity for this factor, consider the case of the lightest charmonium mesons. Even though they have a finite string length, they are kinematically forbidden from decaying by tearing the string and forming a pair of charmed mesons. The part of their decay width that is proportional to the length is multiplied by zero. Now we ask what happens when a hadron is only just above threshold for our type of decay. The most notable example is the \(\phi\) meson, which, at 1020 MeV, is only 40 MeV above the \(K\bar K\) threshold, its one dominant channel of decay, and its width is small as a result of the limited available phase space.

To begin with, we write the decay width of a string proportional to the string's length as
\be \Gamma = \frac\pi2 A T L(M,m_1,m_2,T)\,. \label{eq:fitModel2}\ee
The constant \(A\) is a dimensionless, and it is equal to the asymptotic ratio of \(\Gamma/M\) for large \(L\) (i.e. large \(J\)). 

The string theory calculation of the width via the imaginary part of the self-energy diagram discussed at length in section \ref{sec:decaywidth} is a calculation of the \emph{full} decay width. It implicitly sums over all possible modes of decay. The stringy theory calculation is also one that is expected to be valid for long strings, and is subject to corrections in lower powers of \(L\).

For treating low spin hadrons we need a specific form of correction, that takes account of the limited availability of phase space for decays when the decaying particle is of low mass. This correction would need to be ``informed'' of the masses of the low mass states in the spectrum - pions and kaons notably - so it is clearly not something that can be easily imported from string theory. In any case, we should see phase space as a limiting factor when moving down from high energies to low energies, rather than a factor that measures the increase in available modes when moving in the opposite direction, from low to high energies. Since linearity in the string length is formally a high energy result, we need to take the former view.

To put things in a simpler manner, and take a more practical approach, what we want to do is introduce a suppressive factor \(\Phi(M)\) into the decay width, writing
\be \Gamma = \frac\pi2 A\times\Phi(M)\times TL(M,m_1,m_2,T)\,.\ee
Here, \(\Phi(M)\) is some increasing function of \(M\) that must be smaller than or equal to one at all times. It is most important for hadrons with low spin. We will have to assume that at least part of the observable mesons, which go up to \(J = 6\) and no higher, are veritable high spin hadrons for which the linearity in \(L\) holds well, and so \(\Phi(M)\) reaches quickly enough its high energy value of 1. In the case of the low spin mesons, we will have to normalize the width by dividing it by \(\Phi\). The question of what constitutes low or high spins is perhaps a question that can be left to experiment, as we look for the point the stringy model's prediction of linearity in \(L\) breaks.

In the following, when we need an explicit form of the phase factor, we will use a factor of the form
\be \label{eq:phase_space}
\Phi(M,M_1,M_2) \equiv 2\frac{|p_f|}{M} = \sqrt{\left(1-(\frac{M_1+M_2}{M})^2\right)\left(1-(\frac{M_1-M_2}{M})^2\right)}\,.\ee
This comes from the familiar field theory formula of one-to-two particle decays, where we find the factor of \(|p_f|\), the momentum of each outgoing particle. We divide it by the decaying particle's mass \(M\) and get a dimensionless factor. The factor of 2 in the above is intended to normalize \(\Phi\) to go back to one when \(M\gg M_1+M_2\). We will discuss the choice of the masses of outgoing particles, \(M_1\) and \(M_2\) below, where we start looking at the data from particle decays.

\subsubsection{Time dilation} \label{sec:model_dilation}
The last piece of the puzzle is the time dilation factor affecting the width of the rotating string (sections \ref{sec:decaywidth_rot}--\ref{sec:decaywidth_massive}). It is a function of the endpoint velocities, and for \(m_1 = m_2\) it is
\be f(\beta) = \frac{2}{\pi}\left(\sqrt{1-\beta^2}+\frac{\arcsin\beta}{\beta}\right)\,. \ee
For hadrons containing only \(u\) or \(d\) quarks with their small endpoint masses, the significance of \(f(\beta)\) is mostly in distinguishing between rotating states, where we find that \(\beta > 0.9\) for all \(J > 0\) and then \(f(\beta)\approx1\), from those with \(J = 0\), for which we must multiply the width by \(f(0) = \frac{4}{\pi} \approx 1.27\). For hadrons containing \(s\) or \(c\) quarks we will have to use the exact form of \(f(\beta)\).

There is some ambiguity here when combining this with the notion of the quantum length, especially in the asymmetric case of \(m_1 \neq m_2\). Before introducing the zero point length, we wrote, using the classical expressions
\be \Gamma \propto \sum_i\int_{0}^{\ell_i} d\sigma\sqrt{1-\omega^2\sigma^2} = \sum_i\frac{\ell_i}{2}\left(\sqrt{1-\beta_i^2}+\frac{\arcsin\beta_i}{\beta_i}\right)\,. \ee
Now we have to define a prescription for adding the intercept to each of the \(\ell_i\), or see its effect on the velocities. The simplest solution is again to replace \(J\) with \(J-a\) in the relevant expressions, and substitute the resulting values of \(\ell_i\) and \(\beta_i\) into the above equation.

\subsubsection{Fitting model and parameters} \label{sec:model_parameters}
In the following we will use the fitting formula
\be \Gamma/\Phi(M,M_1,M_2) = \frac{\pi}{2} A T L(J,m_1,m_2,T,a)\,,\ee
We begin, where necessary, by normalizing the decay width by an appropriate phase space factor \(\Phi\), as explained in the last subsection. The width is then taken to be proportional to the string length, which is in turn computed as a function of the orbital angular momentum,\footnote{The orbital angular momentum is in typical notation \(L\). We avoid that notation to prevent confusion with the string length \(L\).} as well as the physical parameters \(m_1\), \(m_2\), \(T\), and \(a\). The length \(L\) is also adjusted for time dilation, as described in the previous section.

The string tension \(T\) is determined once and for all from finding the common slope of the Regge trajectories. The endpoint masses are also expected to be universal. The only fitting parameter that is not taken a priori to be universal, and varies when fitting the widths of mesons on different trajectories, is the dimensionless proportion constant \(A\). It is normalized in such a way that at high mass \(A\) is exactly the asymptotic ratio of decay width over mass for the hadron, \(\Gamma/M\).

The values of the Regge slope, quark masses, and the intercepts are also taken from the fits of \cite{Sonnenschein:2014jwa}, with one difference. The value of the strange quark given in \cite{Sonnenschein:2014jwa} is 220 MeV, and we take a higher value here. This improves the fits to the decay widths, without hurting too much the Regge trajectory fits. The fits to the Regge trajectories are shown in this work in appendix \ref{app:spectrum}. The parameters used for describing the trajectories of light, strange, and charmed mesons are
\be \alp = 0.884 \text{ GeV}^{-2}\,,\qquad m_{u/d} = 60\text{ MeV}\,,\qquad m_s = 400\text{ MeV}\,,\qquad m_c = 1500\text{ MeV}\,.\ee
The intercepts are obtained by fitting the Regge trajectories. The values vary between the trajectories, but are all approximately in the range of \(\approx-0.5\) to \(\approx 0\).

For baryons, we use a somewhat higher slope, \(\alp = 0.950\) GeV\(^{-2}\), and corresponding lower tension, as that is the value obtained from the baryonic Regge trajectories \cite{Sonnenschein:2014bia}.

\subsubsection{Goodness of fit}
The goodness of the fit is defined here by
\be \chi^2/DOF = \frac{1}{N-1}\sum_{i=1}^N \frac{1}{\sigma_i^2} (\Gamma_i^{th}-\Gamma_i^{exp})^2 \ee
To avoid over-emphasis in the fits of the low spin states whose widths are known with great precision, the weights \(\sigma_i^{-2}\) are defined as follows,
\be \sigma_i = \begin{cases}
	\Delta \Gamma_i\,, & \Delta \Gamma_i \geq 10 \mathrm{MeV} \\
	10 \mathrm{MeV}\,, & \mathrm{else} \\ \end{cases}
\ee
where \(\Delta\Gamma_i\) is the experimental uncertainty. The value of 10 MeV was chosen for its practicality, (for the purpose of giving higher weight to the states with smaller \(\Delta\Gamma\), but not so high as to cause the fitting algorithm to completely disregard the higher spin states and the increase in the total decay width). In the plots below the blue error bars will correspond to \(\Delta\Gamma\) while the red bars will show \(\sigma\).

Where \(\Gamma\) is normalized by a phase space factor \(\Phi\) we normalize the error as well, and pick \(\sigma\) after normalization, that is,
\be \sigma = \begin{cases}
	\Phi^{-1}\Delta \Gamma\,, & \Phi^{-1}\Delta \Gamma \geq 10 \mathrm{MeV} \\
	10 \mathrm{MeV}\,, & \textrm{else.} \\ \end{cases}
\ee

In the discussion of the results below we will quote the values obtained for the root mean square error,
\be RMSE \equiv \sqrt{\chi^2/DOF} \ee

\subsection{The decay width of mesons} \label{sec:pheno_mesons}
We look at the behavior of \(\Gamma\) as a function of the orbital angular momentum (or the mass), for several different mesons. The states used throughout are the same that were used in the Regge trajectory fits of \cite{Sonnenschein:2014jwa}, that is mesons that are on their respective leading Regge trajectories. The states on each Regge trajectory are fitted separately from those of other trajectories, and then the results are compared. We look at mesons made up of light quarks as well as those made up of strange and charmed quarks.

We start by discussing the specifics of the fits by looking at the trajectory of the \(K^*\) meson, where some key features of the model and the fitting process can be seen.

Afterward, in section \ref{sec:pheno_mesons}, we write the results for the fits of the rest of the mesons, followed by a discussion of the same for baryons in section \ref{sec:pheno_baryons}.

\subsubsection{A test case: The \texorpdfstring{$K^*$}{K*}} \label{sec:pheno_test}
\paragraph{General considerations:} The trajectory of the \(K^*\) meson is a good place to start testing out the string model. When observing the spectrum, we see five states on a Regge trajectory which is well fitted when adding a small mass correction attributed to the strange quark in the meson. We now want to see if the stringy picture holds when moving on to look at the decays of the \(K^*\) mesons. In doing this, we will outline some of the issues one has to deal with when applying the simple stringy model to real world data, and which will be relevant to all the decay width fits in the following section.

The five states in the \(K^*\) trajectory are detailed in table \ref{tab:Ks}. We list their masses, widths, width to mass ratio, and decay channels, using the data compiled by the PDG \cite{PDG:2016}.

The first state in the trajectory is the \(K^*(892)\) meson. It has a width of 50 MeV, and its sole decay mode is \(K\pi\). This state has zero orbital angular momentum and hence zero classical string length, but there is nothing to suggest that its decay mechanism is different from that of its excited partners. For that reason, we want to attribute its width to a finite length of the string, which is a quantum mechanical effect.

\begin{figure}[ht!] \centering
	\includegraphics[width=0.48\textwidth]{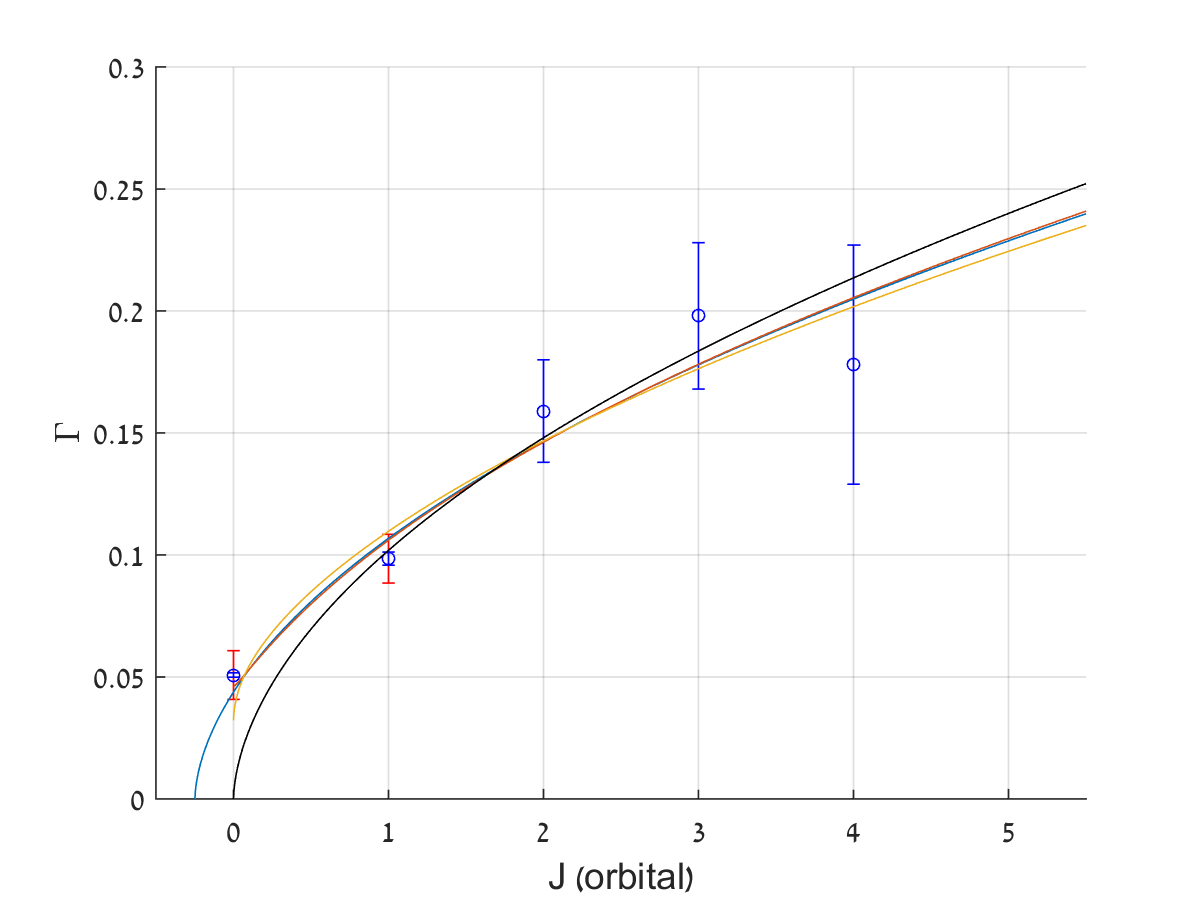}
	\includegraphics[width=0.48\textwidth]{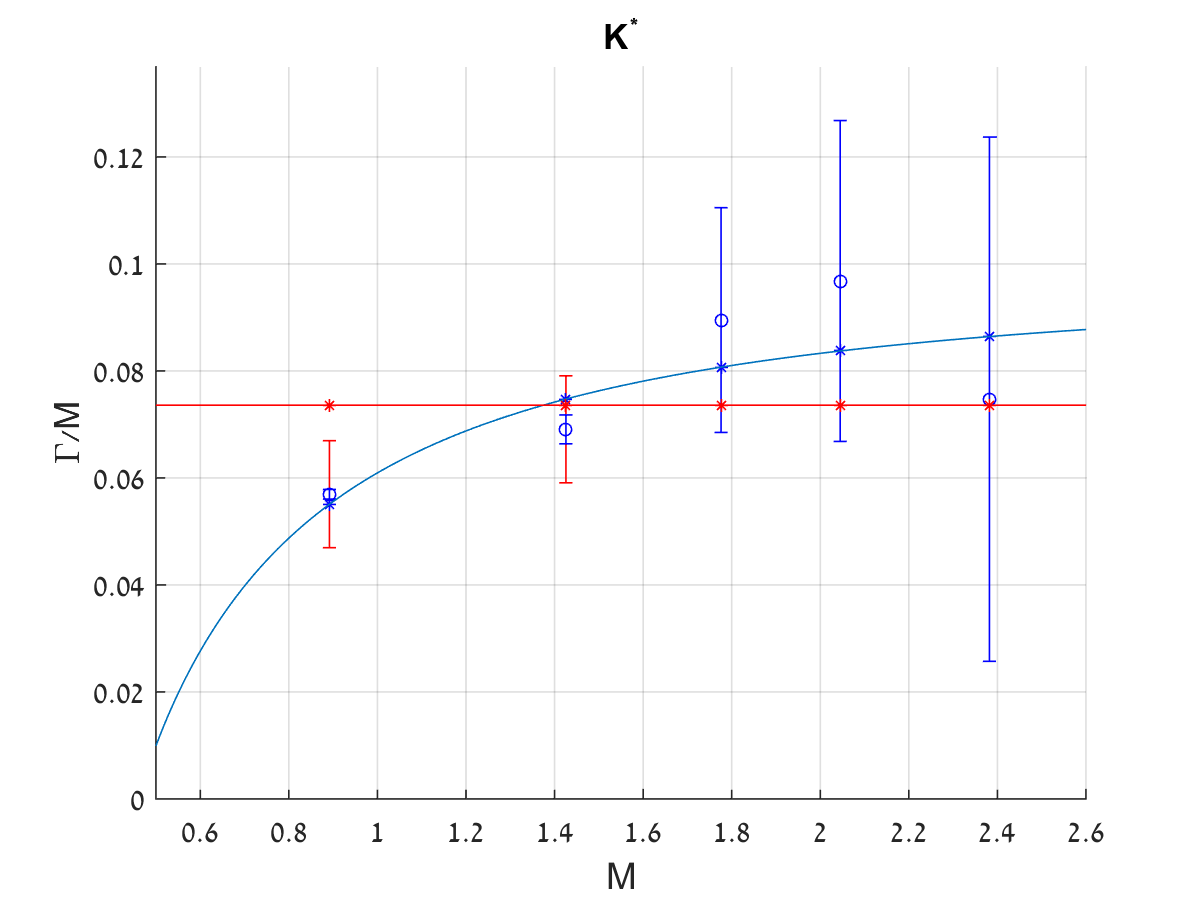} \\
	\includegraphics[width=0.48\textwidth]{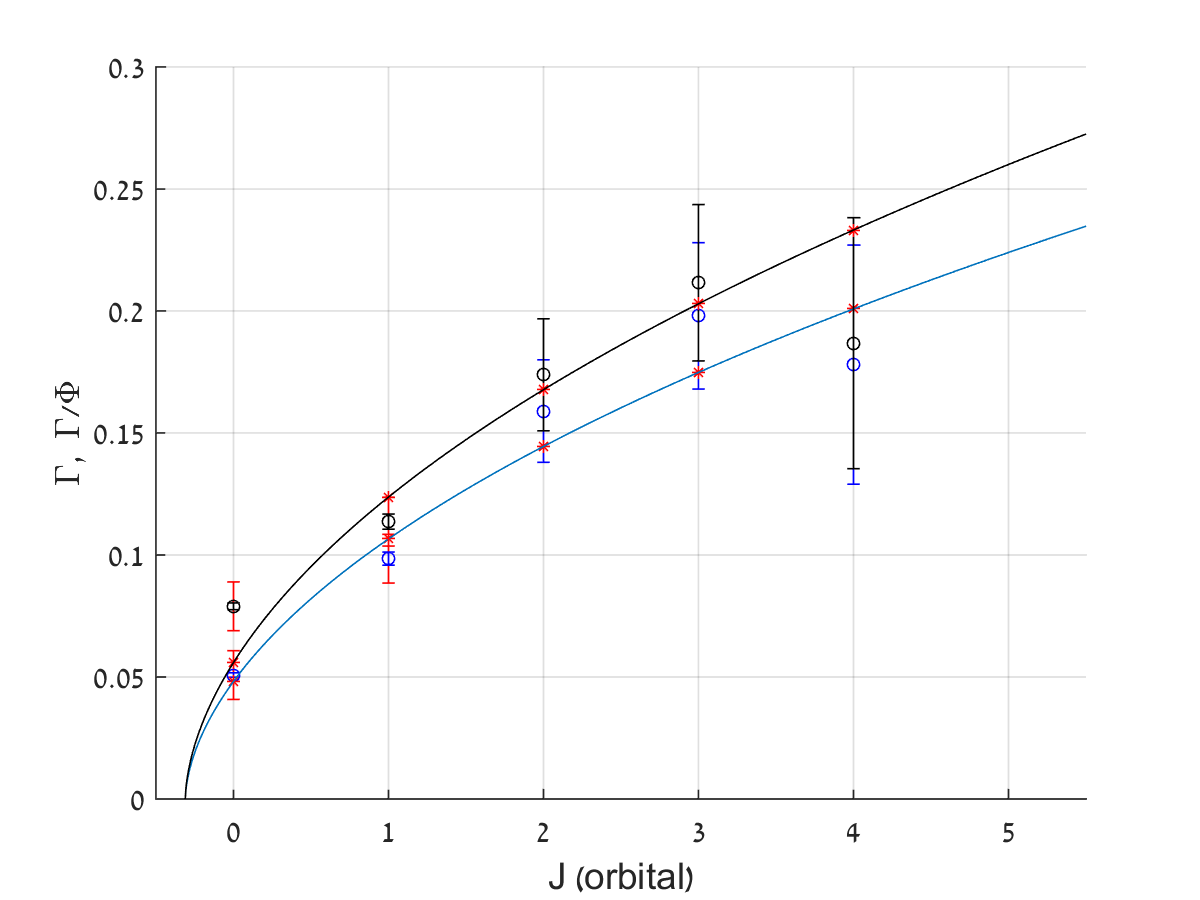}
	\includegraphics[width=0.48\textwidth]{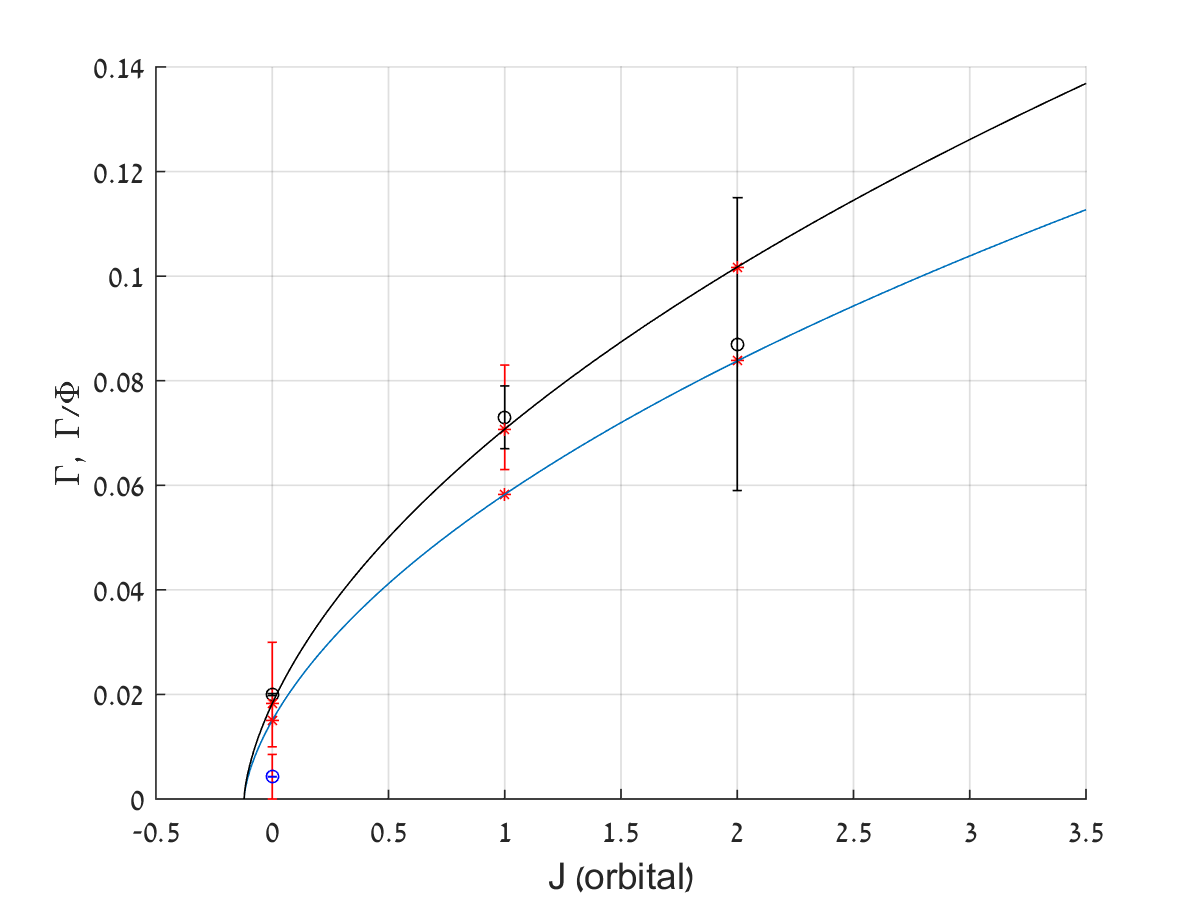} \\
	\caption{\label{fig:model_comparison} Comparison of different fitting procedures for the decay widths of the \(K^*\). \textbf{Top-left:} Three different methods of inserting the intercept are nearly equivalent. Blue line has \(J\to J-a\), red line is when \(L^2\to L^2-2a/T\), and yellow line is with \(L\) determined from modified boundary force equation. We also draw, in black, the fit that would be obtained with \(a=0\) and the classical string length. \textbf{Top-right:} Comparison of fit with \(m_s = 400\) MeV, \(m_{u/d} = 60\) MeV (blue) to one without masses (red). \textbf{Bottom-left:} Comparison of fit where all decay widths are normalized by \(K\pi\) phase space (black), to the fit to the measured widths without normalization (blue). \textbf{Bottom-right:} Comparison with (black) and without (blue) phase space for the \(\phi\) trajectory. Here we normalize only the ground state by dividing it by \(KK\) phase space.}
	\end{figure}

\paragraph{Adding the intercept:} We proposed (section \ref{sec:model_intercept}) three methods of adding the contribution of the intercept to the string length. The simplest is by taking \(J\to J-a\). Another by modifying the boundary equation of the string and adding a repulsive Casimir force, and the third is adding by hand a zero point length to \(L^2\), with a coefficient to be determined from the fits. Here we compare the results of these methods in fitting the \(K^*\) trajectory. The difference between the three methods (in the mass regime of the relevant mesons) is small, and is shown in figure \ref{fig:model_comparison} (top-left). We also draw the case where the intercept is taken to be zero, and then we can fit well the excited states, but not the ground state. Since there are no significant differences between the three methods, we will pick the simplest of them, of adding the zero point length by replacing \(J\to J-a\), to the method we use in all the following fits.

\paragraph{Endpoint masses:} The effect of the endpoint masses can be best seen in a plot of \(\Gamma/M\) as a function of the mass. For a string without endpoint masses, the ratio of width to mass is expected to be constant. With masses, this receives a correction. The width is still assumed to be linear in the length, but the relation between the length and mass is modified.

\begin{table} \centering
\begin{tabular}[ht!]{|l|c|c|c|c|l|}\hline
State & \(J^{P}\) & Mass & Width & \(\Gamma/M\) & Decay modes\footnote{Error estimates in branching ratios are not quoted in this table.} \\ \hline\hline

\(K^*(892)\) & \(1^{-}\) & 891.66\plm0.26 & 50.8\plm0.9 & (5.7\plm0.1)\% & \(K\pi\) (100\%) \\ \hline

\(K^*_2(1430)\) & \(2^{+}\) & 1425.6\plm1.5 & 98.5\plm2.7 & (6.9\plm0.2)\% &  \(K\pi\) (50\%), \(K^*\pi\) (25\%),\\ &&&&& \(K^*\pi\pi\) (13\%), \(K\rho\) (9\%), \(\ldots\) \\ \hline

\(K^*_3(1780)\) & \(3^{-}\) & 1776\plm7 & 159\plm21 & (9.0\plm1.1)\% & \(K\rho\) (31\%), \(K^*\pi\) (20\%), \\ &&&&& \(K\pi\) (19\%), \(K\eta\) (\(\sim\)30\%),\(\ldots\) \\ \hline

\(K^*_4(2045)\) & \(4^{+}\) & 2045\plm9 & 198\plm30 & (9.7\plm1.5)\%& \(K\pi\) (10\%), \(K^*\pi\pi\) (9\%), \\ &&&&& 5 more modes (7\% or less), \(\ldots\) \\ \hline

\(K^*_5(2380)\) & \(5^{-}\) & 2382\plm24 & 178\plm50 & (7.5\plm2.1)\%&\(K\pi\) (6\%), no other measured \\ &&&&& modes. \\ \hline
\end{tabular}
\caption{\label{tab:Ks}The states on the \(K^*\) trajectory and their decay modes. All masses and widths are in MeV. Data from \cite{PDG:2016}.}
\end{table}

The strange \(K^*\) meson has a light quark mass at one endpoint, and a strange quark at the other. The Regge trajectories provided us with a measurement of the quark masses as string endpoints \cite{Sonnenschein:2014jwa}. Fitting the decay widths can give us another measurement. We compare in figure \ref{fig:model_comparison} (top-right) a fit with and without masses. The fit with endpoint masses is preferred, as it has \(\sqrt{\chi^2/DOF} = 0.77\), while the best fit without masses has \(\sqrt{\chi^2/DOF} = 1.32\). When taking into account the widths of the \(\ssb\) \(\phi\) meson, the mass is more important, and the difference in \(\chi^2\) even more significant.

\paragraph{Adding a phase space factor:} We want to see when and where it is necessary to normalize the decay width by a phase space factor, which we insert by hand into the model.

In the case of the \(K^*\) trajectory, we can see, referring back to the data in table \ref{tab:Ks}, that the \(K^*\) meson itself has only one available channel, \(K\pi\). The \(K^*\) is about 260 MeV above the \(K\pi\) threshold. The second state in the trajectory, the \(K^*_2\), is 530 MeV heavier, and already has at least three more participating channels for its decay, in addition to the still dominant \(K\pi\) channel. From \(K^*_3\) onwards there is no single dominant channel, and many decay modes contribute to the total width.

Adding up phase space factors for all the decay modes of the \(K^*_3\) and higher mesons results in overcounting. We do not need this manual counting of available modes to account for the increase in width. In fact, we see from fitting the decay widths of the \(K^*\) to the string model, that it is better in this case \emph{not} to normalize by a phase space factor, and the string length describes the decay width well without further modification. The difference in \(\sqrt{\chi^2/DOF}\) is the difference between 0.8 without phase space factor and 1.6 with them.

This factor will be important for other cases. The case of the \(\phi\) trajectory is the most notable, as the \(\phi\) meson itself is very near to the threshold. There we see a veritable improvement when normalizing the width of the \(\phi\) by the available phase space for \(KK\) decays, with \(\sqrt{\chi^2/DOF}\) improving from 1.8 to 0.5. The decay widths of higher states in the trajectory (\(f^\prime_2(1525)\) and \(\phi_3(1850)\)) do not need this normalization.

The results of the fits are favorable to the assumption that, unless phase space is very limited, there is no need to add factors by hand to the decay width.

\subsubsection{Fit results: the meson trajectories} \label{sec:meson_widths}
\begin{figure}[p!] \centering
	\includegraphics[width=0.48\textwidth]{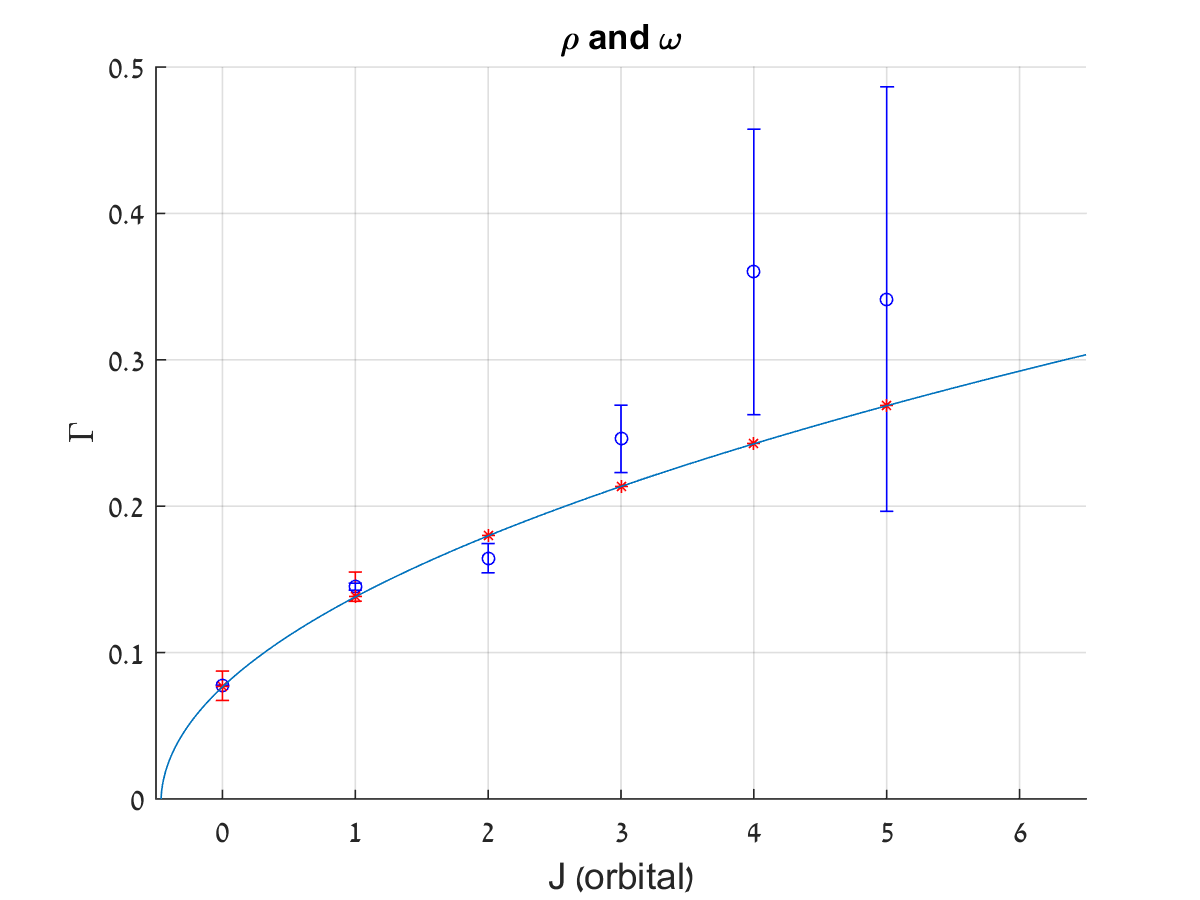}
	\includegraphics[width=0.48\textwidth]{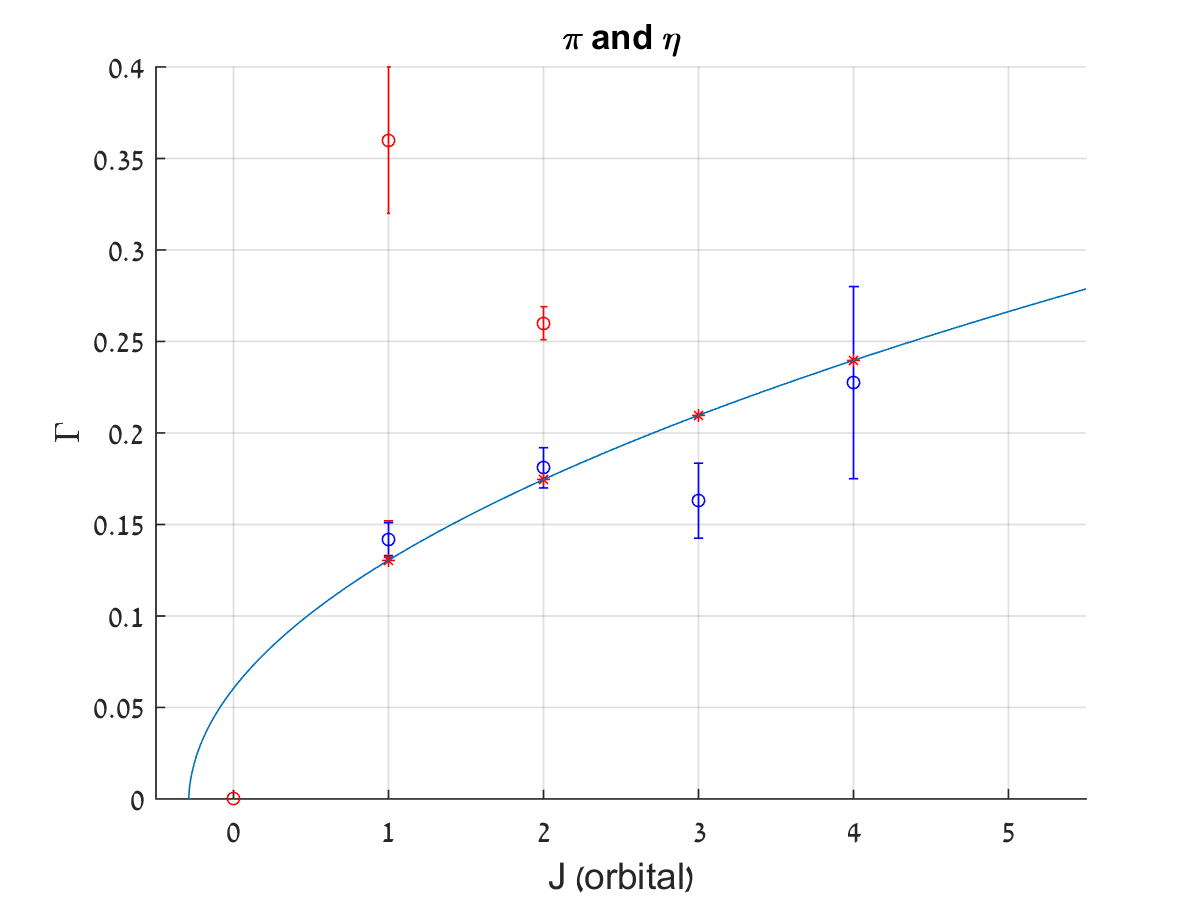} \\
	\includegraphics[width=0.48\textwidth]{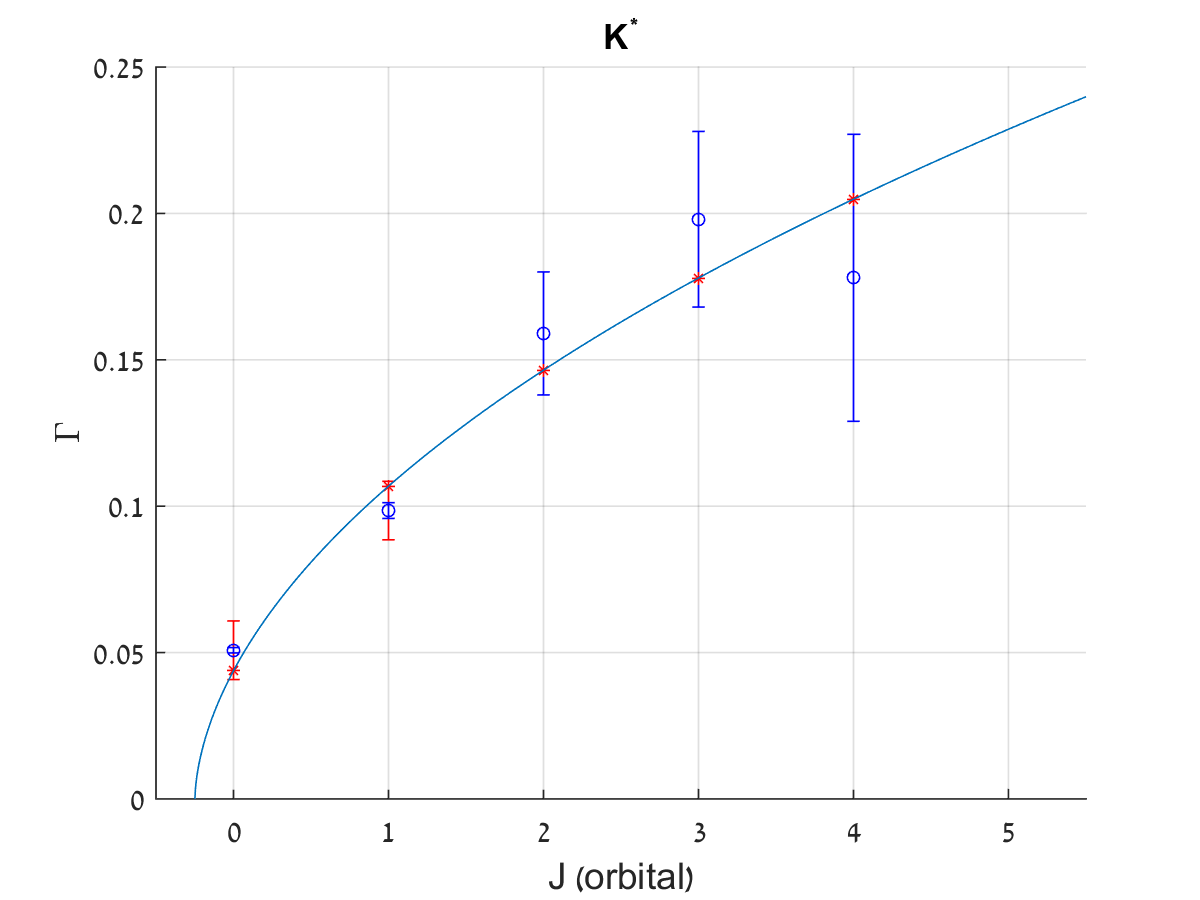}
	\includegraphics[width=0.48\textwidth]{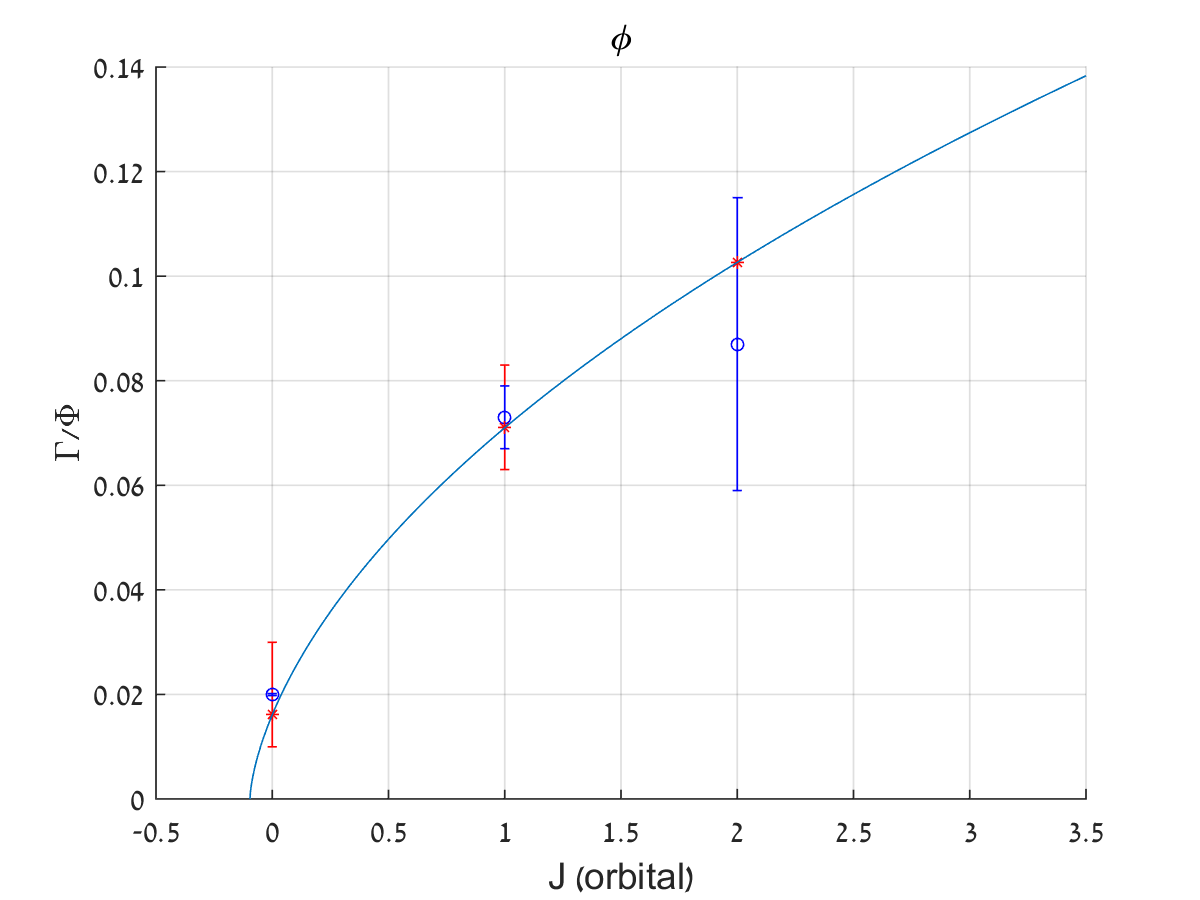} \\
	\includegraphics[width=0.48\textwidth]{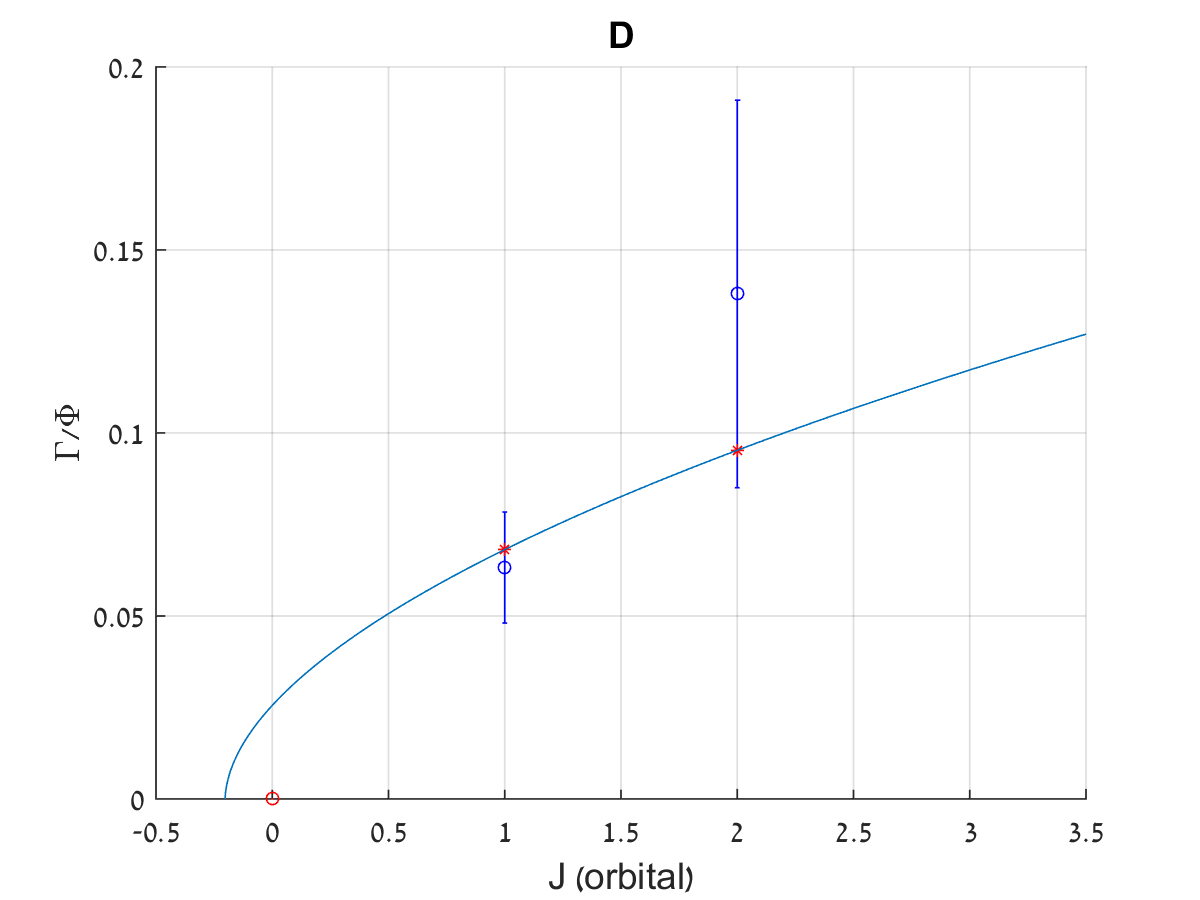}
	\includegraphics[width=0.48\textwidth]{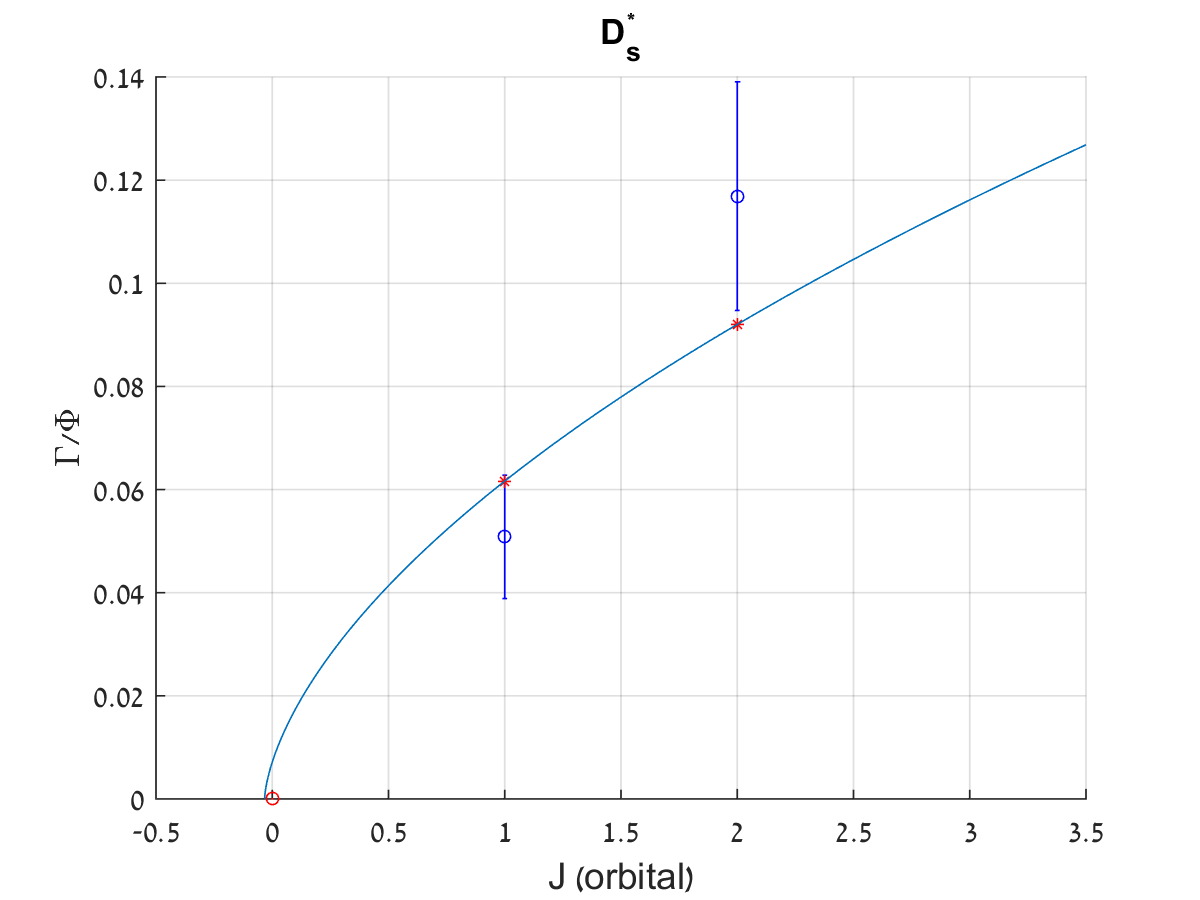}
	\caption{\label{fig:mesons_J} The meson decay widths as a function of their orbital angular momentum. States in red are excluded from the fits.}
	\end{figure}

\begin{figure}[p!] \centering
	\includegraphics[width=0.48\textwidth]{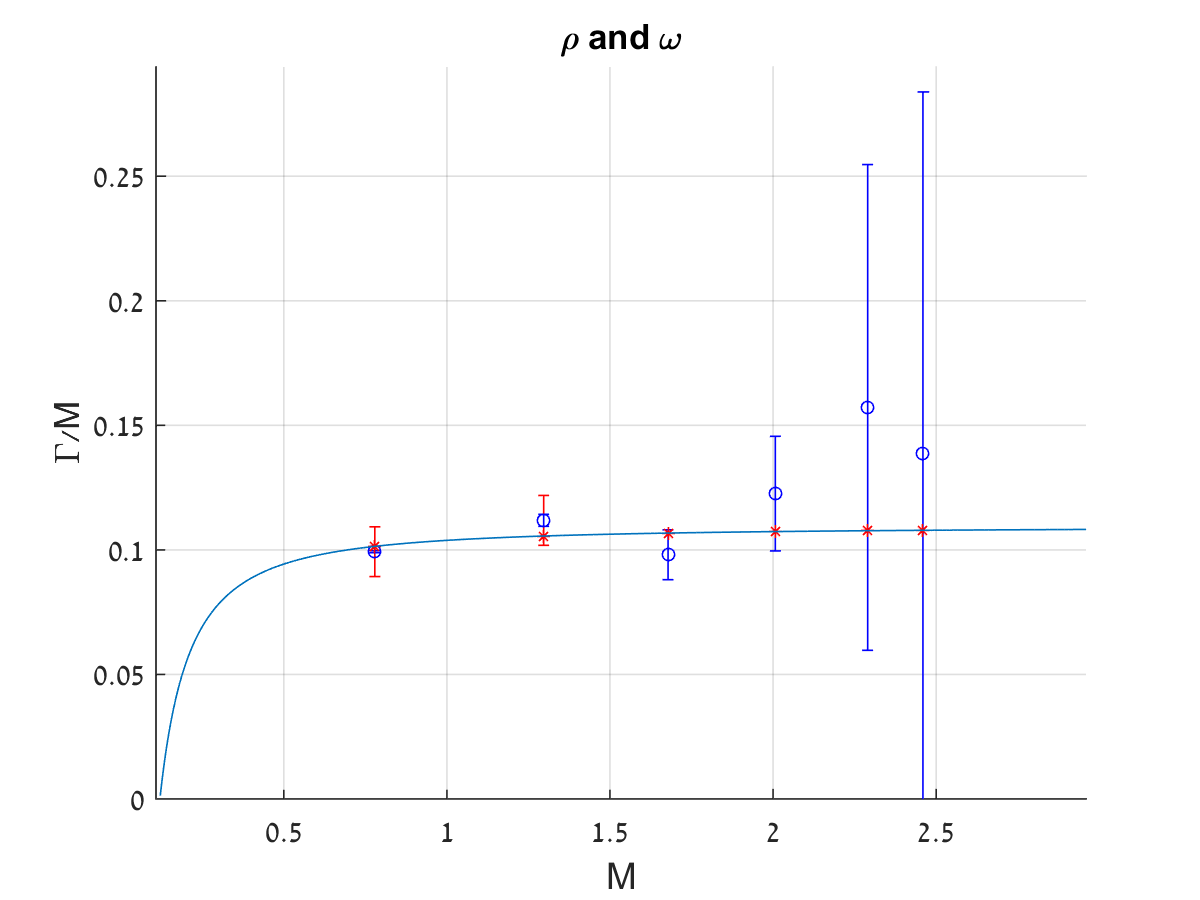}
	\includegraphics[width=0.48\textwidth]{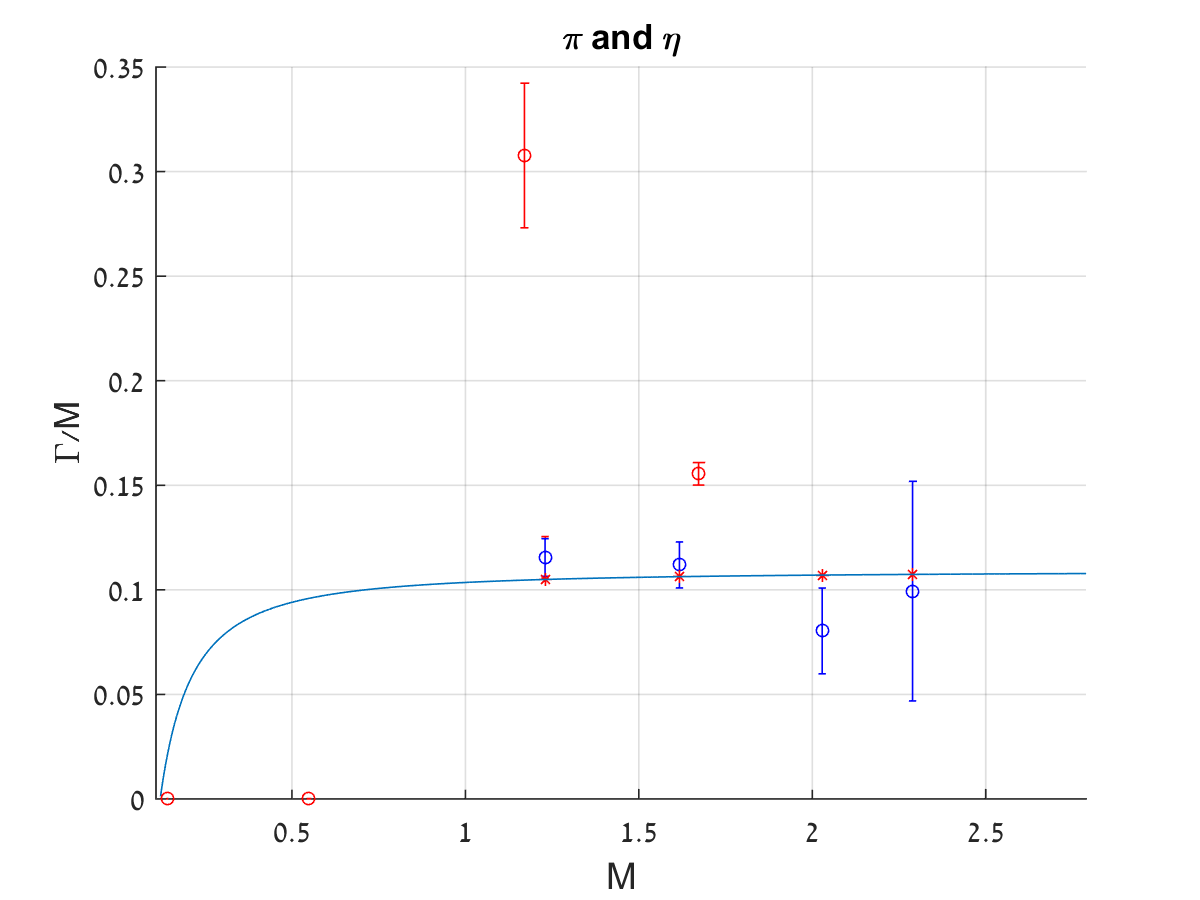} \\
	\includegraphics[width=0.48\textwidth]{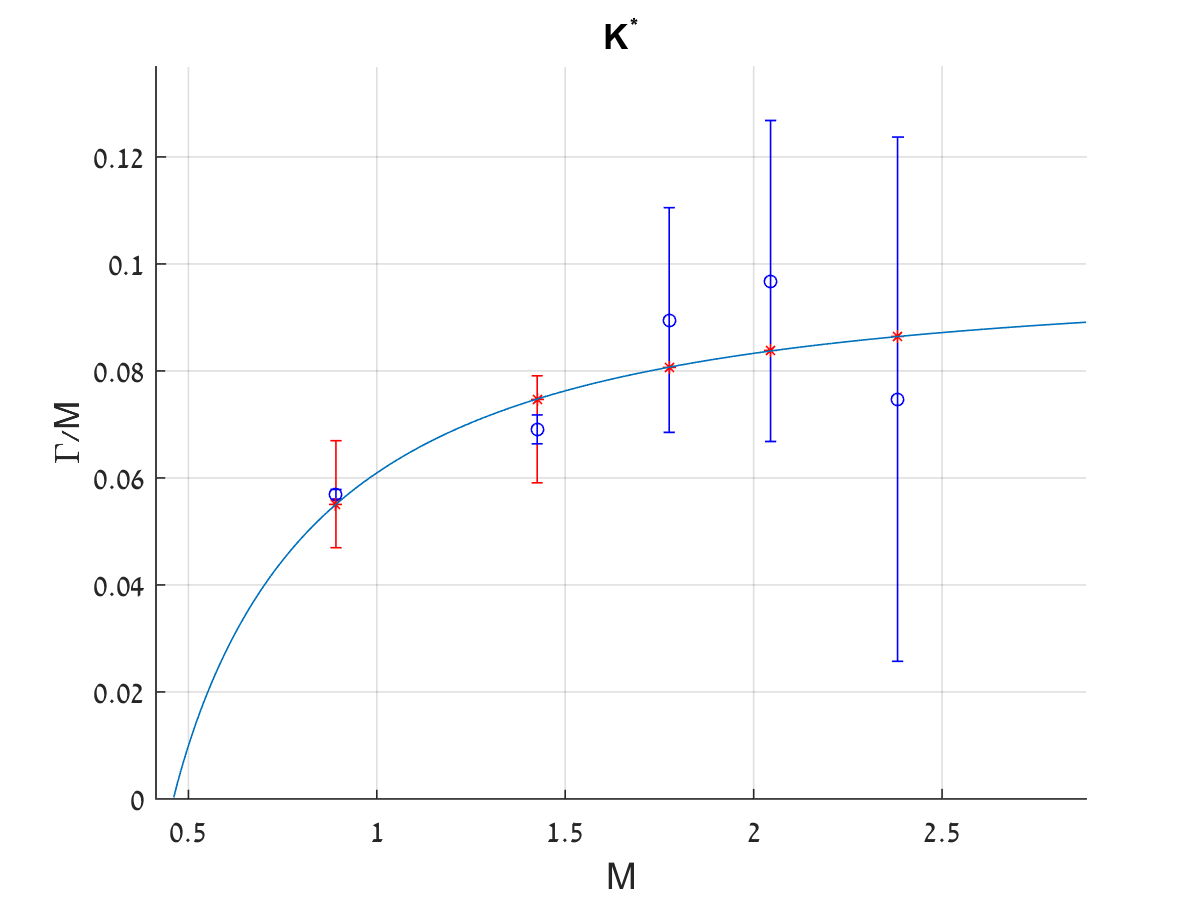}
	\includegraphics[width=0.48\textwidth]{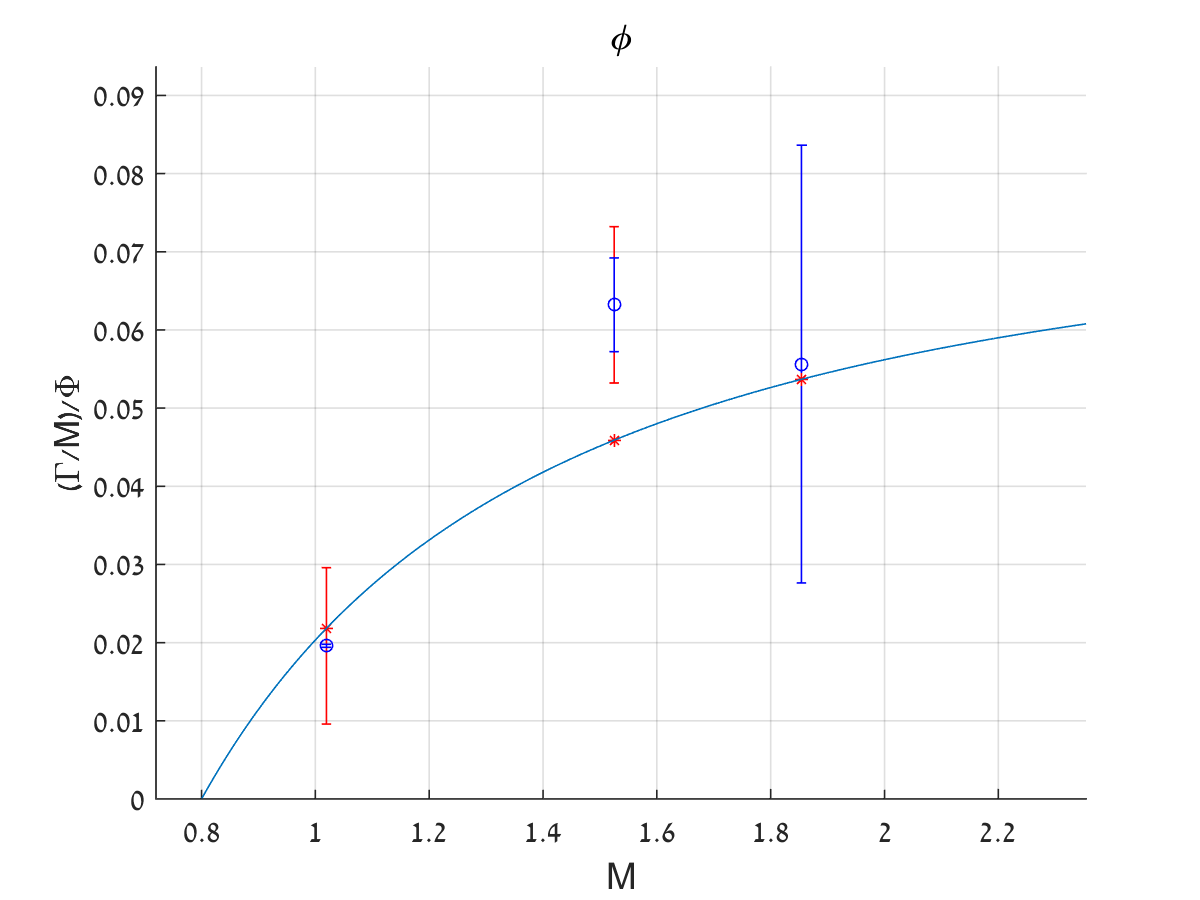} \\
	\includegraphics[width=0.48\textwidth]{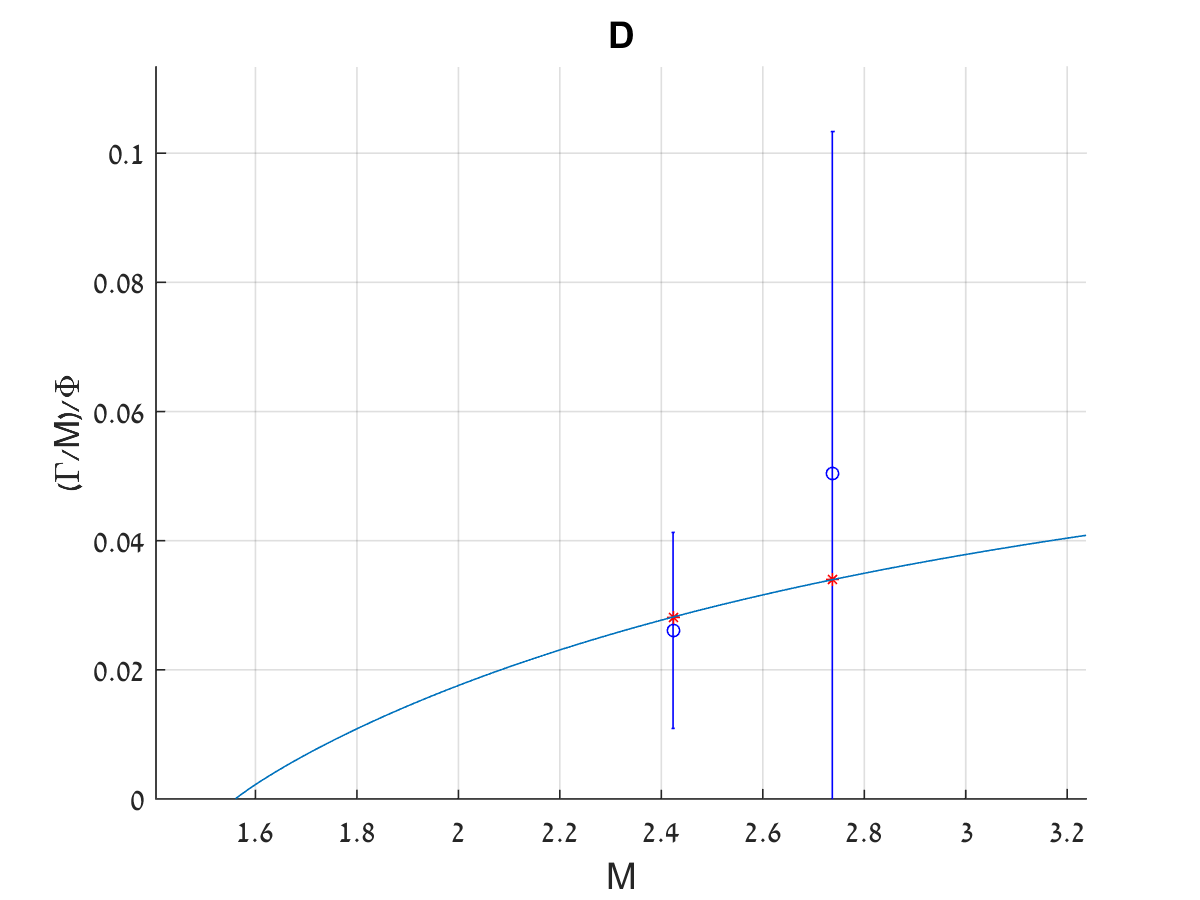}
	\includegraphics[width=0.48\textwidth]{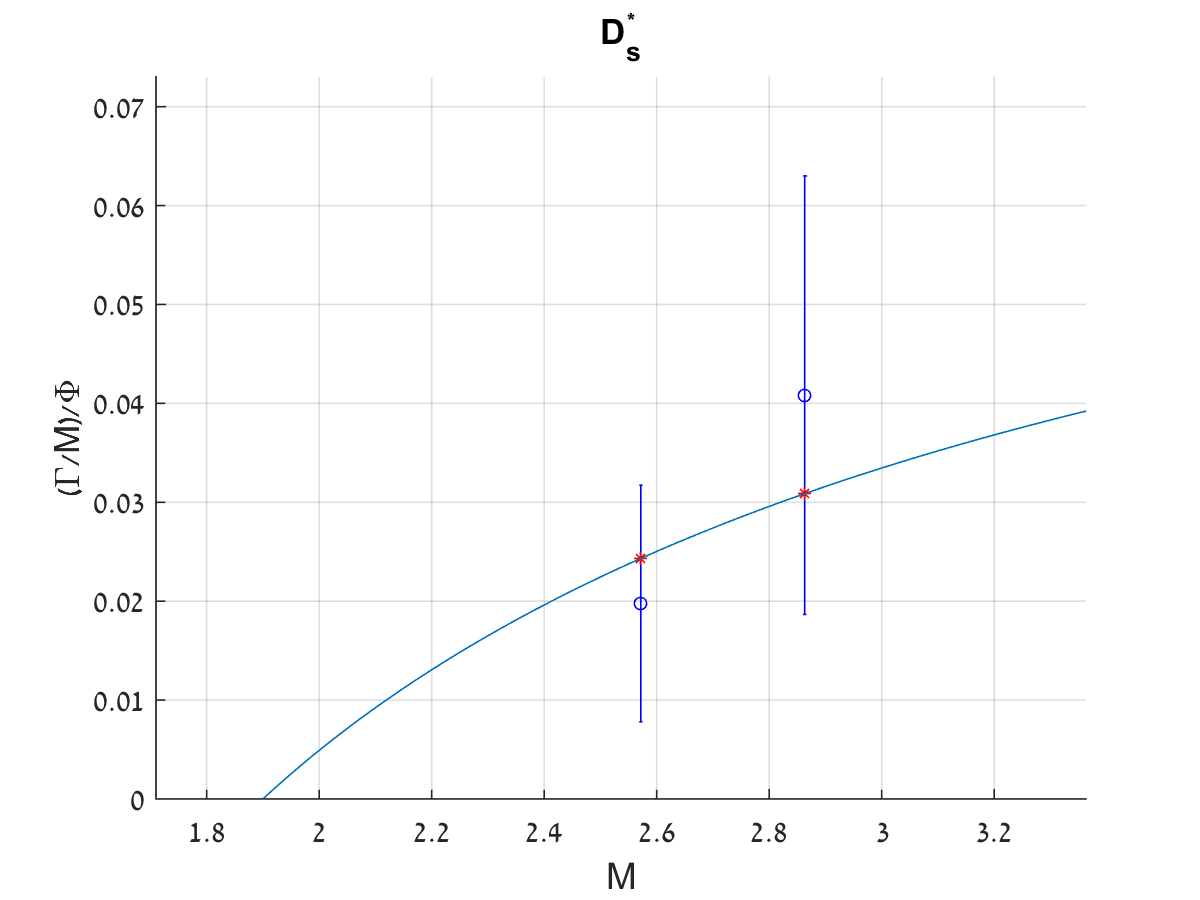}
	\caption{\label{fig:mesons_M} The meson decay widths divided by their masses (\(\Gamma/M\)) as a function of the mass \(M\). States in red are excluded from the fits.}
	\end{figure}

In this section we present the results using the model in which the intercept is added via taking \(J\to J-a\). Phase space factors are included only where specified in the text below. A summary of the results is in table \ref{tab:mesons}. The results of the fits are plotted in figure \ref{fig:mesons_M} as a function of \(M\) and in figure \ref{fig:mesons_J} as a function of the angular momentum. In the first figure, one can see better the effect of the endpoint masses (without masses we would have \(\Gamma/M = Const.\)). In the latter, the effect of the intercept is more visible.

In the following we discuss some of the details of each of the fits. The full specification of all the states used in the fits, with their calculated masses and widths, is in appendix \ref{app:spectrum}.

\begin{table} \centering
	\begin{tabular}{|l|c|c|c|c|c|} \hline
			\multicolumn{2}{|c|}{Trajectory (No. of states)} & \(a\) (from spectrum) & \(A\) (fitted value) & \(\sqrt{\chi^2/DOF}\) \\ \hline
			\(\rho\) & \(5^{[a]}\) & -0.46 & 0.097 & 1.76 \\ \hline
			\(\omega\) & \(5^{[a]}\) & -0.40 & 0.120 & 2.31 \\ \hline
			\(\rho\) and \(\omega\) (avg.) & 6 & -0.46 & 0.108 & 1.14 \\ \hline
			\(\pi\) & \(3^{[a]}\) & -0.34 & 0.100 & 1.66 \\ \hline
			\(\eta\) & \(3^{[a]}\) & -0.29 & 0.108 & 1.56 \\ \hline
			\(\pi\) and \(\eta\) (avg.) & 4 & -0.29 & 0.109 & 1.52 \\ \hline
			\(K^*\) & 5 & -0.25 & 0.098 & 0.77 \\ \hline
			\(\phi\) & 3 & -0.10 & 0.074 & 0.50 \\ \hline
			\(D\) & 2 & -0.20 & 0.072	 & 0.87 \\ \hline
			\(D^*_s\) & 2 & -0.03 & 0.076 & 1.44 \\ \hline
	\end{tabular}
	\caption{\label{tab:mesons} Summary of the meson decay width fits. The slope and masses used are \(\alp = 0.884\) GeV\(^{-2}\), \(m_{u/d} = 60\) MeV, \(m_s = 400\) MeV, \(m_c = 1500\) MeV. The intercepts are obtained from the Regge trajectories. [a] In these trajectories a state was excluded. The number is the number of remaining states. See text below for details, and appendix \ref{app:spectrum} for the full specification of states used and their calculated masses and widths.}
\end{table}

\paragraph{Light mesons, \(S = 1\): the \(\rho\) and \(\omega\):} Here we have the two trajectories with the largest number of points, as we have states going from \(J = 1\) all the way up to \(J = 6\). In the trajectories of light mesons, which can always easily decay into pions, we see that there is no need to normalize the widths by including phase space factors. However, there are some sizable deviations from the simple picture in the widths the ground states. The \(\rho\) meson itself, for instance, is much wider at \(\Gamma\approx 150\) MeV than one would expect - its ratio of \(\Gamma/M\) is significantly larger than that of its excited partners. For this reason, the ground state \(\rho\) meson has to be excluded from the fit. A similar case is the \(\omega\), which is narrower than one would expect. On the other hand, averaging the widths of the \(\rho\) trajectory across isospin with the states of the \(\omega\) trajectory these deviations cancel out, so we can get a good fit of the widths even when we include the ground states.

\paragraph{Light mesons, \(S = 0\): the \(\eta\) and \(\pi\):} In the \(\eta\) and \(\pi\) trajectories, the ground states, which have zero or nearly zero width for strong decays, are excluded. Here we also find large deviations from the model in the cases of some states. Namely these are the \(h_1(1170)\) meson (\(\Gamma = 360\pm40\) MeV), which is the first orbital excitation of the \(\eta\) meson, and the \(\pi_2(1670)\) (\(\Gamma = 260\pm9\) MeV). These states have to be excluded entirely. The deviations here are too large to be rid of by averaging over isospin like in the previous case. What we can do is take the \(J = 1\) state from the \(\pi\) trajectory (\(b_1(1235)\)), the \(J = 2\) state from the \(\eta\) trajectory (\(\eta_2(1645)\)), and average the remaining the \(J = 3\) and 4 states. Then we can place the resulting four points on a line.

\paragraph{Strange mesons: the \(K^*\):} This was discussed in detail in the previous section. We have five points from \(J = 1\) to \(J = 5\) and find that the string model gives a good fit using the quark masses and the intercepts obtained from the Regge trajectories.

\paragraph{\(\ssb\) mesons: the \(\phi\):} Here we first see the need to add the phase space factor. The \(\phi\) meson must decay to two kaons, but it is less than 40 MeV above threshold to do it. Hence, it is narrower than the naive formula, without a suppressive phase space factor, predicts it to be. The next state in the trajectory of the \(\phi\) meson is \(f^\prime_2(1525)\), which no longer suffers that suppression. To fit this trajectory, we first normalized the width of the ground state by dividing it by the phase space available for the \(KK\) decay. The measured width of the \(\phi\) is \(4.27\pm0.03\) MeV. After normalization it becomes 20 MeV, and the widths are well fitted.

\paragraph{Charmed and charmed-strange mesons: \(D\) and \(D^*_s\):} Here the decay products must include a heavy charmed meson, so again phase space restrictions are important. Here we normalize the decay widths of all states in the trajectory with appropriate phase space factors. We take that the \(D\) mesons decay to \(D\pi\) and the \(D^*_s\) to \(DK\). The ground states in these trajectories do not decay strongly and are excluded. We are left with only two points per trajectory, and the test is to see how well the widths are fitted using the parameters previously obtained from the spectrum.

\subsubsection{Zweig suppressed decays and the string length} \label{sec:pheno_Zweig}
The probability of a meson to decay via annihilation of the quark and antiquark forming it is expected (section \ref{sec:decay_Zweig}) to be exponentially suppressed in the string length, so that
\be \Gamma = \Gamma_Z \exp(-T_{Z}L^2/2) \label{eq:Zweig_L}\ee

We can observe this behavior for the \(\Upsilon\) mesons, where we look at
the three \(\Upsilon(nS)\) states that are below the threshold to decay by tearing the string and forming a pair into a pair of \(B\) mesons. The total decay width of the first three \(\Upsilon\) mesons decreases with the mass. The next state is above threshold and is significantly wider. We can attribute the decrease of the width to the increasing string length and the suppression factor of eq. \ref{eq:Zweig_L}.

The excited states \(\Upsilon(2S)\) and \(\Upsilon(3S)\) can also decay by falling to a lower excited state and emitting an \(\eta\) meson or a pair of pions. We do not describe these decays here. We take the partial decay widths for \(\Upsilon \to ggg \to\) hadrons and (the smaller, not wholly significant) width for \(\Upsilon \to \gamma gg \to \gamma+\text{hadrons}\). In the string picture, these would correspond to a decay where the endpoints of the string met, and an intermediate closed string was formed. The full decay width, the measured branching ratios for \(ggg\) and \(\gamma gg\), and the partial decay widths calculated from the width and branching ratios are listed in table \ref{tab:Upsilon_nS}.

\begin{table}[ht!] \centering
	\begin{tabular}{|c|c|c|c|c|c|} \hline
			State & Full width [keV] & \(B(ggg)\) & \(B(\gamma gg)\) & Partial width [keV] & Best fit [keV] \\ \hline
			\(\Upsilon(1S)\) & 54.02\plm1.25 & 81.7\plm0.7\% & 2.2\plm0.6\% & 45.3\plm1.3 & 45.2 \\
			\(\Upsilon(2S)\) & 31.98\plm2.63 & 58.8\plm1.2\% & 1.87\plm0.28\% & 19.4\plm1.7 & 20.6 \\
			\(\Upsilon(3S)\) & 20.32\plm1.85 & 35.7\plm2.6\% & 0.97\plm0.18\% & 7.5\plm0.9 & 7.1 \\ \hline
	\end{tabular}
\caption{\label{tab:Upsilon_nS} The below threshold \(\Upsilon\) mesons and their widths. The partial width for decays where the endpoints meet is extracted from the data and fitted to eq. \ref{eq:Zweig_L}.}
\end{table}

Unlike in the previous two subsections, the relevant states are excited radially rather than orbitally. We do not derive an expression for the length as a function of the excitation number \(n\). We assume here that we can calculate the length by replacing \(J\to J+n-a\) in all relevant equations, as we did when describing the Regge trajectories, including the radial trajectory of the \(\Upsilon\). This trajectory is comprised the above mentioned states as well as three more excited states. Then we proceed by calculating the length in  as we did in the decay width fits. The mass of the \(b\) quark is taken to be 4700 MeV.

The decay widths of table \ref{tab:Upsilon_nS} are then well described by formula \ref{eq:Zweig_L}, using the string tension \(T_Z = 0.10\) GeV\(^2\). The calculated widths are listed in the last column of table \ref{tab:Upsilon_nS}. The fit gives \(\chi^2/DOF = 0.6\). The string tension is not the 0.18 GeV\(^2\) of the spectrum, but it is not very far removed considering the approximations involved. Note that the parameter \(\Gamma_Z \approx \Gamma\left(\Upsilon(1S)\right)\) is what makes the width small, while the exponent describes only the decrease in the width as a function of the mass.

\subsection{The decay width of baryons} \label{sec:pheno_baryons}
We use the same fitting model for the baryons as we did for the mesons. The fitting parameters are the same, but we use a slightly different value for the slope, \(\alp = 0.95\) GeV\(^{-2}\), which is what we obtain from the baryonic Regge trajectories \cite{Sonnenschein:2014bia}. The quark masses are \(m_{u/d} = 60\) MeV and \(m_s = 400\) MeV, as for the mesons.

Diquarks are assumed to have the same mass as the heavier quark in the diquark (i.e. \(m_{ud} = m_u\), \(m_{us} = m_s\) etc.).  This is because the mass of the endpoint in the mapping from the holographic string to the flat spacetime string with massive endpoints is determined by the length of the vertical segment of the string, and this length is identical for quark and diquark (see figure \ref{holtoflat2}). A priori there would also be a contribution from the baryonic vertex, but that was found to be negligible \cite{Sonnenschein:2014bia}, at least in the case where one of the quarks in the diquark is a \(u\) or \(d\) quark.

The intercepts are determined from fitting the Regge trajectories, when using the above values of the masses and slope. Again the intercept is included by replacing \(J\to J-a\) in the classical expressions. We do not normalize widths by phase space factors except where specified.

While there are many identified baryon resonances, we have less available data here for the widths. We fit the widths of states on the trajectories of the light quark baryons \(N\) and \(\Delta\), and the strange \(\Lambda\) and \(\Sigma\). For baryons containing heavier quarks there is no trajectory with enough strongly decaying baryons to be fitted.

\begin{table}[ht!] \centering
	\begin{tabular}{|l|c|c|c|c|c|} \hline
			\multicolumn{2}{|c|}{Trajectory (No. of states)} & \(a\) (from spectrum) & \(A\) (fitted value) & \(\sqrt{\chi^2/DOF}\) \\ \hline
			\(N\) (even)& 2 & -0.77 & 0.080 & 3.33 \\ \hline
			\(N\) (odd) & 3 & -1.11 & 0.082 & 2.43 \\ \hline
			\(\Delta\) (even) & 3 & -1.37 & 0.101 & 1.90 \\ \hline
			\(\Lambda\) & 4 & -0.46 & 0.041 & 2.33 \\ \hline
			\(\Sigma\) (\(S=1/2\)) & 2 & -0.95 & 0.052 & 0.96 \\ \hline
			\(\Sigma\) (\(S=3/2\)) & 3 & -1.22 & 0.100 & 1.57 \\ \hline
	\end{tabular}
\caption{\label{tab:baryons}  Summary of the baryon decay width fits. The slope and masses used are \(\alp = 0.950\) GeV\(^{-2}\), \(m_{u/d} = 60\) MeV, and \(m_s = 400\) MeV. The intercepts are obtained from the Regge trajectories. See appendix \ref{app:spectrum} for the full specification of states used and their calculated masses and widths.}
\end{table}

The results are summarized in table \ref{tab:baryons} and plotted in figure \ref{fig:baryons_J}. In appendix \ref{app:spectrum} we also list all the states used in the fits together with their calculated widths and masses. We find that the fits to the decay widths for the baryons are significantly worse than those of the mesons, especially for the \(N\) and \(\Delta\) baryons. The decay widths of the excited \(N/\Delta\) baryons are much larger than one would expect. Their decay width seems closer to linear in \(J\) or \(M^2\) than linear in the length.

\paragraph{Light baryons, \(N\) and \(\Delta\):} One of the differences between baryons and mesons is the ``even-odd effect'' in the Regge trajectory of light baryons \cite{Sonnenschein:2014bia}. For the \(N\) and \(\Delta\) trajectories, even and odd orbital angular momentum states have their own separate trajectories, lying on parallel lines with different intercepts. Neglecting corrections from the endpoint masses, the trajectories are split as
\be J = \begin{cases}
	\alpha M^2 + a_{e}\,, & J \text{ (orbital) even} \\
	\alpha M^2 + a_{o}\,, & J \text{ (orbital) odd}\,. \\ \end{cases}
\ee
This also affects the decay widths. For the Regge trajectories, this even-odd effect manifests itself as a difference between the two intercepts of the even orbital angular momentum states and that of the odd states.

The width of the \(N\) and \(\Delta\) baryons rises too quickly with \(J\) to be linear in the string length. However, the measurements of the widths of large spin baryons are far from accurate. For instance, the PDG estimates the width of the \(N(2190)\) (\(J^P = 7/2^{-}\)) to be 300 to 500 to 700 MeV \cite{PDG:2016}, based on several conflicting measurements. Our model would prefer the width to take the lowest values reported.

\paragraph{Strange baryons, \(\Lambda\) and \(\Sigma\):} Here we have three trajectories. We do not see the even-odd effect  in the Regge trajectories of strange baryons, so here there is no need to separate even and odd states. The trajectories are one of the \(\Lambda\), and two of the \(\Sigma\). Of the \(\Sigma\) trajectories, one begins with the \(\Sigma\) ground state baryon (\(J^{P} = 1/2^{+}\)) and another with the \(\Sigma(1385)\) (\(J^P = 3/2^+\)). The ground state \(\Lambda\) and \(\Sigma\) baryons have no strong decays and are excluded. The \(\Sigma(1385)\) decays to \(\Lambda\pi\) or \(\Sigma\pi\) with limited phase space for a total measured width of 36\plm5 MeV. We normalize it by the \(\Lambda\pi\) phase space to 120\plm10 MeV. The rest of the decay widths used are the measured widths with no normalization.

\begin{figure}[p!] \centering
	\includegraphics[width=0.48\textwidth]{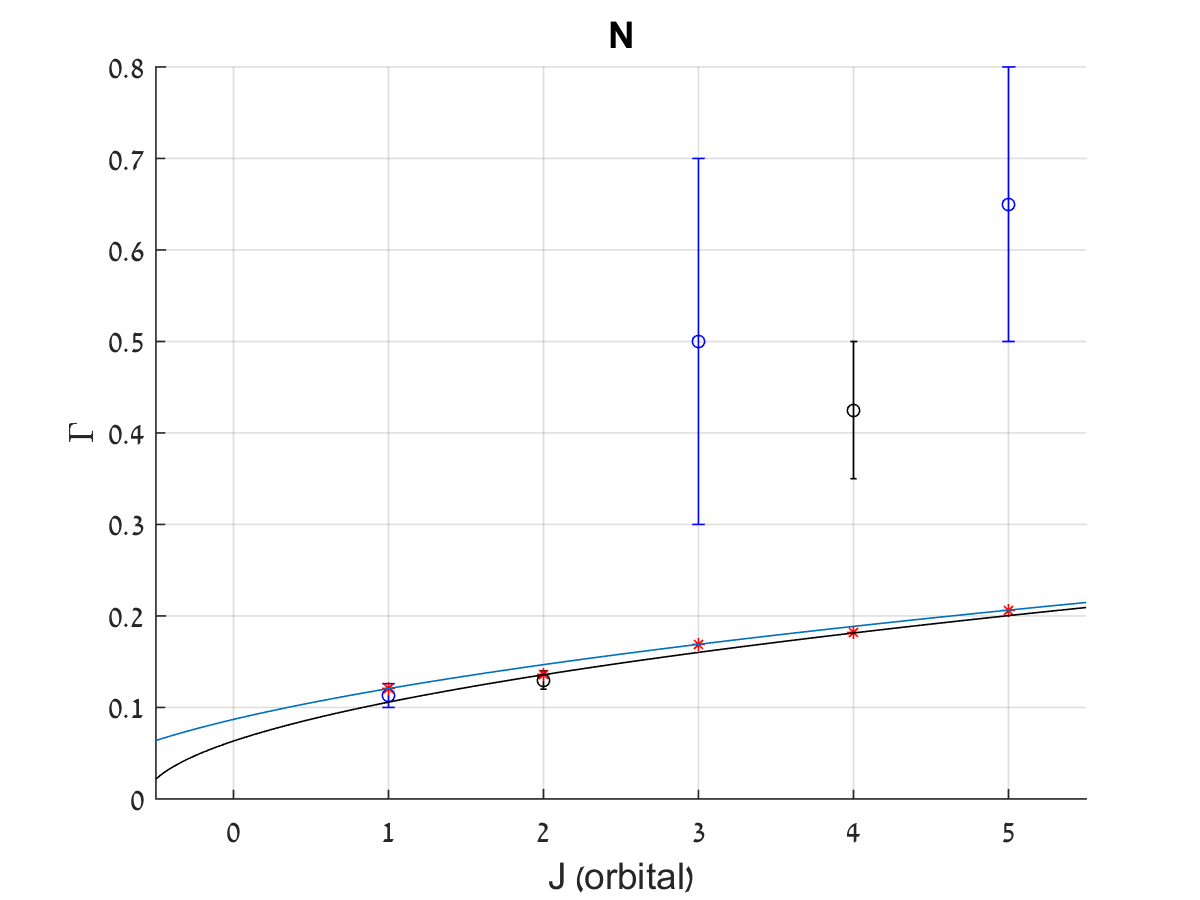}
	\includegraphics[width=0.48\textwidth]{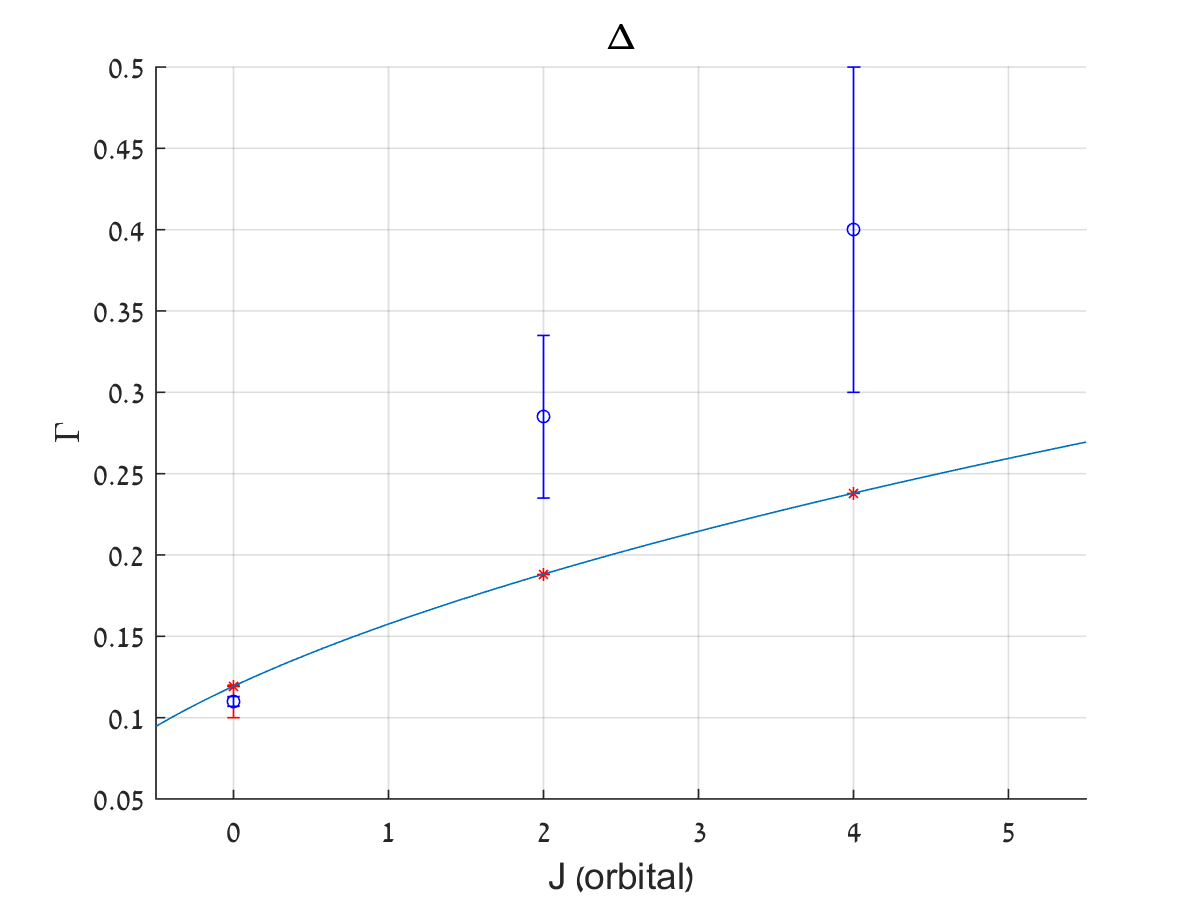} \\
	\includegraphics[width=0.48\textwidth]{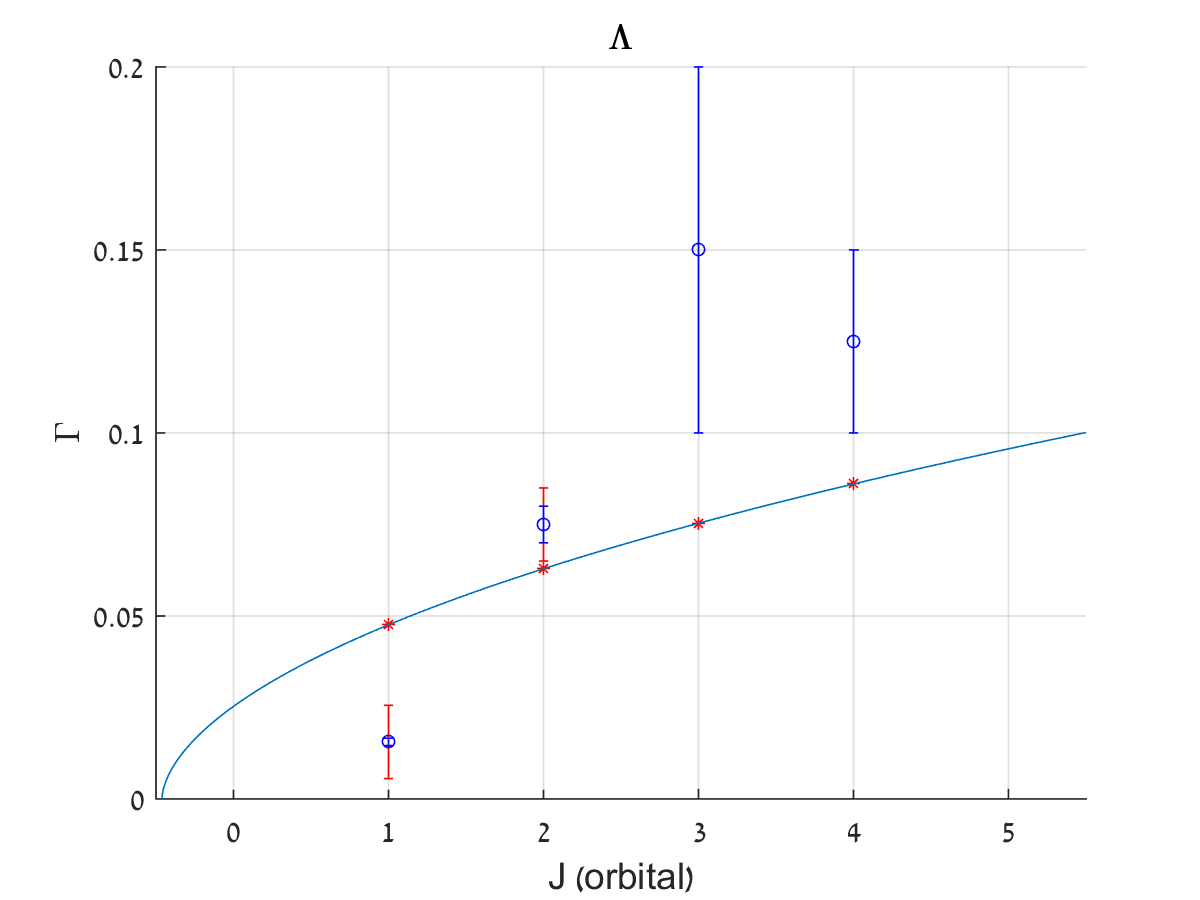} \\
	\includegraphics[width=0.48\textwidth]{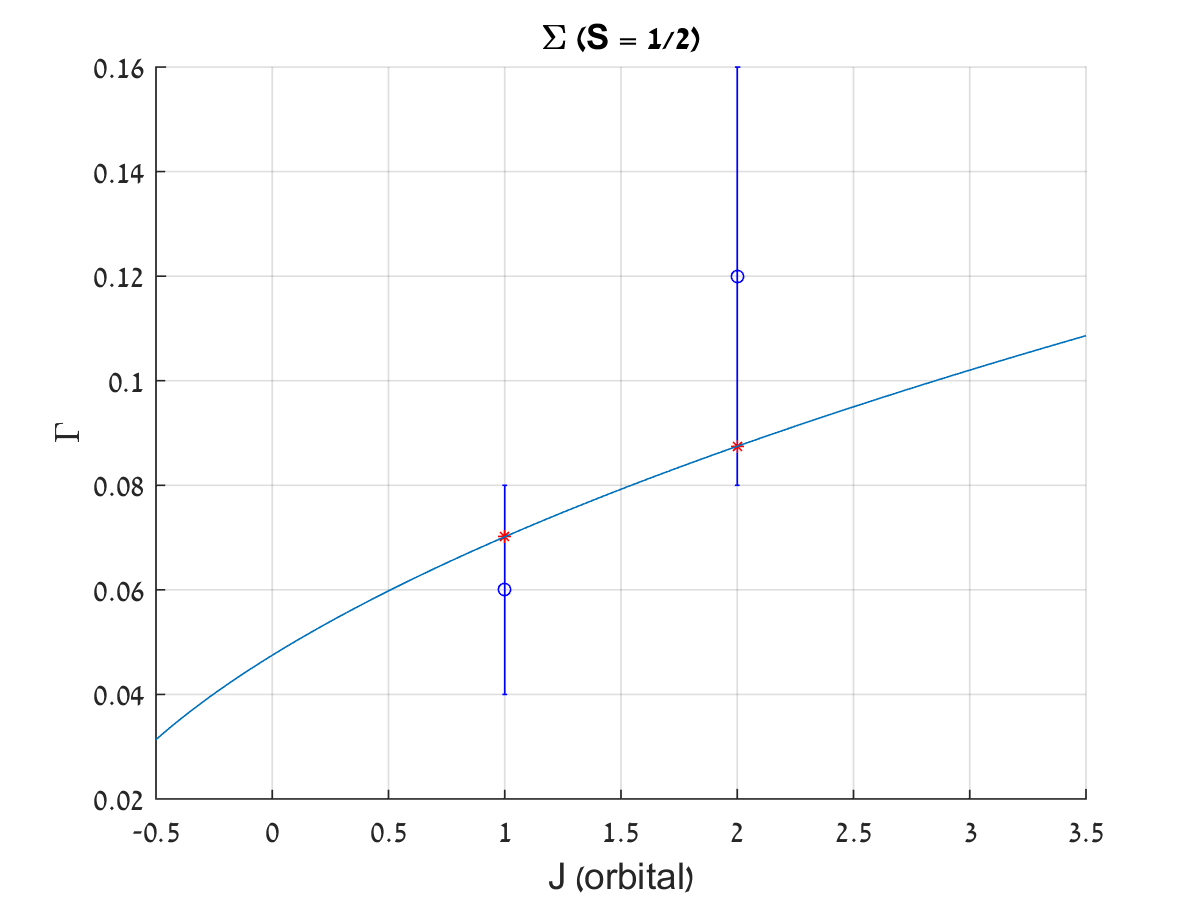} 
	\includegraphics[width=0.48\textwidth]{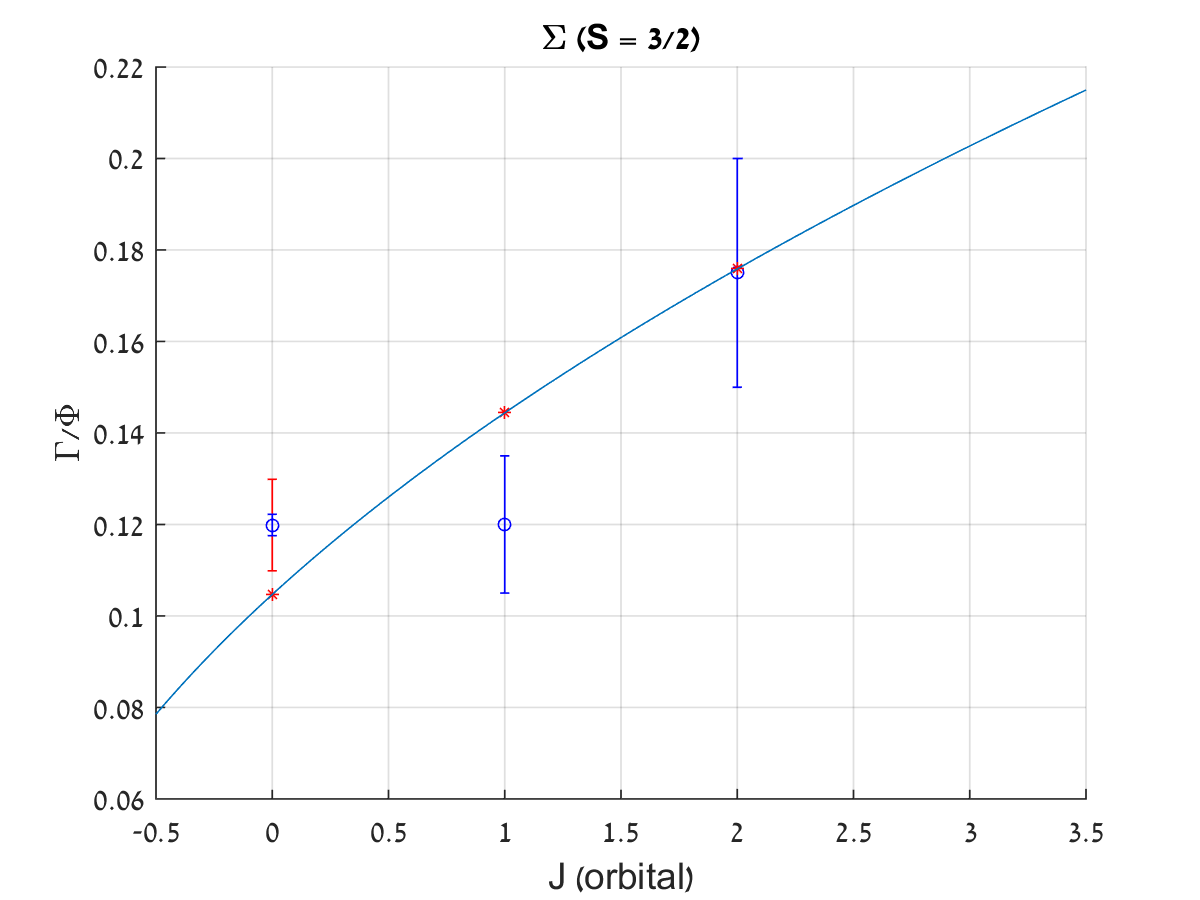}
	\caption{\label{fig:baryons_J} The baryon decay widths as a function of the orbital angular momentum.}
	\end{figure}

\subsubsection{Baryon decays and the quark diquark structure} \label{sec:baryon_structure}
As was discussed in section \ref{sec:decay_baryons} the decays of baryons can shed light on their structure. Since the baryons form Regge trajectories they are described in the HISH model as a single string with a quark at one endpoint and a diquark at the other \cite{Sonnenschein:2014bia}, with the motivation for this being chiefly that the Regge trajectories of the baryons have the same, or nearly the same slope as the meson trajectories.

The issue then is, out of the three quarks, which pair forms the diquark and which quark is at the other end of the string. When the baryon decays, the leading mechanism is assumed to be the same as for meson decays. The string between the diquark and quark tears, and a lower mass baryon and meson are produced. The diquark is then preserved in the decay.

In \cite{Sonnenschein:2014bia} we extracted the endpoint masses by fitting the Regge trajectories of baryons. This provides us with some information on the structure of the baryon, but without the decays we cannot extract information on the arrangement of the quarks within the baryon.

To form a notion of the diquark structure, it is best to start with baryons containing strange quarks. If we look at an excited doubly strange baryon \(\Xi\), it is possible that both strange quarks are in the diquark, or that only one of them is in the diquark and the other is located at the other end of the string. In the former case, \((ss)q\), the \(\Xi\) would decay predominantly to a lower doubly strange baryon and a light meson, \(\Xi\pi\) most likely, as was drawn in figure \ref{fig:ssddecay}. In the latter case, \((qs)s\), the \(\Xi\) would decay mainly to two strange hadrons, \(\Lambda K\) or \(\Sigma K\). This also seems to break the \(SU(3)\) flavor symmetry, as one could expect both types of channels to exist based on group theoretical factors. In section \ref{sec:isospin} it was maintained that the baryons made up of light quarks only have to be linear combinations of the possible quark-diquark configurations in order to preserve isospin. With \(SU(3)\), on the other hand, we can simply claim that the quark-diquark structure breaks the \(SU(3)\) symmetry, as the strange quark mass already breaks it in decay processes by introducing the exponential suppression factor discussed extensively in section \ref{sec:decay_exponent}.

Looking at the decay modes of known \(\Xi\) baryons, for which there are sufficient measurements of the relevant branching ratios, it appears that they are all consistent with the \((qs)s\) configuration. That is, they decay mostly to \(\Sigma K\) or \(\Lambda K\). One exception to the above is the first excited doubly strange baryon, \(\Xi(1530) \jp{3}{+}\). The \(\Xi(1530)\) decays predominantly to \(\Xi\pi\), but it does not have the phase space for any other strong decays. The width of the \(\Xi(1530)\) is 9.1\plm0.5 MeV. An estimate for the width of a baryon of that mass in the \((ss)q\) configuration, based on its string length, gives 15--25 MeV. This is wider than what is measured, but not sufficiently wide to draw conclusions from, especially since our estimates of the baryon widths are not always accurate. Had the calculated value been much wider, we could have excluded the \((ss)q\), on those grounds. In the case of the \((qs)s\) configuration we do not have an estimate of the width, except that we expect it to be narrower.

For baryons containing a single \(s\) quark the two possible configurations are \((qq)s\), decaying mainly to a strange meson and a non-strange baryon, or \((qs)q\), which should decay to a light meson and a strange baryon. Unlike for the doubly strange baryons, here we see that both options are present for most baryons. The \(\Lambda\) baryons tend to decay to both \(NK\) and \(\Sigma\pi\), with no dominant channel identifiable. Similarly, the \(\Sigma\) can decay to either of the channels \(NK\), \(\Sigma\pi\), and \(\Lambda\pi\).

It is possible than the strange baryons are mixtures of \((qq)s\) and \((qs)q\). In the HISH picture, the masses of both configurations are expected to be the same, as \(m_{(qs)} \approx m_s\). Then the different branching ratios can be explained as a function of this mixing.

Note that for the doubly strange mesons, where the mixing between \((ss)q\) and \((qs)s\) might be smaller because of the initial mass differences between the two configurations, we found that \((qs)s\) is clearly preferred for the observed \(\Xi\) baryons. The question then is whether one could observe also baryons that are most likely to be purely \((ss)q\).

Another option is the decay of the Y-shaped baryon (section \ref{sec:decay_Y}), where the probability for each channel is proportional to the length of the leg of the string that tears in that decay mode. In the case of a \(\Lambda\) or \(\Sigma\) baryon, this means that the strange quark should have a higher probability of remaining in the baryon, since the string connecting a quark to the baryonic vertex is shorter for the heavier quarks. We do not see any indication that this might be the case for the different strange baryons.

Another case that is worth commenting on is that of the newly discovered \(\Omega_c\) excited baryons, with the quark content \(css\). In \cite{Aaij:2017nav} were reported five new resonances in the range 3000--3119 MeV, noted for their very small decay widths, of a few MeV each. The five resonances are expected to be the first orbitally excited states of the \(\Omega_c^0\) \jp{1}{+} and \(\Omega_c(2770)^0\) \jp{3}{+}, with the quantum numbers \(J^P = \jp{1}{-}\) (two states), \jp{3}{-} (two states), and \jp{5}{-} \cite{Karliner:2017kfm}. In fact, these states at the given mass range are what one would expect also from the HISH picture, where they can be identified as the first excited states on the Regge trajectories of \(\Omega_c^0\) and \(\Omega_c(2770)^0\).

The narrow excited \(\Omega_c\) were discovered in their decays to \(\Sigma_c+K^-\), a charmed-strange baryon and a strange meson. This would be the natural decay mode of the \((cs)s\) configuration. On the other hand, all five of these resonances are below threshold for the decay that would be dominant if they were \((ss)c\), which is \(\Xi D\). It is possible that these new resonances are in fact \((ss)c\), and that the narrow width is due to the fact that the diquark has to break for the baryon to decay. Their width if they were \((cs)s)\) should be at least 15--20 MeV based on their estimated string length. If there no mixing between the two configurations, then \((ss)c\) is the configuration that would explain why these new states are so narrow.


\subsection{Exponential suppression of pair creation} \label{sec:pheno_lambdas}
In this part we look in the data for the exponential factor suppressing pair creation of massive quarks, namely strange quarks as the heavier quarks are too massive, and we compare it to the creation of light \(u/d\) quarks.

In the holographic string model (section \ref{sec:decay_exponent}) the strangeness suppression factor is a result of the distance between the holographic wall and the flavor brane associated with the \(s\) quark. The probability that the string fluctuate and reach the brane is exponentially suppressed with this distance squared. The quark masses, as the vertical segments of the holographic string, are in turn proportional to the same distance. The probability to create a quark pair of mass \(m_q\) generally behaves like
\be \exp\left(-2\pi C m_q^2/T_{\text{eff}}\right)\,.\ee

What we would like to measure is the ratio between the probability to create an \(\ssb\) pair and the probability of light pair creation, noted \(\lambda_s\). Using the expression above it is expressible as
\be \lambda_s = \exp\left(-2\pi C (m_s^2-m_{u/d}^2)/T_{\text{eff}}\right)\,. \ee
The ``canonical'' value of \(\lambda_s\) is \(\approx0.3\) \cite{PDG_fragmentation}. This value is used in event generators such as Pythia \cite{Sjostrand:2006za}, which base themselves on the string fragmentation picture of the Lund model \cite{Andersson:1983ia}. In \cite{Abreu:1996na}, the value of \(\lambda_s = 0.3\) was obtained by comparison with \(e^+e^-\to\) hadrons data from LEP. Other measurements favoring \(\lambda_s \approx 0.2\) have also been performed, with dependence on center-of-mass energy also investigated \cite{Aid:1996ui}. A more recent measurement is \cite{Chekanov:2006wz}.

The value \(\lambda_s = 0.3\) can be obtained from theory by inserting constituent quark masses - \(m_u = m_d = 330\) MeV and \(m_s = 500\) MeV - into the pair production probability of the Schwinger mechanism (eq. \ref{e:singledecay}). This also means taking \(C = 1/4\) in the equation above. The tension is that obtained from the Regge trajectories, \(T_{\text{eff}} = 0.18\) GeV\(^2\).

We argue that the masses that are relevant are not the constituent masses, but those of the quarks as string endpoints. Since the ratio depends on the difference \(m_s^2-m_{u/d}^2\) then it is not sensitive to the definition of the masses by themselves. We can lower the masses to their values as we obtain them and maintain the same strangeness suppression factor. For example, \(m_s = 400\) MeV (as used in the decay width fits) and \(m_{u/d} \approx 100\) MeV, reproduces the same value. The bigger discrepancy is with the prefactor in the exponent, where \(C = 1\) as obtained in the holographic models of section \ref{sec:decay_exponent}. If we want to use \(C=1\) we need to combine it with smaller values of the strange quark mass, around 200 MeV. We can still describe the spectrum and decay widths using \(m_s = 200\) MeV, but this leads to suboptimal results.

\subsubsection{In strong decays of hadrons}
The holographic model of the decays predicts that \(\lambda_s\) should play a role in simple one to two hadron decays, as it always enters into the probability that the meson tear and decay (section \ref{sec:decay_exponent}). This is different from the results cited above, where \(\lambda_s\) is obtained from processes where a large number of hadrons is produced, and determines the ratio between strange (\(K\) or \(\Lambda\)) and non-strange light hadrons (\(\pi\) or \(N\)). Even though the mechanism supposed is identical, of a string breaking with the probability of an \(\ssb\) pair suppressed compared with light quarks, we take a more ``microscopic'' view here, by looking at each hadron's decay patterns separately, rather than describing ratios obtained from multi-particle events.

In the following we try and isolate the strangeness suppression factor \(\lambda_s\) in hadron decays, by comparing branching ratios of decay channels differing only by the mass of the quark-antiquark pair created. Other quantum numbers, like the spins of the outgoing particles, should be the same in both channels. For example, a decay into \(\omega \pi\), a vector and pseudoscalar, can be compared to a decay to \(K^* K\), their strange counterparts.

To summarize, the partial widths of a channel where an \(\ssb\) pair is created and one with light quarks are taken to be related by
\be \frac{\Gamma_s}{\Gamma_l} = \lambda_s \times \frac{p_s}{p_l}. \ee
where \(p_i\) are the momenta of the outgoing hadrons in the relevant decays. This is the same phase space factor as in eq. \ref{eq:phase_space}.

In other contexts, the ratio between these types of partial decay widths can be predicted using the \(SU(3)\) flavor group \cite{Samios:1974tw,Guzey:2005vz}. On the other hand, the strangeness suppression factor \(\lambda_s\) badly breaks the \(SU(3)\) symmetry and these ratios are no longer expected to hold. Isospin Clebsch-Gordan coefficients may play a role, but for the decays examined, when we do not care for differences in charges, they add up to one.

The kinematic factors in the above are taken to be proportional to the momentum of the outgoing particles. There is a second option, where one takes \(\phi_i \propto p_i^{2\ell+1}\), with \(\ell\) the orbital angular momentum in the final state. This is the convention used in \cite{Samios:1974tw}. The extra factor of \(p_i^{2\ell}\) is from the ``barrier penetrationary factor''. We find that when using this form of the kinematic factor, our measurements of \(\lambda_s\) are not compatible with the expected value from jet experiments, and often giving values larger than one, albeit with very large error bars. 

\begin{table}[ht!] \centering
	\begin{tabular}{|lc|ll|ll|c|c|} \hline
			Hadron & \(J^P\) & \multicolumn{2}{|l|}{Light channel} & \multicolumn{2}{|l|}{\(\ssb\) channel} & Ratio & \(\lambda_s\) \\ \hline\hline
			
			\(\rho_3(1690)\) & \(3^{-}\) & \(\omega\pi\) & 16\plm6\% & \(K\bar K\pi\) & 3.8\plm1.2\% & 0.24\plm0.12 & 0.30\plm0.15\\ \hline
			
			\(\rho_3(1690)\) & \(3^{-}\) & \(\pi\pi\) & 23.6\plm1.3\% & \(K\bar K\) & 1.58\plm0.26\% & 0.07\plm0.01 & 0.08\plm0.01 \\ \hline
			
			
			\(K^*_4(2045)\) & \(4^{+}\) & \(K^*\pi\pi\pi\) & 7\plm5\% & \(\phi K^*\) & 1.4\plm0.7\% & 0.20\plm0.17 & 0.32\plm0.28\\ \hline\hline
			
			\(N(1650)\) & \(1/2^{-}\) & \(N\pi\) & 50--70\% & \(\Lambda K\) & 5--15\% & 0.17\plm0.09 & 0.57\plm0.30  \\ \hline
			
			\(N(1710)\) & \(1/2^{+}\) & \(N\pi\) & 5--20\% & \(\Lambda K\) & 5--25\% & ? & large? \\ \hline
			
			\(N(1720)\) & \(3/2^{+}\) & \(N\pi\) & 11\plm3\% & \(\Lambda K\) & 4--5\% & 0.41\plm0.12 & 0.87\plm0.26  \\ \hline
			
			
			\(N(1880)\) & \(1/2^{+}\) & \(\Delta\pi\) & 29\plm12\% & \(\Sigma K\) & 17\plm7\% & 0.59\plm0.34 & 0.81\plm0.47 \\ \hline
			
			
			\(N(2190)\) & \(7/2^{-}\) & \(N\pi\) & 10--20\% & \(\Lambda K\) & 0.5\plm0.3\% & 0.03\plm0.02 & 0.04\plm0.03 \\ \hline
			
			\(\Delta(1910)\) & \(1/2^{+}\) & \(\Delta\pi\) & 34--66\% & \(\Sigma K\) & 4--14\% & 0.18\plm0.12 & 0.25\plm0.16 \\ \hline
			
	\end{tabular}
\caption{\label{tab:lambda_s}  Measurements of \(\lambda_s\) from various hadrons.}
\end{table}	

We list some measurements of \(\lambda_s\) using the above formula in table \ref{tab:lambda_s}. As can be seen, the values vary greatly between different hadrons, although in many cases there are also large uncertainties involved. It should be noted that the PDG estimates are often imprecise, as a result of large uncertainties and sometimes conflicting results from experiment.

On the other hand averaging over the different results gives \(\lambda_s = 0.40\pm0.21\), which is close to what one would expect. More precise measurements are presented in the next subsection, where \(\lambda_s\) is measured in the radiative decays of heavy quarkonia. The result there will agree with the average measured here.

To comment on some of the individual hadrons' decay channels from table \ref{tab:lambda_s}, we note that in the case of the \(K^*_4(2045)\) we compare the decay to \(\phi K^*\) to that to \(K^*\pi\pi\pi\). Essentially, we assume that three pions come from the decay of an \(\omega\) meson. It is then that the channels are comparable. Similarly, for the \(\rho_3(1690)\) we compare the channel \(\omega\pi\) to \(K\bar K\pi\) (\(\phi\pi\)). In both of these cases the value of \(\lambda_s\) is close to its expected value.

A particularly interesting case is the \(N(1710)\), which seems to prefer to decay to \(\Lambda K\). The ratios listed in table \ref{tab:lambda_s} are careful estimates by the PDG, but some measurements are more conclusive. In \cite{Anisovich:2011fc} a branching ratio of 23\plm7\% is reported for \(N(1710)\to\Lambda K\), and only 5\plm4\% to \(N\pi\). Additionally, there is a large \(N\eta\) branching ratio, 17\plm10\%. This could indicate that there is a large \(\ssb\) component in the \(N(1710)\) itself, with it perhaps being an exotic state, specifically a pentaquark. This possibility has been raised before \cite{Jaffe:2003sg}.

\subsubsection{In radiative decays} \label{sec:pheno_rad}
Another place where we can find the strangeness suppression factor is the radiative decays of heavy quarkonia. This type of decays involves a different mechanism to the one discussed in the previous section, as it involves the annihilation of the quark and antiquark in the meson, as well as the emission of a photon. Part of the process involves a closed string which tears (as seen in section \ref{sec:decay_Zweig}), introducing the exponential suppression factor we want to measure.

We can measure the factor by comparing the decays \(J/\Psi\to\gamma f_2(1270)\) and \(J/\Psi\to\gamma f_2^\prime(1525)\), where the \(f_2(1270)\) is a spin-two isoscalar meson made up of light quarks, and the \(f_2^\prime(1525)\) is its \(\ssb\) counterpart. The measured ratio is \cite{PDG:2016}
\be \frac{\Gamma(J/\Psi\to\gamma f_2^\prime(1525))}{\Gamma(J/\Psi\to\gamma f_2(1270))} = 0.31\pm0.06\,. \ee
For the \(\Upsilon(1S)\), the ratio between the same two channels is
\be \frac{\Gamma(\Upsilon\to\gamma f_2^\prime(1525))}{\Gamma(\Upsilon\to\gamma f_2(1270))} = 0.38\pm0.10\,. \ee

Including two-body phase space corrections (whose effect is negligible in the latter case), the strangeness suppression factor measured from these processes is
\be \lambda_s = 0.34\pm0.07 \ee
from the \(J/\Psi\) decays, and
\be \lambda_s = 0.38\pm0.10 \ee
from the \(\Upsilon\) decays.

One could look for additional measurements in the radiative decays of the excited states \(\Psi(2S)\) and \(\Upsilon(2S)\). The \(\Psi(2S)\) already has been observed to decay to \(\gamma f_2(1270)\), and it should therefore have the second decay mode \(\gamma f_2^\prime(1525)\), with a similar ratio of \(\approx0.36\) between the the two channels. 

\subsection{The decays of exotic hadrons} \label{sec:pheno_exotic}
\subsubsection{Glueball decays} \label{sec:pheno_glueball}
Using the measured value of \(\lambda_s\), we can give a prediction for some branching ratios of glueball decays, as was presented in section \ref{sec:decay_glueball}. We use the value of \(\lambda_s\) obtained in section \ref{sec:pheno_rad} from the radiative decays of quarkonia, since there the decay mechanism also involves a closed string tearing. For the glueball, the closed string must tear twice to decay, and so we can create zero, one, or two pairs of strange quarks in the decay.

To give an example, consider a glueball candidate of a mass of 2.2 GeV, where we assume a specific value of the mass in order to evaluate the phase space factors explicitly. A glueball of this mass could be an excited scalar glueball, or a tensor \cite{Sonnenschein:2015zaa}. It is above threshold for decays into two \(\ssb\) mesons. So we can predict the existence of several decay channels and the relative ratios between them. For instance, for decays into a pair of chargeless vector mesons, we predict the ratio between the decay widths
\be \Gamma(GB\to\omega\omega) : \Gamma(GB\to K^{*0}K^{*0}) : \Gamma(GB\to\phi\phi) = 1 : 0.83\times \lambda_s : 0.53\times \lambda_s^2\,. \ee
Taking \(\lambda_s = 0.36\) from the last subsection, the numbers are
\be \Gamma(GB\to\omega\omega) : \Gamma(GB\to K^{*0}K^{*0}) : \Gamma(GB\to\phi\phi) = 1 : 0.30 : 0.07. \ee
The presence of these three types of two body decays, with the given hierarchy between them, is unique to closed strings, and hence could be useful in identifying glueballs.

\subsubsection{The decays of tetraquarks} \label{sec:pheno_tetra}
In section \ref{sec:decayexotic} we have described a possible decay mechanism of  tetraquark states built from a string connecting a diquark and an anti-diquark. This proposed mechanism \cite{Sonnenschein:2016ibx} followed from the experimental observation that there exists a charmonium-like state, $Y(4630)$, that  decays primarily into \(\Lambda_c\overline{\Lambda}_c\). In \cite{Sonnenschein:2016ibx} we made the claim that if indeed the state is a tetraquark then  there should be higher excited  tetraquark states residing on a trajectory with the $Y(4630)$ state. These states should also predominantly decay into a pair of $\Lambda_c\overline{\Lambda}_c$. Assuming the linearity property of the decay width we can now estimate the width of the higher excited states on the trajectory of the $Y(4630)$ as given in table \ref{tab:pred_y_c}.
\begin{table}[h!] \centering
	\begin{tabular}{|c|c|c|} \hline
		\(n\)	&	 Mass 	&	 Width \\ \hline\hline
		``-2'' & 4060\plm50 & Narrow \\ \hline
		``-1'' & 4360\plm50 & Narrow \\ \hline\hline
		0 & \(\bf{4634^{+9}_{-11}}\)	&	\(\bf{92^{+41}_{-32}}\)	\\ \hline\hline
		1 &	4870\plm50 	&	 150--250	\\ \hline
		2	&	5100\plm60 	&	 200--300	\\ \hline
		3	&	5305\plm60 	&	 220--320	\\ \hline
		4	&	5500\plm60 	&	 250--350	\\ \hline
	\end{tabular} \qquad\qquad
	\begin{tabular}{|c|c|c|} 
		\multicolumn{3}{c}{} \\ \hline
		\(J^{PC}\) &	  Mass 	&	 Width \\ \hline\hline
		\(0^{++}\) & 4485\plm40					& Narrow \\ \hline\hline
		\(1^{--}\) & \(\bf{4634^{+9}_{-11}}\)					& \(\bf{92^{+41}_{-32}}\)	\\ \hline\hline
		\(2^{++}\) & 4800\plm40 	&	 150--250	\\ \hline
		\(3^{--}\) & 4960\plm40 	&	 180--280	\\ \hline
		\(4^{++}\) & 5100\plm45 	&	 200--300	\\ \hline
		\(5^{--}\) & 5260\plm45 	&	 250--350	\\ \hline
	\end{tabular} \caption{\label{tab:pred_y_c} Trajectories of the \(Y(4630)\). Based on the experimental mass and width of the \(Y(4630)\) we extrapolate to higher excited states on the trajectory. Uncertainties are based on both experimental errors and uncertainties in the fit parameters. The excited states are expected to decay into \(\Lambda_c\overline{\Lambda}_c\). Some possible lower states on the trajectory, with masses below the \(\Lambda_c\overline{\Lambda}_c\) threshold, are also included. \(Y(4360)\), observed to decay to \(\Psi(2S)\pi^+\pi^-\), is a candidate for the \(n = -1\) state \cite{Sonnenschein:2016ibx}.}
\end{table}

In the spectroscopy of heavy quarkonia, one usually expects analogies to exist between the \(\ccb\) and \(\bbb\) spectra. Tetraquarks and other exotics should be no exception. Therefore, we proposed to search for an analogous state to the \(Y(4630)\), which we will call the \(Y_b\), that decays primarily to \(\Lambda_b\bar{\Lambda}_b\). We will assume that, like the \(Y(4630)\), the \(Y_b\) state is located a little above the relevant baryon-antibaryon threshold of \(\Lambda_b\overline{\Lambda}_b\).

The mass of the \(\Lambda_b^0\) is \(5619.51\pm0.23\) MeV. As an estimate, take a mass of
\be \begin{split} & M(Y_b) \approx 2M(\Lambda_b^0)+40\MEV \\
	& \Delta M(Y_b) \approx 40\MEV \end{split} \ee
The \(Y_b\) would also have its own Regge trajectory, so, based on the mass we choose for it, we can predict the rest of the states that lie on the trajectory. The trajectories are calculated using the slopes of the \(\Upsilon\) trajectories, which are \(\alp \approx 0.64\GEVm\) in the \((J,M^2)\) plane, and \(\alp\approx0.46\GEVm\) in the \((n,M^2)\) plane. The string endpoint mass of the \(b\) quark is \(m_b = 4730\) MeV.\footnote{Values taken from the Regge trajectory fits of \cite{Sonnenschein:2014jwa}.} The resulting masses are listed in table \ref{tab:pred_y_b}.


\begin{table}[h!] \centering
	\begin{tabular}{|c|c|} \hline
		\(n\)	&	 Mass 	\\ \hline\hline
		``-2'' & 10870\plm50 \\ \hline
		``-1'' & 11080\plm50 \\ \hline\hline
		0 & 11280\plm40	\\ \hline
		1 &	11460\plm40 \\ \hline
		2	&	11640\plm40 \\ \hline
		3	&	11810\plm40 \\ \hline
		4	&	11980\plm40 \\ \hline
	\end{tabular} \qquad\qquad
	\begin{tabular}{|c|c|} \hline
		\(J^{PC}\) &	  Mass 	\\ \hline\hline
		\(1^{--}\) & 11280\plm40	\\ \hline
		\(2^{++}\) & 11410\plm40 	\\ \hline
		\(3^{--}\) & 11550\plm40 	\\ \hline
		\(4^{++}\) & 11670\plm40	\\ \hline
		\(5^{--}\) & 11800\plm40 	\\ \hline
	\end{tabular} 
	\caption{\label{tab:pred_y_b} Predictions for the states of the \(Y_b\), a tetraquark containing \(\bbb\) and decaying to \(\Lambda_b\bar{\Lambda}_b\). The mass of the first state is taken near threshold, masses of higher states are on the Regge trajectories that follow from the ground state. Like in table \ref{tab:pred_y_c}, we can continue the radial trajectory backwards. The observed resonance \(Y_b(10890)\) is a potential match to be the ``\(n = -2\)'' state \cite{Sonnenschein:2016ibx}.}
\end{table}

We also raised the possibility that  the  \(s\) quarks are heavy enough to accommodate tetraquarks. Then we again predict a state that would decay to \(\Lambda\bar{\Lambda}\), and a trajectory of its excited states. We then predict again a trajectory of states beginning with a near-threshold state. The masses are in table \ref{tab:pred_y_s}.

For the predictions, we take a ground state mass of
\be M(Y_s) \approx (2M(\Lambda)+40)\pm40\MEV \ee
with the usual slope for light quark (\(u\), \(d\), \(s\)) hadrons of 0.9 \GEV in the \((J,M^2)\) plane. In the \((n,M^2)\) the slope is lower, and we take 0.8 \GEV. The mass of the \(s\) quark is taken to be 220 MeV.

\begin{table}[h!] \centering
	\begin{tabular}{|c|c|} \hline
		\(n\)	&	 Mass 	\\ \hline\hline
		``-2'' & 1580\plm40	\\ \hline
		``-1'' & 1960\plm40	\\ \hline\hline
		0 & 2270\plm40	\\ \hline
		1 &	2540\plm40 		\\ \hline
		2	&	2780\plm40 		\\ \hline
		3	&	3000\plm40 		\\ \hline
		4	&	3210\plm40 		\\ \hline
	\end{tabular} \qquad\qquad
	\begin{tabular}{|c|c|} \hline
		\(J^{PC}\) &	  Mass 	 \\ \hline\hline
		\(1^{--}\) & 2270\plm40			\\ \hline
		\(2^{++}\) & 2510\plm40 		\\ \hline
		\(3^{--}\) & 2730\plm40 		\\ \hline
		\(4^{++}\) & 2930\plm40			\\ \hline
		\(5^{--}\) & 3120\plm40 		\\ \hline
	\end{tabular} 
	
	\caption{\label{tab:pred_y_s} Predictions for the states of the \(Y_s\), a tetraquark containing \(\ssb\) and decaying to \(\Lambda\bar{\Lambda}\).}
\end{table}
 

\section{Summary and open questions} \label{sec:summary}
In spite of  five decades of research, the story of the strong decays of mesons and baryons has not yet been fully deciphered. One does not know who to determine the decay width from the underlying theory of QCD. There are several theoretical frameworks like the quark model, potential models, chiral Lagrangians, and the Skyrme model, but it is fair to say that there is no well defined prescription to compute the total or partial decay width of the strong decays of hadrons. Much more is known about weak decays.

The aim of this paper has been to check to what extent can a stringy model of hadrons account for their strong decay processes. The spectra of hadrons and the corresponding (modified) Regge trajectories are a very strong signal of the stringy nature of hadrons. The question has been whether one can reach a similar conclusion from the analysis of the decay processes. We believe, though not in the same strength as for the spectrum, that the moral of this paper is that the decays of hadronic states tell us that indeed hadrons are strings.

This statement is based on three ingredients: (i) The linearity relation between the decay width and the length of the string. (ii) The exponential suppression factor associated with the creation of a pair that accompanies the breaking of the string into two strings. (iii) The constraints due to approximated symmetries like isospin baryon number and flavor $SU(3)$ which are realized in the stringy description.  It is the combination of the three ingredients that indicates that the basic structure of a hadron both a meson and a baryon is that of the string.
 
In this paper we have used two string frameworks. We make use of strings residing in a holographic confining background in critical dimensions, and also of the HISH model of strings in flat four spacetime dimensions. The holographic string is the parent of the HISH. For  certain properties like the linearity with the length we can use both but for the exponential suppression we can derive it only in the holographic setup and then assume it in the HISH model. From the HISH perspective a hadron is a system composed of a string and particles on its ends. Therefore one can contemplate decays of the string and decays of the endpoints. The main theme of this paper is that strong decays of hadrons are the decays of the string or more precisely the breaking apart of the string. On the other hand, since the electromagnetic and flavor charges are been carried by the endpoints, the electromagnetic and weak decays seem to correspond to processes where the string is unchanged and only the endpoints undergo certain transition. In section \ref{sec:decay_vertical} we have speculated that these processes may be related to decays of the vertical segments of the holographic strings. This will be examined in our future research.

Since the spin and flavor degrees of freedom are carried by the endpoints, the decays of the hadronic strings are affected by the spin and flavor configuration of the endpoints.

As was stated in the introduction we have made a special effort to confront any theoretical prediction we have made with experimental data. For getting results in agreement with the data is the purpose of the theoretical analysis. Moreover, we would like to argue that there should be a dual relation and that the experimental picture should serve as guiding lines for constructing the theoretical model. 

Comparing the neat theoretical picture of linearity in the string length to real world hadrons is not a straightforward task. The basic assumption of linearity is a long string result, which we expect to hold best at high spins/energies. On the other hand, we see that low spin hadrons still display stringy behavior in their belonging to Regge trajectories. Thus we are motivated to include hadrons down to the ground state in our fits of the decay widths.

To do that we need to introduce two components to the theory. First is the notion of a zero point length to the string. This can be seen as the effect of the negative quantum intercept, which also had to be introduced in describing the Regge trajectories. We saw that the effect of the intercept is can be thought of as a repulsive Casimir force, giving it a finite length, and hence finite mass and width, even when it is not rotating.

Incorporating a ``zero-point length'' of the string, in the form of a contribution from the quantum intercept to the string length, is necessary in fitting the decay width of the various hadrons, just as its introduction is necessary in describing the experimental Regge trajectories. In the case of the trajectories, we need a negative intercept to ensure there are no tachyonic states, and strings with zero angular momentum have mass. Here, in the case of the decay widths, we need the same negative intercept to account for the non-zero decay width of these same zero angular momentum mesons. Another option is to exclude the zero angular momentum states, but for many of these states, such as the \(\rho\), \(\omega\), or \(\phi\) mesons, there is no reason to suppose that the mechanism by which they decay is not that of a string of finite length tearing. We have shown, by fitting the data, that a good and simple way to incorporate this zero-point length is by replacing \(J\) with \(J-a\) in all the classical expressions, just as it is possible to describe the Regge trajectories after making the same replacement.

The inclusion of the intercept is well motivated by both the Regge trajectories and the fits to the decay widths, but we had to take a rather heuristic approach by adding it by hand. Moreover, we needed in some cases to insert a suppressive kinematic factor to account for the limited phase space some low spin hadrons have to decay (notably the \(\phi\) meson). The phase space factor can be thought of as picking a specific form of corrections for short strings.

The decay widths of meson and baryons were fitted using the model where they are proportional to the length of a string with endpoint masses. We used parameters obtained from the Regge trajectories for the quark masses, string tension (related to the Regge slope), and quantum intercepts. These results are reproduced here in appendix \ref{app:spectrum}. 

The results of the decay width fits are divided in terms of the goodness of the fits obtained. For the mesons, the fits, whose results were summarized in table \ref{tab:mesons}, yielded some good results, excluding some exceptionally wide states. The states that were excluded, \(\rho\), \(h_1(1170)\), and \(\pi_2(1670)\), were much wider relative to their masses compared with other states on their respective Regge trajectories.

An interesting result of the fits is the potential universality of the dimensionless factor \(A\), defined to be equal to the asymptotic ratio of width to mass, that is the ratio \(\Gamma/M\) at large \(J\). This ratio was the only fitting parameter that was not determined from the spectrum.

One can examine the values of \(A\) obtained from the fits and compare them to one another, hoping to find a more or less universal width to string length ratio. This ratio would be determined ultimately by the string coupling constant of QCD, up to numerical constants. For the light mesons and the \(K^*\), the fitted value of \(A\) is always between 0.10 and 0.12. For the heavier \(\phi\), \(D\), and \(D^*_s\), there is a preference for lower values, as \(A\) is 0.07--0.08. Averaging the results from the meson trajectories we find
\be A = 0.095\pm0.015\,.\ee
The cases where \(A\) differs from the mean are also the cases where we had to normalize the widths by phase space factors. It is possible that by tweaking the input values of endpoint masses and corresponding intercepts, or doing the phase space normalization differently, we could reduce the variance in \(A\). It should be noted that the phase space factors we insert do help in reducing this variance. If we had done the fits to the widths without normalizing by these factors, we would get even smaller values of \(A\) for the \(\phi\), \(D\), and \(D^*_s\).

The baryons were fitted with the same fitting model as the mesons, but with less success. In particular, it seems that the decay width of  baryons with light ``quarks'', namely the excited \(N\) and \(\Delta\) baryons, rises much faster with the mass than the string length. The string model cannot account for the large reported width of the high spin baryons. On the other hand, the measurements of these widths are far from precise. For example, the PDG estimates the width of the \(N(2600)11/2^{-}\) to be ``500 to 650 to 800'' MeV, based on two cited measurements claiming widths of 400\plm100 MeV and 1311\plm996 MeV. The string model with the parameters given in section \ref{sec:pheno_baryons} would predict the width of the \(N(2600)\) to be 210 MeV.

The second component investigated was the exponential suppression factor of pair creation. In particular, we looked at the strangeness suppression factor \(\lambda_s\), defined as the ratio between the probability that an \(\ssb\) pair will be created when the string tears, and the probability for light quark pair creation. The canonical value from jet experiments and used in event generators based on the stringy Lund model is \(\lambda_s \approx 0.3\) \cite{PDG_fragmentation}. From looking at branching ratios of stringy hadronic decays we found
\be \lambda_s = 0.40\plm0.21\,.\ee
This was the value obtained after averaging the results from multiple hadrons in section \ref{sec:pheno_lambdas}. We hoped to see \(\lambda_s\) as a constant between any two channels differing only by the mass of the created \(q\bar q\) pair, as the string model predicts. This is not what we see, as we find a large variance in the different measurements, although there are again large uncertainties in most measurements. The above result was supplemented by a more precise measurement from the radiative decays of \(\Psi\) and \(\Upsilon\) mesons, where we measure
\be \lambda_s = 0.36\plm0.09\,. \ee
Using this value we also predict a distinguishing decay pattern of the glueball, giving numerical estimates (section \ref{sec:pheno_glueball}) for the ratio between channels in which two, one, or no \(\ssb\) pairs are created in the decay of the closed string. Since glueball decays are usually assumed to be flavor blind, this pattern can serve as a test to the stringy nature of the glueball.

It is evident that the study of the decay width of hadrons via a stringy mechanism has still a long way to go. There are many follow up questions to the research presented in this paper. Here we list several of them.
\begin{itemize}
\item
As was summarized above the fits with experimental data indeed admit  for mesons a  linear dependence of the decay width on the length. For baryons the story is more subtle. Whereas for the $\Sigma$ and $\Lambda$ baryons the fits are reasonably good, for the $\Delta$ and even more for the nucleons it seems that the states measured have a larger width than the theoretical model predicts. This is a puzzle for us and currently we do not know how to explain it. We intend in the future to study the methods used to extract the width from the experimental measurements  and to study what are the predictions of other theoretical models.   
\item
Confronting the exponential suppression  of the decay width for creation of a quark-antiquark pair with experimental data was found in section \ref{sec:pheno_lambdas} to be quite limited due to the fact that there are very few cases of decays of hadrons both by creating a pair of light quarks and by creating a pair of medium (i.e. strange) quarks. However, as was discussed in  subsection \ref{sec:decay_multi}, there is another type of experimental phenomenon for which the mechanism of string breaking  is relevant and this is string  fragmentation and the creation of jets. It is well known that event generators such as Pythia incorporate the exponential suppression factors \cite{Sjostrand:2006za}. As was mentioned in subsection \ref{sec:decay_multi}, though  the exponential suppression   is  common to the  different multi-breaking mechanisms, there are also major differences. In particular the question of the direction of the momenta of the stringy fragments relative to the coordinate along which the string is stretched. The implications of these difference have to be studied together with an attempt to find better characteristics to the various types of jet, namely gluon versus quark jets. We intend to study these questions and  develop a model for the jet  formation based on the HISH model. 
\item
It was mentioned in the introduction and also in \ref{sec:decay_vertical} that the exponential suppression is based on the breaking apart of only  the horizontal segment of the holographic string. This immediately raises the question of what is the role of possible breaking  of the vertical segments. In \ref{sec:decay_vertical} we sketched several possible  hadronic decay processes that associates with the breaking of the vertical segment. One very basic property of this breaking is that the  hadronic string state before and after the vertical 
breaking do not have the same flavor properties. This is obviously the nature of weak decays but  the determination of  whether  they can be related and if yes in what way   is left for a future work.
\item
The stringy decays correspond mainly to strong interaction decays. In section \ref{sec:pheno_rad} we have also discussed their implications on radiative decays but we have nothing to say about weak interactions. If, as we believe, hadrons and in particular excited ones are indeed strings, there should be also a stringy description of the weak decays of hadrons. Very little, if at all, was done about this question. It might be related to the previous open question.
\item
So far the HISH program has tackled the issues of the spectra of hadrons and in this paper the decay processes of them. A very important question that has not been addressed yet and is in fact related to the decay processes is that of the creation mechanisms of the hadrons. One has to develop an understanding of the creation of hadron resonances both via hadron collider like the LHC and $e^+ e^-$ machines. In the QCD picture the former is taken to be  collisions that involve gluons and quarks. In the HISH picture it is a scattering of two strings producing many strings. In this manner string resonances can be produced and these are the hadron states whose decay widths we have analyzed in this paper. For $e^+  e^-$ collisions as a creator of a hadrons the HISH program has to be enlarged to  involve also leptons.  
\item
As was emphasized several times the main idea behind this paper is the fact that hadrons can be described by a single string  and correspondingly its decay associates with the breaking of the string. As was discussed in \cite{Sonnenschein:2016ibx} a priori there could be other types of stringy hadronic matter built from more complicated string configurations. A simple example, that was discussed in section \ref{sec:decay_Y} is the Y-shape baryon. We have argued that all the states that furnish nicely the modified Regge trajectories do not have such a structure. However, there are certain states that do not fit the trajectories and for them one should also examine the possibility of different string layouts. The analysis of section \ref{sec:baryon_structure} can be extended to further investigate the possibility to extract essential information on the structure of a given baryon from its decay channels and branching ratios.

\end{itemize}
 
\section*{Acknowledgments}
We would like to thank Ofer Aharony, Michael Green, Vadim Kaplunovsky, Marek Karliner,  Shmuel Nussinov, Kasper Peteers,  Amit Sever, Shimon Yankielowicz, and  Marija Zamaklar for useful discussions. We thank Ofer Aharony for his remarks on the manuscript. J.S would like to thank the theory group of Imperial College London and the Leverhulme trust for supporting his stay at Imperial College where part of this work has been carried out. This work was supported in part by a center of excellence supported by the Israel Science Foundation (grant number 1989/14), and by the US-Israel bi-national fund (BSF) grant number 2012383 and the Germany Israel bi-national fund GIF grant number I-244-303.7-2013.

\bibliographystyle{JHEP}
\bibliography{Decays}

\clearpage
\appendix

\clearpage
\section{The states used in the fits} \label{app:spectrum}
In this appendix we reproduce the results of the Regge trajectory fits of \cite{Sonnenschein:2014jwa} and \cite{Sonnenschein:2014bia}, and list 
the masses and decay widths calculated for the different hadrons from the string model.

The widths are calculated using the model of section \ref{sec:fitting_model}. The masses are calculated using the Regge trajectory equations for the rotating string with massive endpoints of section \ref{sec:rotating_string}, with the intercept added by taking \(J \to J-a\) in the classical expressions. This is the fitting model that was used in \cite{Sonnenschein:2014jwa,Sonnenschein:2014bia}

The Regge slope is \(0.88\) GeV\(^{-2}\) in all the meson fits, and \(0.95\) GeV\(^{-2}\) iin the baryon fits. The quark masses are common to all hadrons and are:
\be \qquad m_{u/d} = 60 \MEV\,,\qquad m_s = 400 \MEV\,,\qquad m_c = 1500 \MEV\,. \ee
For baryons it is assumed that a diquark mass containing some quark \(q\)and light quark is \(m_{qd} = m_{qu} = m_q\).

The remaining parameters are the different intercepts \(a\) and the proportion constant between width and length \(A\). The vary per trajectory and were listed in tables \ref{tab:mesons} (for mesons) and \ref{tab:baryons} (for baryons).

\begin{figure}[ht!]
			\centering
				\includegraphics[ width=0.85\textwidth]{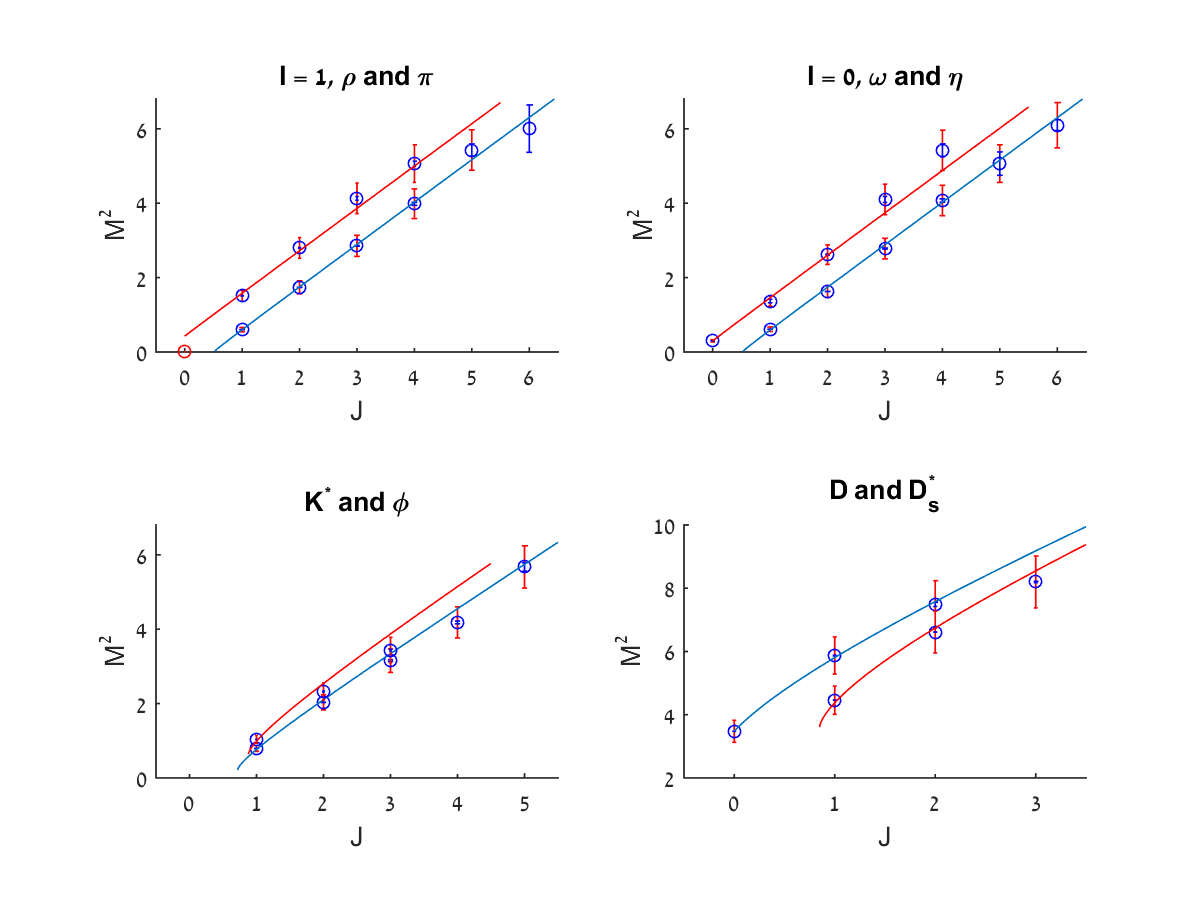}
				\caption{\label{fig:spectrum_mesons} The Regge trajectories of mesons fitted using the common slope \(\alp = 0.88\) GeV\(^{-2}\) and quark masses \(m_{u/d} = 60\) MeV, \(m_s = 400\) MeV, \(m_c = 1500\) MeV. \textbf{Top-left:} \(\rho\) (blue) and \(\pi\) (red) trajectories. Note that the pion itself is excluded. \textbf{Top-right:} \(\omega\) (blue) and \(\eta\) (red). \textbf{Bottom-left:} \(K^*\) (blue) and \(\phi\) (red). \textbf{Bottom-right:} \(D\) (blue) and \(D^*_s\) (red).}
	\end{figure}

 The Regge trajectories of mesons are plotted in figure \ref{fig:spectrum_mesons}. The results of the Regge trajectory and decay width fits are in tables \ref{tab:all_mesons} and \ref{tab:all_baryons}.

	\clearpage

\begin{table}[tpb] \centering
	\begin{tabular}{|c|c|l|l|l|l|l|} \hline

		Traj. & \(J^{PC}\) & State & Mass & Mass & Width & Width \\
		 &  &  & (Exp.) & (Calc.) & (Exp.) & (Calc.) \\ \hline\hline
		
		\(\pi/b\) & \(1^{+-}\) & \(b_1(1235)\) & 1229.5\plm3.2 & 1257 & 142\plm9 & 122\\

		          & \(2^{-+}\) & \(\pi_2(1670)\) & 1672.2\plm3.0 & 1650 & (260\plm9) & 136 \\

		          & \(3^{+-}\) & \(b_3(2030)\) & 2032\plm12 & 1965 & 181\plm11 & 193\\

							& \(4^{-+}\) & \(\pi_4(2250)\) & 2250\plm15 & 2236 & 215\plm15 & 220 \\ \hline

		\(\rho/a\)& \(1^{--}\) & \(\rho\) & 775.5\plm0.3  & 773 & (146.2\plm0.7) & 68 \\

							& \(2^{++}\) & \(a_2(1320)\) & 1318.3\plm0.6 & 1322 & 105.0\plm1.9 & 123\\

							& \(3^{--}\) & \(\rho_3(1690)\) & 1688.8\plm2.1 & 1700 & 161\plm10 & 160 \\

							& \(4^{++}\) & \(a_4(2040)\) & 1996\plm10 & 2008 & 255\plm28 & 190\\

							& \(5^{--}\) & \(\rho_5(2350)\) & 2330\plm35 & 2273 & 400\plm100 & 216\\

							& \(6^{++}\) & \(a_6(2450)\) & 2450\plm130 & 2511 & 400\plm250 & 239 \\ \hline

		\(\eta/h\)& \(0^{-+}\) & \(\eta\) & 547.9\plm0.0 & 548 & 0.0 & - \\

							& \(1^{+-}\) & \(h_1(1170)\) & 1170\plm20 & 1207 & (360\plm40) & 128 \\

							& \(2^{-+}\) & \(\eta_2(1645)\) & 1617\plm5 & 1613 & 181\plm11 & 171\\

							& \(3^{+-}\) & \(h_3(2025)\) & 2025\plm20 & 1934 & 145\plm30 & 206 \\

							& \(4^{-+}\) & \(\eta_4(2330)\) & 2328\plm38 & 2209 & 240\plm90 & 235 \\ \hline

		\(\omega/f\)&\(1^{--}\) & \(\omega\) & 782.6\plm0.1 & 769 & (8.5\plm0.1) & 78 \\

							& \(2^{++}\) & \(f_2(1270)\) & 1275.1\plm1.2 & 1320 & 185.1\plm2.9 & 149 \\

							& \(3^{--}\) & \(\omega_3(1670)\) & 1667\plm4 & 1698 & 168\plm10 & 196 \\
							
							& \(4^{++}\) & \(f_4(2050)\) & 2018\plm11 & 2006 & 237\plm18 & 234 \\

							& \(5^{--}\) & \(\omega_5(2250)\) & 2250\plm70 & 2272 & 320\plm95 & 266 \\

							& \(6^{++}\) & \(f_6(2510)\) & 2469\plm29 & 2510 & 283\plm40 & 295 \\ \hline
							
		\(K^*\) & \(1^-\)    & \(K^*\) & 891.7\plm0.3  & 876 & 50.8\plm0.9 & 44 \\
           
					& \(2^+\)    & \(K^*_2(1430)\) & 1425.6\plm1.5 & 1451 & 98.5\plm2.7 & 107 \\

            & \(3^-\)   & \(K^*_3(1780)\)  & 1776\plm7 & 1828 & 159\plm21& 147\\

           & \(4^+\)    & \(K^*_4(2045)\) & 2045\plm9 & 2133 & 198\plm30& 178\\ 

         &	 \(5^-\)    & \(K^*_5(2380)\) & 2382\plm24 & 2396 & 178\plm49 & 205\\ \hline

	   \(\phi/f'\) & \(1^{--}\) & \(\phi(1020)\) & 1019.5\plm0.0 & 995 & 4.3\plm0.0 & 3.5\\

	            	& \(2^{++}\) & \(f_2^\prime(1525)\) & 1525\plm5 & 1594 & 73\plm6 & 71 \\

           	& \(3^{--}\) & \(\phi_3(1850)\)  & 1854\plm7 & 1967 & 87\plm28 & 103 \\ \hline

	    	\(D\)   & \(0^-\)    & \(D\) & 1867.2\plm0.07 & 1879 & 0.0 & - \\

	            	& \(1^+\)    & \(D_1(2420)\) & 2422\plm1.5 & 2388 & 28.4\plm4.3 & 27 \\ 

           	& \(2^-\)    & \(D_J(2740)^0\) & 2737\plm12 & 2715 & 73\plm28 & 50 \\ \hline

		  	\(D^*_s\) & \(1^-\)    & \(D^*_s(2112)^\pm\) & 2112.1\plm0.4 & 2106 & $<1.9$ & - \\

	            	& \(2^+\)    & \(D^*_{s2}(2573)\) & 2569.1\plm0.8 & 2577 & 16.9\plm0.8 & 21 \\

	            	& \(3^-\)    & \(D^*_{s3}(2860)\) & 2860.5\plm7.1 & 2888 & 53\plm10 & 46 \\ \hline
	\end{tabular}
	\caption{\label{tab:all_mesons} The mesons used in the decay width fits trajectory fits of section \ref{sec:pheno_mesons}, with their calculated widths and masses (in MeV). Values marked with () in the exp. columns were not fitted. Values absent from the calc. column are not described by our model. All data from PDG \cite{PDG:2016}, except the data for \(D_J(2740)^0\) \cite{Aaij:2013sza}. Further explanations on the selection of states for the fits are in \cite{Sonnenschein:2014jwa}.}
\end{table}

\begin{table}[tpb] \centering
	\begin{tabular}{|c|c|l|l|l|l|l|} \hline

		Traj. & \(J^{P}\) & State & Mass & Mass & Width & Width \\
		 &  &  & (Exp.) & (Calc.) & (Exp.) & (Calc.) \\ \hline\hline

		\(N\) (even) & \jph{1}{+}  & \(n/p\) & 939.8 & 936 & 0 & - \\ 
					 & \jph{5}{+}  & \(N(1680)\) & 1680--1690 & 1734 & 120--140 &  136 \\ 
					 & \jph{9}{+}  & \(N(2220)\) & 2200--2300 & 2264 & 350--500 & 179 \\ \hline
					
		\(N\) (odd)	& \jph{3}{-}  & \(N(1520)\) & 1510--1520 & 1517 & 100--125 & 121	\\
		 & \jph{7}{-}  & \(N(2190)\) & 2100--2200 & 2103 & 300--700 & 169 \\ 
					 & \jph{11}{-} & \(N(2600)\) & 2550--2750 & 2557 & 500--800 & 207 \\ \hline
		
		\(\Delta\) (even) & \jph{3}{+}  & \(\Delta\) & 1232\plm2 & 1232 & 117\plm3 & 119 \\ 
		 & \jph{7}{+}  & \(\Delta(1950)\) & 1915--1950	& 1909 & 235--335 & 188 \\ 
					& \jph{11}{+} & \(\Delta(2420)\) & 2300--2500 & 2400 & 300--500 & 238 \\ \hline

		\(\Lambda\)	& 	\jph{1}{+}  & \(\Lambda(1116)\) & 1115.7 & 991 & 0 & - \\
		  & \jph{3}{-}  & \(\Lambda(1520)\) &  1519.5\plm1.0 & 1484 & 15.6\plm1.0 & 48	\\
					 & \jph{5}{+}  & \(\Lambda(1820)\) & 1815--1825 & 1831 & 70--90 & 63 \\
					 & \jph{7}{-}  & \(\Lambda(2100)\) & 2090--2110 & 2116 & 100--250 & 75	\\
					 & \jph{9}{+}  & \(\Lambda(2350)\) & 2340--2370 & 2364 & 100--250 & 86 \\ \hline
					
				\(\Sigma\) & 	\jph{1}{+} & \(\Sigma\) & 1193.2 & 1260 & 0 & -\\
					&  \jph{3}{-} & \(\Sigma(1670)\) & 1665--1685 & 1663 & 40--80 &70  \\
					& \jph{5}{+} & \(\Sigma(1915)\) & 1900--1935 & 1975 & 80--160 &88 \\ \hline
					
					\(\Sigma\) & \jph{3}{+} & \(\Sigma(1385)\) & 1384.6\plm0.6 & 1386 & 37.1\plm2.6 & 31 \\
					& \jph{5}{-} & \(\Sigma(1775)\) & 1770--1780 & 1755 & 105--135 & 145 \\
					& \jph{7}{+} & \(\Sigma(2030)\) & 2025--2040 & 2052 & 150--200 & 176 \\ \hline
	\end{tabular}
	\caption{\label{tab:all_baryons} The baryons used in the decay width fits trajectory fits of section \ref{sec:pheno_baryons}, with their calculated widths and masses (in MeV). Values absent from the calc. column are not described by our model. All data from PDG \cite{PDG:2016}. Further explanations on the selection of states for the fits are in \cite{Sonnenschein:2014bia}.}
\end{table}

\end{document}